\documentclass[a4paper,11pt]{article}
\pdfoutput=1 
\usepackage{jheppub} 
\usepackage[T1]{fontenc} 
\usepackage{placeins}
\usepackage{upgreek}
\usepackage{multirow}
\usepackage{multicol}
\usepackage{makecell}
\usepackage{amsmath}
\usepackage{mathtools}
\usepackage{float}
\usepackage{bm}
\usepackage{sectsty}
\usepackage{array}

\title{\boldmath Measurement of the central exclusive production of charged particle pairs in proton-proton collisions at $\sqrt{s} = 200$ GeV with the STAR detector at RHIC}
\collaborationImg{\hspace*{370pt}\includegraphics[height=40pt]{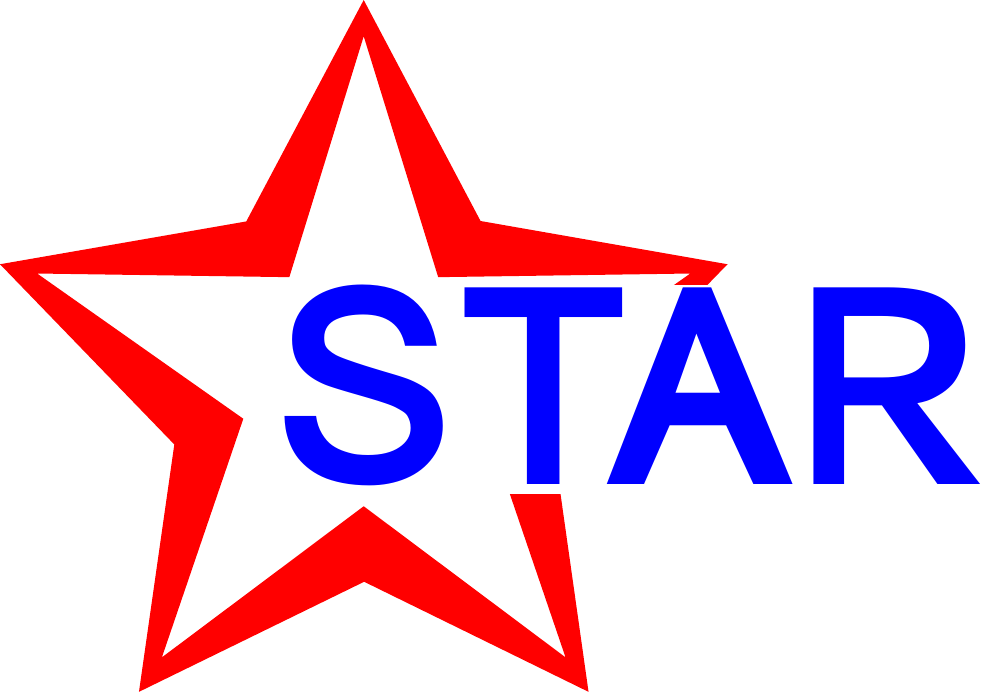}}
\collaboration{The STAR Collaboration\vspace*{-2em}}

\emailAdd{leszek.adamczyk@agh.edu.pl}

\allsectionsfont{\normalfont\large\bfseries\boldmath}
\newcolumntype{L}[1]{>{\raggedright\let\newline\\\arraybackslash\hspace{0pt}}m{#1}}
\newcolumntype{C}[1]{>{\centering\let\newline\\\arraybackslash\hspace{0pt}}m{#1}}
\newcolumntype{R}[1]{>{\raggedleft\let\newline\\\arraybackslash\hspace{0pt}}m{#1}}
\newcolumntype{P}[1]{>{\raggedright\arraybackslash}p{#1}}
\newcolumntype{T}[1]{>{\raggedleft\arraybackslash}p{#1}}

\abstract{We report on the measurement of the Central Exclusive Production of charged particle pairs $h^{+}h^{-}$ ($h = \pi, K, p$) with the STAR detector at RHIC in proton-proton collisions at $\sqrt{s} = 200$ GeV. The charged particle pairs produced in the reaction $pp\to p^\prime+h^{+}h^{-}+p^\prime$ are reconstructed from the tracks in the central detector and identified using the specific energy loss and the time of flight method, while the forward-scattered protons are measured in the Roman Pot system. Exclusivity of the event is guaranteed by requiring the transverse momentum balance of all four final-state particles. Differential cross sections are measured as functions of observables related to the central hadronic final state and to the forward-scattered protons. They are measured in a fiducial region corresponding to the acceptance of the STAR detector and determined by the central particles' transverse momenta and pseudorapidities as well as by the forward-scattered protons' momenta. This fiducial region roughly corresponds to the square of the four-momentum transfers at the proton vertices in the range $0.04\; \mbox{GeV}^2 < -t_1 , -t_2 < 0.2\; \mbox{GeV}^2$, invariant masses of the charged particle pairs up to a few GeV and pseudorapidities of the centrally-produced hadrons in the range $|\eta|<0.7$.
The measured cross sections are compared to phenomenological predictions based on the Double Pomeron Exchange (DPE) model. Structures observed in the mass spectra of $\pi^{+}\pi^{-}$ and $K^{+}K^{-}$ pairs are consistent with the DPE model, while angular distributions of pions suggest a dominant spin-0 contribution to $\pi^{+}\pi^{-}$ production. 
For $\pi^+\pi^-$ production, the fiducial cross section is extrapolated to the Lorentz-invariant region, which allows decomposition of the invariant mass spectrum into continuum and resonant contributions. The extrapolated cross section is well described by the continuum production and at least three resonances, the $f_0(980)$, $f_2(1270)$ and $f_0(1500)$, with a possible small contribution from the $f_0(1370)$.  
Fits to the extrapolated differential cross section as a~function of $t_1$ and $t_2$ enable extraction of the exponential slope parameters in several bins of the invariant mass of $\pi^+\pi^-$ pairs. These parameters are sensitive to the size of the interaction region.} 
\begin{document} 
\maketitle
\flushbottom
\section{Introduction}
\label{sec:intro}
The study of exclusive production of meson and baryon pairs has long been recognised as an important ground for Quantum Chromodynamics (QCD) tests. Exclusive production of pion and kaon pairs has been studied both theoretically~\cite{brodsky,terazawa,boyer} and experimentally, in two-photon collisions at lepton colliders~\cite{tpc,topaz,cleo,aleph} and via  photoproduction~\cite{pho_1, pho_2,rho_STAR,rho_CMS} and deep inelastic scattering~\cite{dis_1, dis_2} in lepton-proton and heavy-ion experiments. 
Exclusive production of meson and baryon pairs belongs to the class of Central Exclusive Production (CEP) processes. In hadron-hadron collisions, CEP processes can proceed via Double Pomeron Exchange (DPE), photon-Pomeron exchange or  photon-photon exchange, where the Pomeron is a colour-singlet object with internal quantum numbers of the vacuum, see, e.g.,~\cite{book_1, book_2}. 
Although several properties of diffractive scattering at high energies are described by the phenomenology of Pomeron exchange in the context of Regge theory~\cite{book_3}, the exact nature of the Pomeron still remains elusive.\\
\indent
This paper presents a measurement of CEP of $\pi^+\pi^-$, $K^+K^-$ and $p\bar{p}$ pairs in $pp$ collisions at a centre-of-mass energy of $\sqrt{s} = 200$ GeV with the STAR detector at RHIC. Differential and integrated cross sections are measured in a fiducial region and compared to phenomenological predictions. The fiducial region roughly corresponds to the square of the four-momentum transfers at the proton vertices in the range $0.04\,\mbox{GeV}^2 < -t_1 , -t_2 < 0.2\,\mbox{GeV}^2$ and invariant masses of the charged particle pairs up to a few GeV. Throughout the paper the convention $c=\hbar=1$ is used.
\section{Theoretical framework and current experimental situation}
\label{sec:theory}
\begin{figure}[tb]
\centering
\includegraphics[width=.44\textwidth]{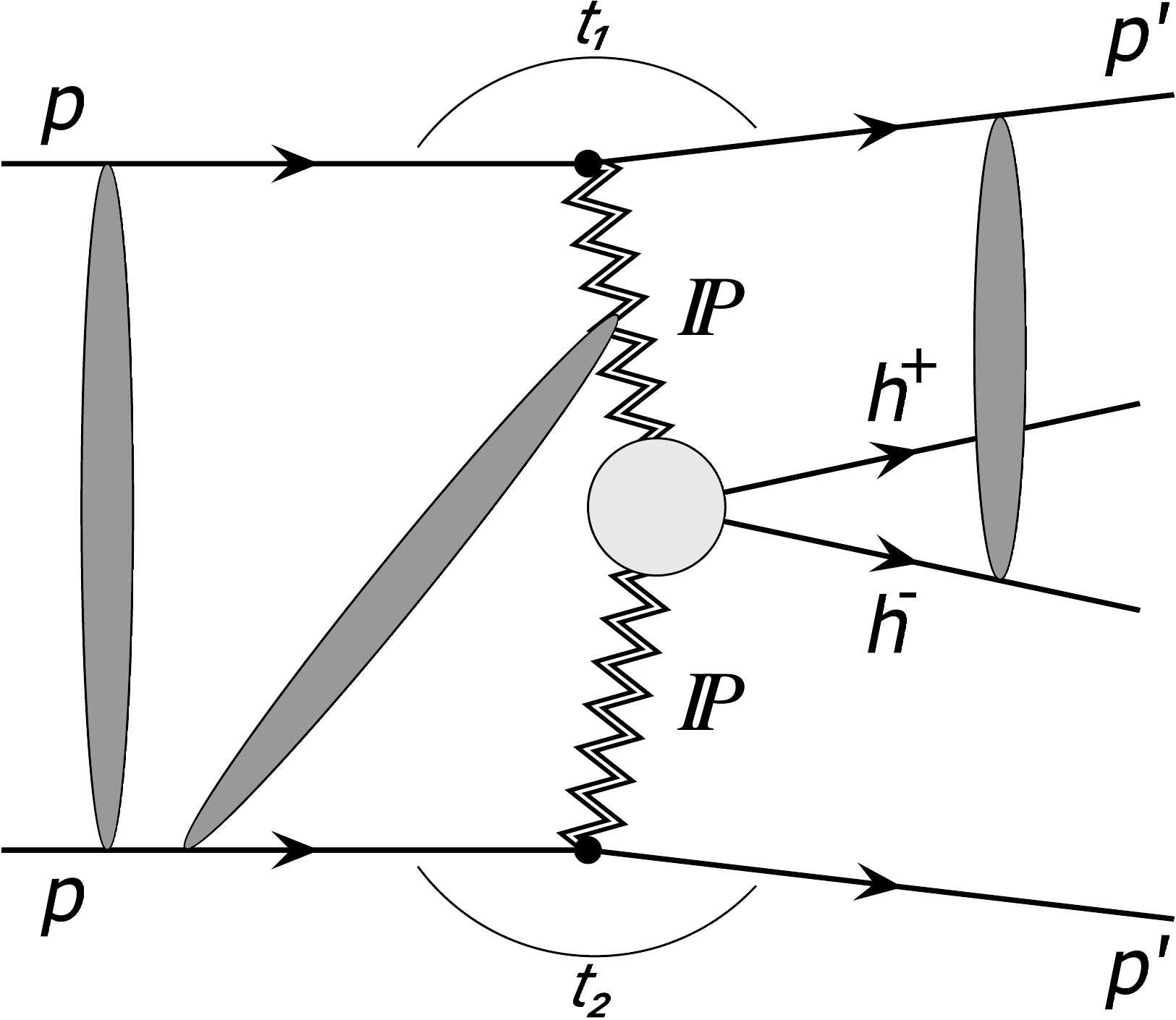}
\hfill
\includegraphics[width=.44\textwidth]{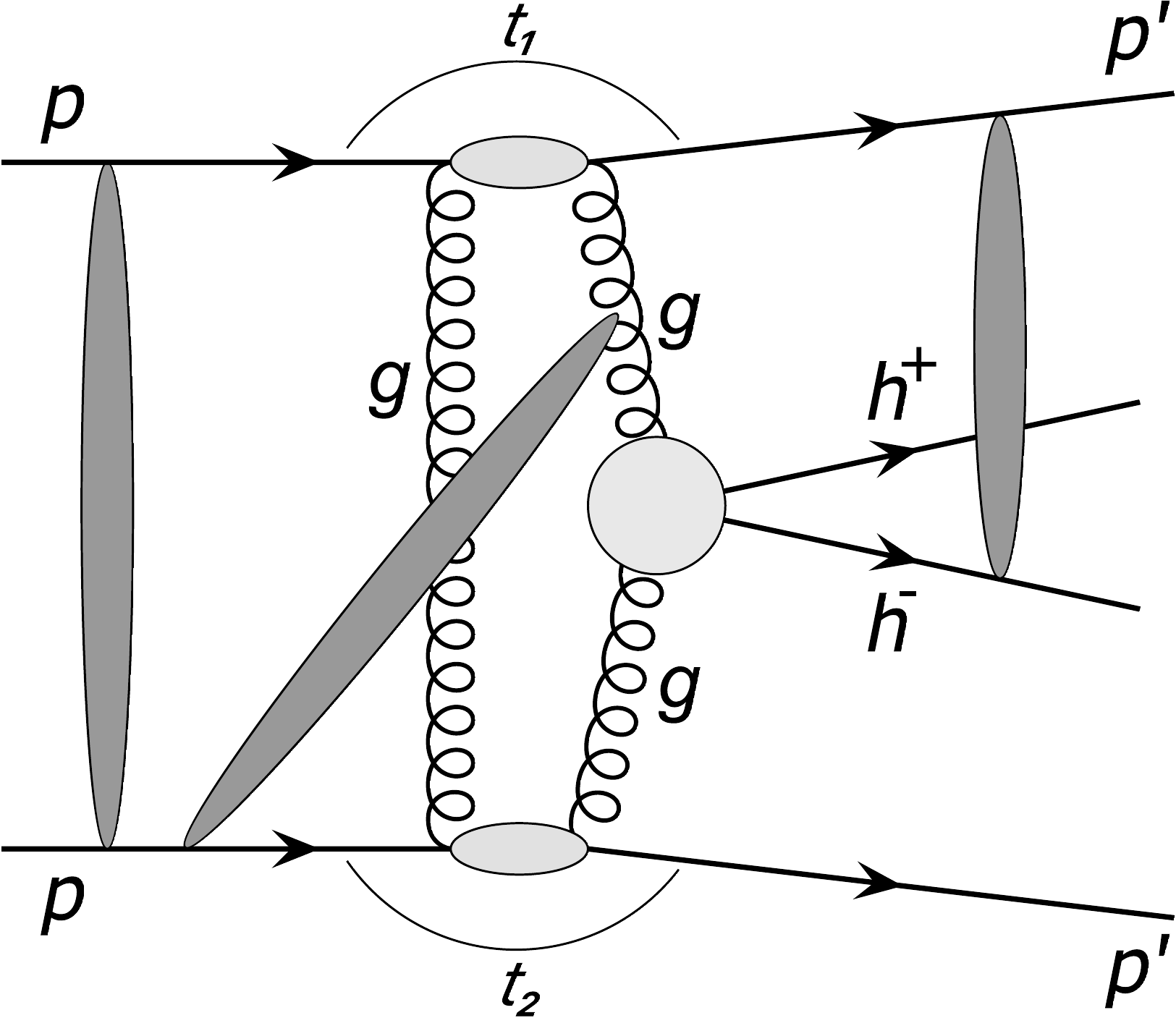}
\caption{(left) Generic diagram of CEP of $h^+h^-$ in DPE model. The scattered beam protons emerge intact from the collision and the charged particle pair is produced in the central rapidity region. (right) The two-gluon approximation of DPE in pQCD. The grey ovals represent some of the possible absorptive corrections.} 
\label{feyn_diagrams}
\end{figure}

Over the last decade, one could observe a renewal of interest in studies of CEP processes in high energy proton-(anti)proton collisions (see,  e.g.,~\cite{harland_lang_1,dime,albrow_1} for review and further references).
CEP processes in hadron-hadron collisions provide an especially clean environment to study the nature and quantum numbers (spin, parity) of centrally-produced resonance states~\cite{harland_lang_1}. In proton-proton collisions, the CEP reaction may be written in the form
\begin{equation}
pp\to p^\prime \oplus X \oplus p^\prime,
\end{equation}
where the $\oplus$ symbols denote the presence of large rapidity gaps which separate the final state system $X$ from the diffractively-scattered protons. This process in the DPE mode and with the hadronic final state, $X$, consisting of just an oppositely-charged particle pair is schematically shown in Fig.~\ref{feyn_diagrams}(left). In Fig.~\ref{feyn_diagrams}(right), its representation within perturbative QCD (pQCD) is shown in the two-gluon approximation. The scattered protons emerge intact from the collision at small polar angles with respect to the incoming beams and can be detected with special tagging devices. The final state, $X$, can be fully measured at central rapidities. The upper limit of the invariant mass, $M_X$, of the system $X$ 
depends on the energy of the colliding particles. At the LHC, this upper limit can reach above 100 GeV and the central production of even the Higgs or supersymmetric particles might be possible~\cite{CEPatLHC}. Recently, CEP of pion pairs in $pp$ collisions at $\sqrt{s}=5.02, 7$ and 13 TeV has been reported by the CMS experiment~\cite{cms_pipi,cms_pipi_2}. The LHCb experiment has measured photo-induced CEP of $J/\psi$ and $\psi(2S)$ states in $pp$ collisions at $\sqrt{s} = 13$ TeV~\cite{lhcb} and $\Upsilon(nS)$ $(n=1,2,3)$ states in $pp$ collisions at $\sqrt{s} = 7$ and 8~TeV~\cite{lhcb_2}. At the Tevatron, it was possible to study CEP of $\pi^+\pi^-$ pairs with invariant masses up to a few GeV in $p\bar{p}$ collisions at several centre-of-mass energies up to $\sqrt{s} =1.96$ TeV~\cite{albrow_1, cdf}. The above measurements were performed without forward-proton tagging. The experiments at CERN at the ISR~\cite{afs} and the SPS~\cite{wa91} have provided measurements of many CEP processes with forward proton tagging, however at significantly smaller centre-of-mass energies (62~GeV for ISR and 30~GeV for SPS).\\
\indent
In the present analysis, $X$ stands for either continuum or resonant production of $\pi^+\pi^-$, $K^+K^-$ or $p\bar{p}$ pairs in the non-perturbative regime up to around $M_X = 3$~GeV. 
At the relatively high centre-of-mass energy of $\sqrt{s}=200$ GeV, only small contributions from Reggeon exchange are expected. Contributions from photon-Pomeron and photon-photon processes are also not significant and are additionally suppressed at $-t>0.04$ GeV$^2$. Therefore, the DPE process is expected to be dominant. The DPE process can be regarded as a spin and parity filter, i.e., the $h^+h^-$ system must have even spin and positive parity, so only exclusive production of $f_0$ and $f_2$ resonances are allowed on top of the continuum in the studied $h^+h^-$ production. In general, the resonance and continuum contributions may interfere. Calculations of the hadron mass spectrum in this domain were first done for only the continuum production~\cite{lebiedowicz_2, LSModelKK, harland_lang_1} of $\pi^+\pi^-$ or $K^+K^-$ pairs using an approach based on Regge theory. In these models, the parameters of the Pomeron and sub-leading Reggeon exchanges were adjusted to describe the total and elastic $\pi p$ or $Kp$ scatterings. In this approach, the amplitude for the $p+p \rightarrow p^\prime+\pi\pi(KK)+p^\prime$ process is expressed in terms of the product of two amplitudes describing the interaction of each of the two protons with one of the two mesons. The intermediate meson form factor is parameterised with one of three functions: an exponential, $\exp{\left[(\hat{t}-m_h^2)/\Lambda^2_{of\!f}\right]}$, an Orear-like function, $\exp{\smash[b]{\left[-b(\sqrt{\smash[b]{-\hat{t}+m_h^2+a^2}}-a)\right]}}$, or a power-like function, $1/(1-(\hat{t}-m_h^2)/a_0)$, where $\hat{t}$ is the square of the four momentum transfer at the Pomeron-meson vertex and $\Lambda_{of\!f}^2$, $b$, $a$ and $a_0$ are free parameters. 
The models can be supplemented to include absorption effects (shown symbolically on Fig.~\ref{feyn_diagrams}) which are needed to calculate the `survival probability' for no additional soft re-scatterings between the colliding protons or the final state mesons. Absorption corrections are related to the non-perturbative interaction in the initial or final state of the reaction. The re-scattering leads to suppression of the cross section and distortion of the distributions of the kinematic variables. 
The suppression factor depends on the collision energy. They usually reduce the cross sections, even by a factor of 5 at RHIC energy and a factor of 10 at LHC energies~\cite{LSAbsorption}. Recently, the production of a variety of resonances: the $f_0(500)$, $f_0(980)$ and $f_2(1270)$ decaying to $\pi^+\pi^-$, the $f_0(980)$, $f_0(1500)$, $f_0(1710)$, $f_2(1270)$ and $f_2^\prime(1525)$  decaying to $K^+K^-$, and the $f_0(2020)$, $f_0(2100)$ and $f_0(2200)$ decaying to $p\bar{p}$, were studied theoretically ~\cite{lebiedowicz_1,lebiedowicz_4,lebiedowicz_5} including interference effects between resonant and non-resonant amplitudes. The calculations are based on a tensor Pomeron model.
The amplitudes for the processes are formulated in terms of vertices, respecting the standard crossing and charge-conjugation relations of Quantum Field Theory. 
In recent work~\cite{schiecker}, the authors also consider resonant CEP of $\pi^+\pi^-$ through Pomeron-Pomeron fusion ignoring the spin effects in the Pomeron-Pomeron-resonance vertices.\\
\indent
At relatively large $M_X$ ($>2$ GeV), perturbative QCD (pQCD) calculations can be performed~\cite{harland_lang_1}. In Fig.~\ref{feyn_diagrams}(right), the relevant diagram is shown, where the hard sub-process $gg\rightarrow X$ is initiated by gluon-gluon fusion and the second gluon is needed to screen the colour flow across the rapidity gap intervals. 
The cross section is calculated based on the generalised (skewed) unintegrated gluon densities of the protons~\cite{gpd}. The skewed unintegrated density can be obtained from the conventional integrated gluon densities~\cite{updf}. The hard scale needed for pQCD calculations should be larger than $\Lambda_\mathrm{QCD}$ and is typically given by the mass of the produced state~\cite{harland_lang_1}.  The pQCD predictions include the contribution from $\pi^+\pi^-$($K^+K^-$) produced both directly and from $\chi_{c0}$ decay.\\
\indent 
One important motivation for measuring DPE processes is to search for gluonic bound states, called glueballs, which are predicted by QCD due to its non-Abelian nature.
The properties of these compound objects offer a unique insight into the strong interaction, since the gluon self-interaction is exclusively responsible for the mass of glueballs. The search for these exotic states, and determining their possible role within the family of 
mesons, is a long-standing quest in hadron spectroscopy~\cite{glueball_2}. Glueballs are preferentially produced in gluon-rich processes such as $p\bar{p}$ annihilation~\cite{glueball_3, glueball_4}, the radiative decay of the $J/\psi$-meson~\cite{glueball_5}, and CEP processes~\cite{albrow_1, afs, harland_lang_1, harland_lang_3} in $pp(\bar{p})$. The absence of valence quarks in the production process makes CEP a favorable place to look for hadronic production of glueballs.
Lattice QCD calculations have predicted~\cite{glueball_2} the lowest-lying scalar glueball state in the mass range of $1000-1700$~MeV, and tensor and pseudo-scalar glueballs in the range of $2000-2500$~MeV. Experimentally measured candidates for scalar glueball states are the $f_0(1500)$ and the $f_0(1710)$ observed in central production as well as in other gluon-rich reactions. The glueballs are expected to be unstable and decay in diverse ways, yielding typically two or more mesons. The $f_0(1710)$ state decays into $K\bar{K}$ and the $f_0(1500)$ into $\pi\pi$ and $4\pi$. For the tensor meson sector, $\mathrm{I^GJ^{PC} = 0^+2^{++}}$, neither the established $f_2(1950)$ nor less well-established states, such as the $f_2(1910)$ and $f_2(2150)$, have been thoroughly explored. This is partly due to a small production cross section and partly due to not being able to clearly separate the Reggeon contribution. 
%
\section{Experimental setup}
\label{sec:detector}
The data used in this analysis were collected by the STAR experiment at RHIC~\cite{rhic} in 2015 in proton-proton collisions at $\sqrt{s} = 200$ GeV and correspond to an integrated luminosity of 14.2~pb$^{-1}$. A detailed description of the STAR detector is given in Ref.~\cite{star}.\\
\begin{figure}[t]
\centering
\includegraphics[width=.95\textwidth]{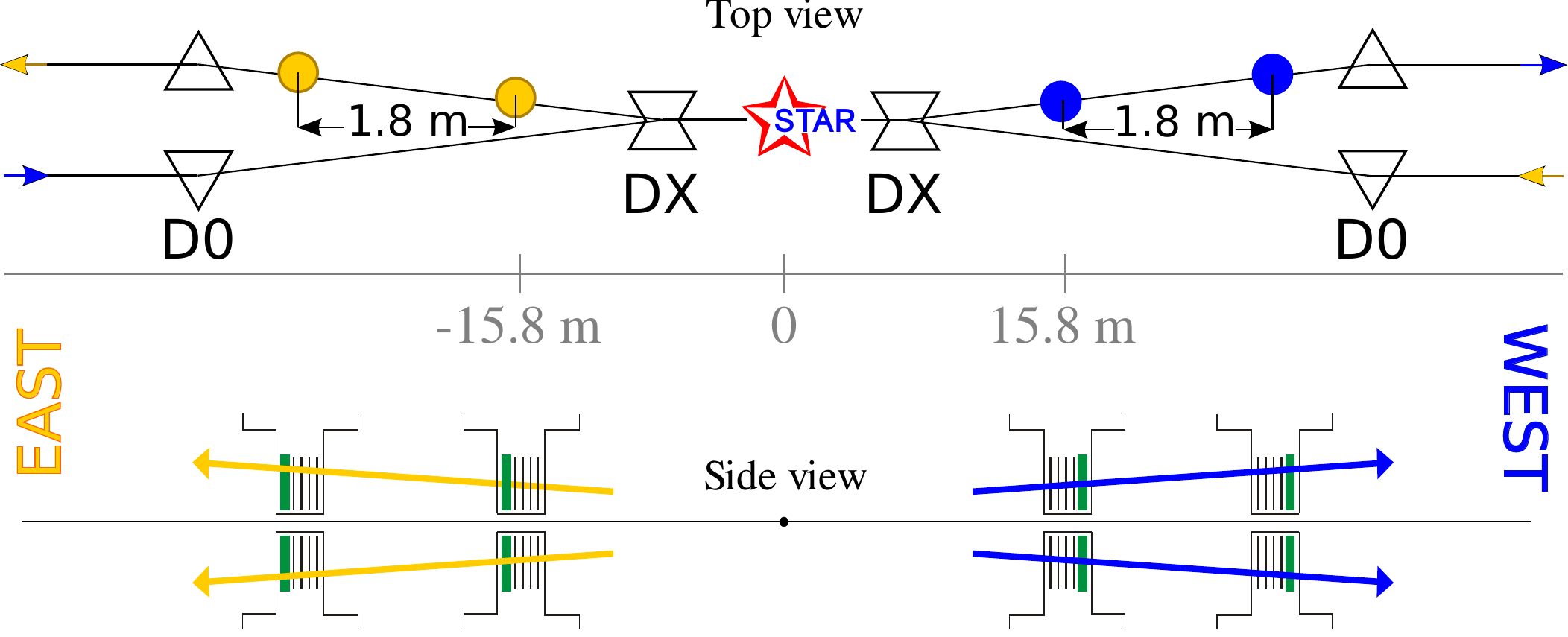}
\caption{(top) The layout of the beam-line elements outside of the STAR main detector (not to scale). Two sets of RPs are installed between the DX and D0 dipole magnets, at 15.8~m~and~17.6~m on each side of the nominal IP, respectively. (bottom) Each station consists of two RPs, one~above and one below the beam-line, housing four planes of silicon strip detectors and a scintillation counter.}
\label{rp_layout}
\end{figure}
\indent
The forward-scattered protons are measured in the Roman Pot (RP) system adopted from the pp2pp experiment \cite{pp2pp} at RHIC. This is schematically shown in Fig.~\ref{rp_layout}(top), where the locations of the RPs are displayed together with the beam line elements. Silicon strip detectors are located in two stations on each side of the interaction point (IP) at distances of 15.8 m and 17.6 m, respectively. Each station has two RPs, one placed above and one below the beam-line, see Fig.~\ref{rp_layout} (bottom). The RPs are situated downstream of the DX dipole magnets responsible for head-on targeting of the incoming beams and for bending outgoing beams back into the respective accelerator pipelines. The constant and uniform magnetic field of the DX magnet works as a spectro\-meter, and thus knowledge of the scattered proton's trajectory allows reconstruction of its momentum. 
Each RP houses a package of 4 silicon strip detector planes - two with vertical and two with horizontal orientation of the strips - allowing measurement of the position of a proton in the transverse plane. The strip pitch is about 100~$\upmu$m, resulting in a spatial resolution of about 30~$\upmu$m. Scintillator counters placed inside each RP station allow for triggering on forward protons and also provide timing information with 0.5~ns \mbox{resolution.}\\
\indent
Measurement of a pair of charged particles produced in the final state at central rapidity is performed using the Time Projection Chamber (TPC)~\cite{star_tpc}, which provides tracking for charged particles in the 0.5~Tesla solenoidal magnetic field. The TPC covers the pseudorapidity range of $|\eta|<1.8$ in full azimuthal angle.\footnote{STAR uses a right-handed coordinate system with its origin at the nominal IP in the centre of the detector and the $z$-axis along the beam pipe. The $x$-axis points from the IP to the outside of the RHIC ring, and the $y$-axis points upward. Cylindrical coordinates $(r, \upvarphi)$ are used in the transverse plane, $\upvarphi$ being the azimuthal angle around the beam pipe. The pseudorapidity is defined in terms of the polar angle $\uptheta$ as $\eta = -\ln{[\tan{(\uptheta/2)}]}$. Transverse momentum is defined as $p_\mathrm{T}=p\sin{\uptheta}$.}
The TPC is used to determine the momenta of the charged particles, and also helps in locating the position of the collision vertex. The tracking efficiency is $\sim 85\%$ for $|\eta|<1$, but falls to $50\%$ at $|\eta| \sim 1.3$.
The measurement of the specific energy loss in the TPC gas, $dE/dx$, is used for particle identification. Furthermore, to extend the particle identification power of the STAR detector performed by the TPC, a~Time-Of-Flight (TOF) detector~\cite{star_tof} is placed around the TPC covering a pseudorapidity range of $|\eta|<0.9$.  
The TOF detector is a system of adjacent Multi-gap Resistive Plate Chambers. In addition to being a precise timing detector, it is used to measure event multiplicity at the trigger level and to discriminate TPC tracks arrived in preceding/posterior bunch crossings (out-of-time pile-up) from the in-time tracks. Using the TOF timing information together with the momentum and path length reconstructed in the TPC allows for the particle mass determination.\\
\indent
For the TOF efficiency study, an unbiased sample with tracks reconstructed using Heavy Flavor Tracker (HFT)~\cite{hft} hits in addition to TPC hits was analysed. This sample provides a clean source of in-time tracks. The HFT is a system of multi-layer silicon pixel and strip detectors. It improves the impact parameter resolution of the STAR tracking system and enables reconstruction of secondary decay vertices of open heavy flavour hadrons. \\
%
%
\indent
To suppress non-exclusive background, the Beam-Beam Counters (BBCs)~\cite{star_bbc} and the Zero-Degree Calorimeters (ZDCs)~\cite{star_zdc} are used. The BBCs are scintillator detectors placed in the endcap regions of the STAR detector and covering pseudorapidity ranges of $2.1<|\eta|<3.3$ (large BBC tiles) and $3.3<|\eta|<5$ (small BBC tiles).
The ZDCs are used to tag neutral particles which leave the interaction region close to the beam direction.
\FloatBarrier
\section{Event reconstruction}
\label{sec:reconstruction}
In order to reconstruct the position and momentum of the scattered protons, a clustering procedure is applied for each RP detector plane separately. A cluster is formed by a continuous series of Si strips with signals well above the pedestal. Pairs of matched clusters found in detector planes measuring the same coordinate define the $(x,y)$ coordinates of space points for a given RP. Correlating space points between RP stations, and reconstructing the proton kinematics, relies on an alignment procedure, which is carried out using elastic scattering data for each run separately, as described in Ref.~\cite{elastic_paper}.
A track is formed based on one or two points reconstructed in the two RP  detector stations on the same side of the IP. Using elastic scattering events reconstructed with tracks formed from the two points, the average transverse position of the primary vertex was measured to be $\langle x_\mathrm{\scriptscriptstyle IP}\rangle = 0.42\pm 0.04$~mm and $\langle y_\mathrm{\scriptscriptstyle IP}\rangle = 0.46\pm 0.05$~mm. 
With the average transverse position of the vertex and two-point proton tracks, the proton transverse momentum ($p_x, p_y)$ can be reconstructed, and hence the value of the Mandelstam variable $t$. For single point proton tracks, the transverse momentum is reconstructed assuming that the scattered proton energy is equal to the beam energy. Such an approximation is justified because the proton loses on average less than 1\% of its initial energy for events with $M_X<2$~GeV.\\
\indent
Particle pair identification is performed using the combined information from the TPC and TOF detectors for both tracks simultaneously.  
%
%
The compatibility of the track's $dE/dx$ with that expected for a given particle ($h=\pi$, $K$, $p$) is determined using the quantity
\protect \begin{equation}\label{eq:nSigmaDef} n\sigma_{h} = \frac{\ln{\left[(dE/dx)
/ (dE/dx)_{h}
\right]}}{\sigma}, \end{equation}%
where 
$(dE/dx)_{h}
$ 
is the Bichsel~\cite{Bichsel} expectation for particle type $h$ and $\sigma$ is the relative resolution of $dE/dx$ for a given track. From $n\sigma_{h}$ for each of the two tracks, the $\chi^{2}$ statistic for a $hh$ pair hypothesis is calculated:
\protect \begin{equation}\label{eq:chiSqDef}\chi^{2}_{dE/dx}(hh) = \left(n\sigma_{h,1}\right)^{2} + \left(n\sigma_{h,2}\right)^{2}. \end{equation}
The time at which a particle is detected in the TOF system is used to reconstruct its squared mass $m^{2}_\mathrm{\scriptscriptstyle TOF}$. For this purpose, the time of the primary interaction is required; however, it is not  known for CEP events. Instead, the unknown time of the primary interaction can be eliminated by assuming that both tracks present in an event are of the same type. In that case, the measured TOF time difference between particles is given by $\Delta t = L_1\sqrt{1+m_\mathrm{\scriptscriptstyle TOF}^2/p_1^2} - L_2\sqrt{1+m_\mathrm{\scriptscriptstyle TOF}^2/p_2^2}$, where $p_{1,2}$ are tracks' momenta and $L_{1,2}$ are the lengths of the helical paths between the primary vertex and TOF hit associated with them. $m^{2}_\mathrm{\scriptscriptstyle TOF}$ can be then calculated per event. If $m^{2}_\mathrm{\scriptscriptstyle TOF}$ is negative, due to the detector's resolution, then it is set to zero.
\FloatBarrier
\section{Monte Carlo simulation}
\label{sec:mc}
Monte Carlo (MC) simulations are used for the modelling of background contributions, unfolding of detector effects, calculation of systematic uncertainties and comparisons of models with the hadron-level cross section measurements.\\
\indent
The GenEx~\cite{genex} and DiMe~\cite{dime} event generators are based on simple phenomenological models~\cite{lebiedowicz_2, LSModelKK, harland_lang_1} of continuum production of $\pi^+\pi^-$ or $K^+K^-$ pairs.\\
%
\indent
In the DiMe event generator, four models for absorption are available. The prediction from "model 1", which is most consistent with data, is used in this analysis. DiMe predictions are also sensitive to the choice of meson form factor. Three different parameterisations of meson form factor are implemented. We chose an exponential form using $\Lambda_{of\!f}^{2}=1.0~\textrm{GeV}^{2}$, which fits the present data better. A larger value of $\Lambda_{of\!f}^{2}$ was used~\cite{harland_lang_1} to fit DiMe predictions to ISR data, however the 50\%  normalisation uncertainty of the ISR data does not exclude $\Lambda_{of\!f}^{2}=1.0~\textrm{GeV}^{2}$.\\ 
\indent
In GenEx, the absorption corrections are not taken into account. However, the model developers estimated the suppression factor to be of the order of $2-5$ ($\pi^+\pi^-$) and 2 ($K^+K^-$)~\cite{LSAbsorption}. To account for absorption, the $\pi^+\pi^-$ cross sections obtained from GenEx are scaled by 0.25 to fit DiMe predictions for masses above 0.8~GeV, while the $K^+K^-$ cross sections from GenEx are scaled by 0.45 to fit DiMe predictions for masses above 1.2~GeV. Above these limits, the absorption effects only weakly depend on pair mass. In the GenEx generator, we also use an exponential form for the meson form factor using $\Lambda_{of\!f}^{2}=1.0~\textrm{GeV}^{2}$.
Therefore the differences between GenEx and DiMe are almost entirely due to the absorption effects.\\
%
%
%
\indent
The MBR model~\cite{mbr_pythia8} implemented in PYTHIA8~\cite{pythia8} was tuned to describe the inclusive cross section for central diffraction (CD), $p+p\rightarrow p^\prime+X+p^\prime$, measured by the CDF experiment. In this model, the exclusive $h^+h^-$ final state occurs from fragmentation and hadronisation of the central state based on the Lund string model. The MBR model implemented in PYTHIA8.244 allows generation of the central state starting from the mass threshold of 
$0.5$~GeV, however the suggested value is 1.2~GeV. Therefore, PYTHIA8 expectations for very low masses are in question, but are shown for \mbox{completeness.} The obtained cross sections from PYTHIA8 are scaled by an arbitrary value of 0.25 for easier comparison with the data. \\ 
%
%
\indent
Single particle 
MC embedded into zero-bias data events were used to calculate the TPC and TOF reconstruction and matching efficiencies separately for: $\pi^+,\pi^-,K^+,K^-,p$ and $\bar{p}$. The GenEx sample of $p+p\rightarrow p^\prime+\pi^+\pi^-+p^\prime$ embedded into zero-bias data events was used to calculate the RP reconstruction efficiency for forward-scattered protons and also for closure tests of the full analysis chain. The inclusive CD and Minimum Bias (inelastic) PYTHIA8  samples without embedding were used for calculation of the impact of systematic uncertainties in the subtraction of non-exclusive background, as well as for the comparisons with the data. Prior to the embedding, MC samples were passed through a detailed GEANT3~\cite{geant3} simulation of the STAR central detector and the GEANT4~\cite{geant4} simulation of the beam optics and RP detectors. All MC samples were then subjected to the same reconstruction and analysis software as applied to the data.\\
\indent
A fast and simplified MC simulation was used for estimation of the central pair particle identification (PID) efficiency and misidentification probability. 
In this simulation, the $dE/dx$ and the times of detection of particles in the TOF detector were generated according to parameterisations obtained from the inclusive data and the full TPC/TOF simulations, while the amount of exclusive $\pi^+\pi^-$, $K^+K^-$  and $p\bar{p}$ were chosen to describe the data. 
\section{Data sample and event selection}
\label{sec:analysis}
\begin{figure}[tbp]
\centering
\includegraphics[width=.485\textwidth,page=1]{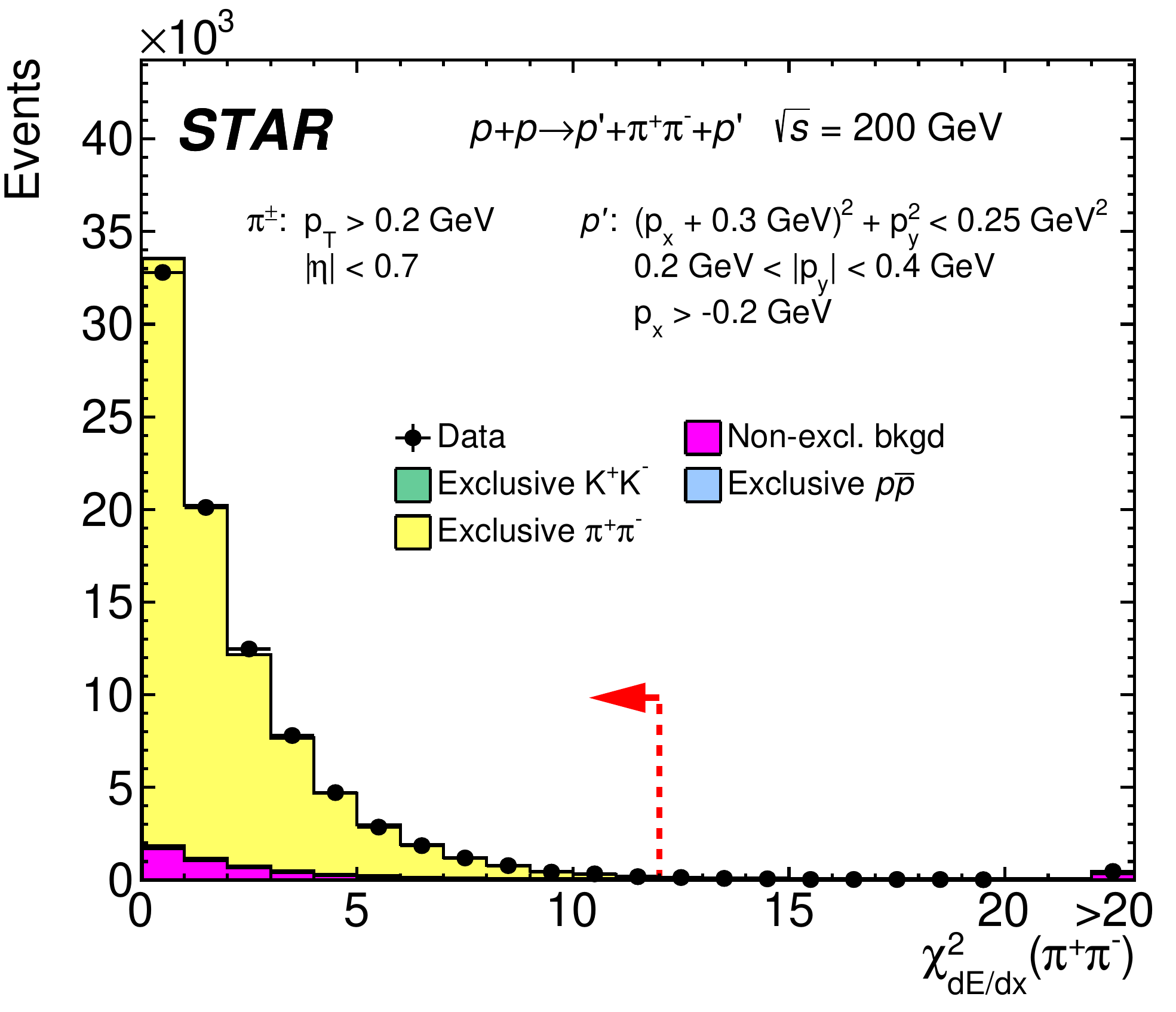}
\hfill
\includegraphics[width=.485\textwidth,page=1]{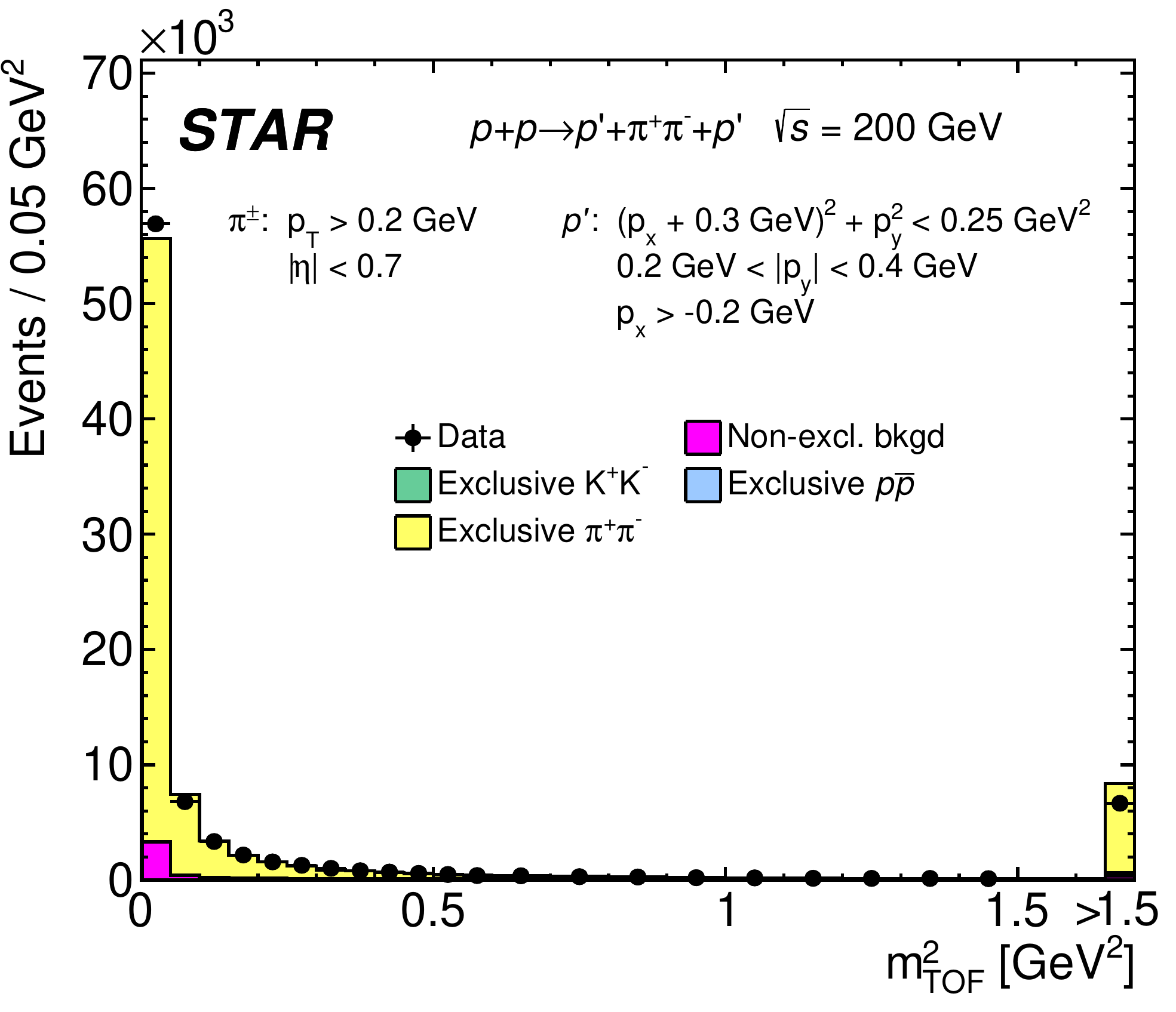}
\newline
\includegraphics[width=.485\textwidth,page=1]{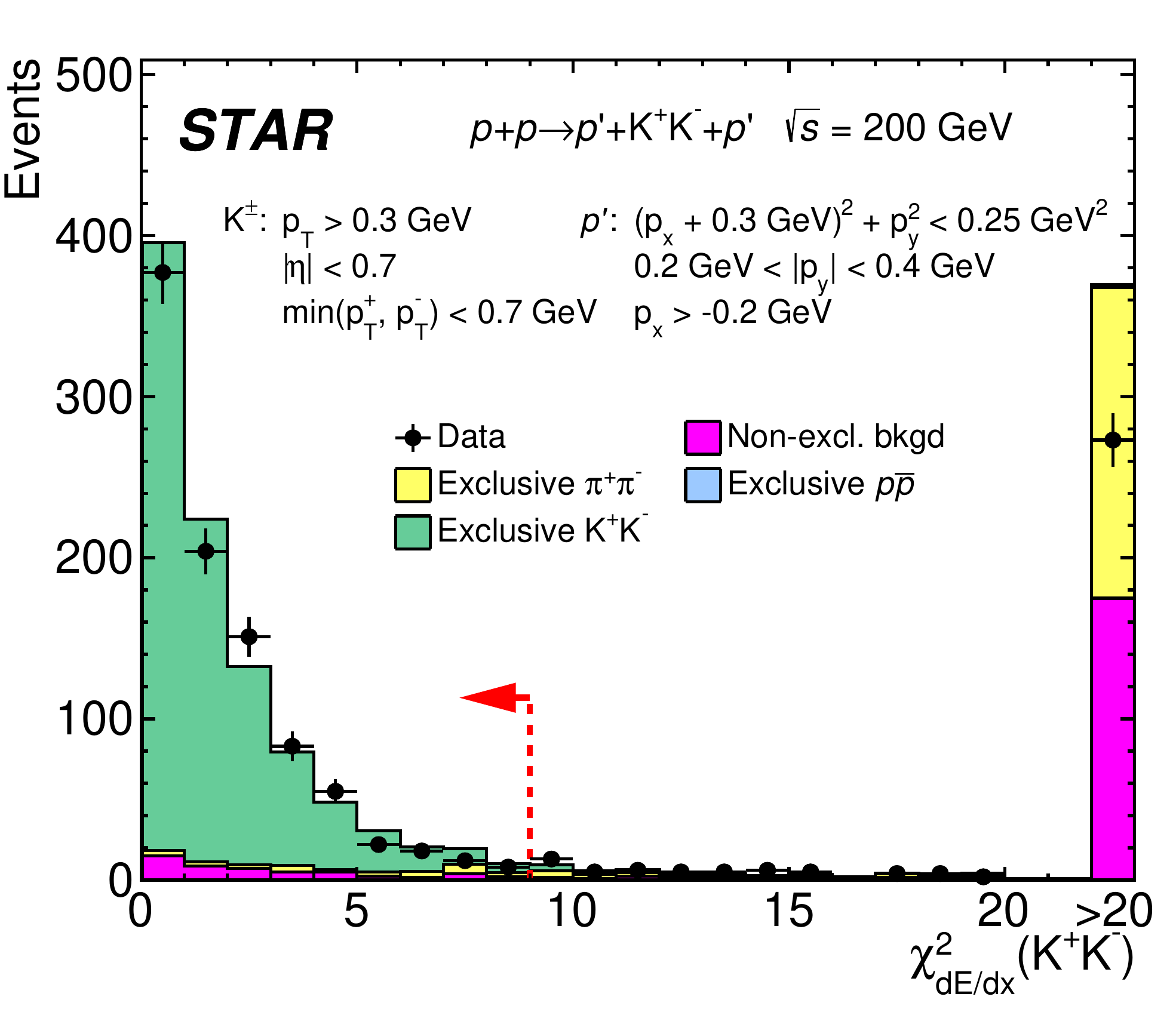}
\hfill
\includegraphics[width=.485\textwidth,page=1]{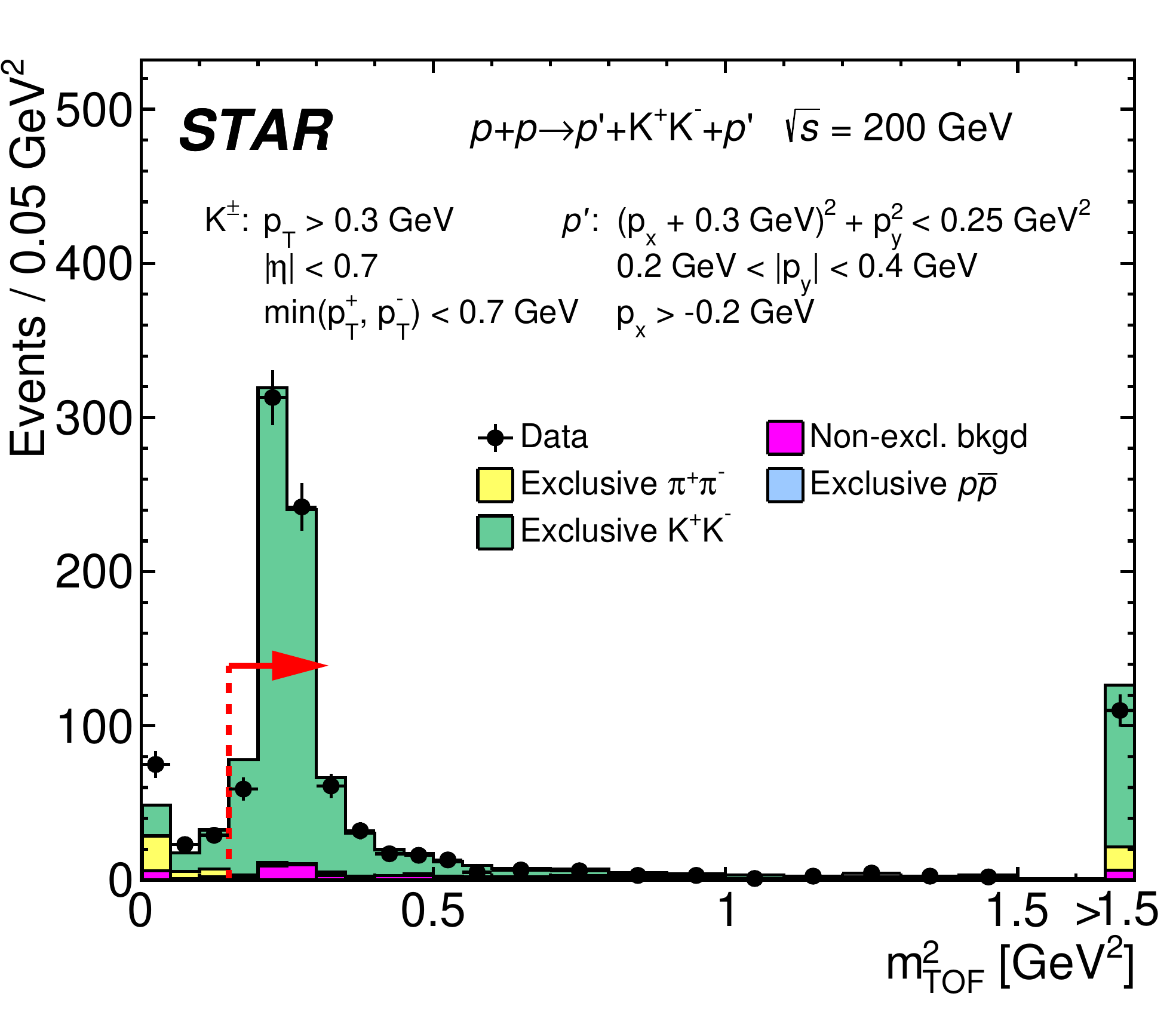}
\newline
\includegraphics[width=.485\textwidth,page=1]{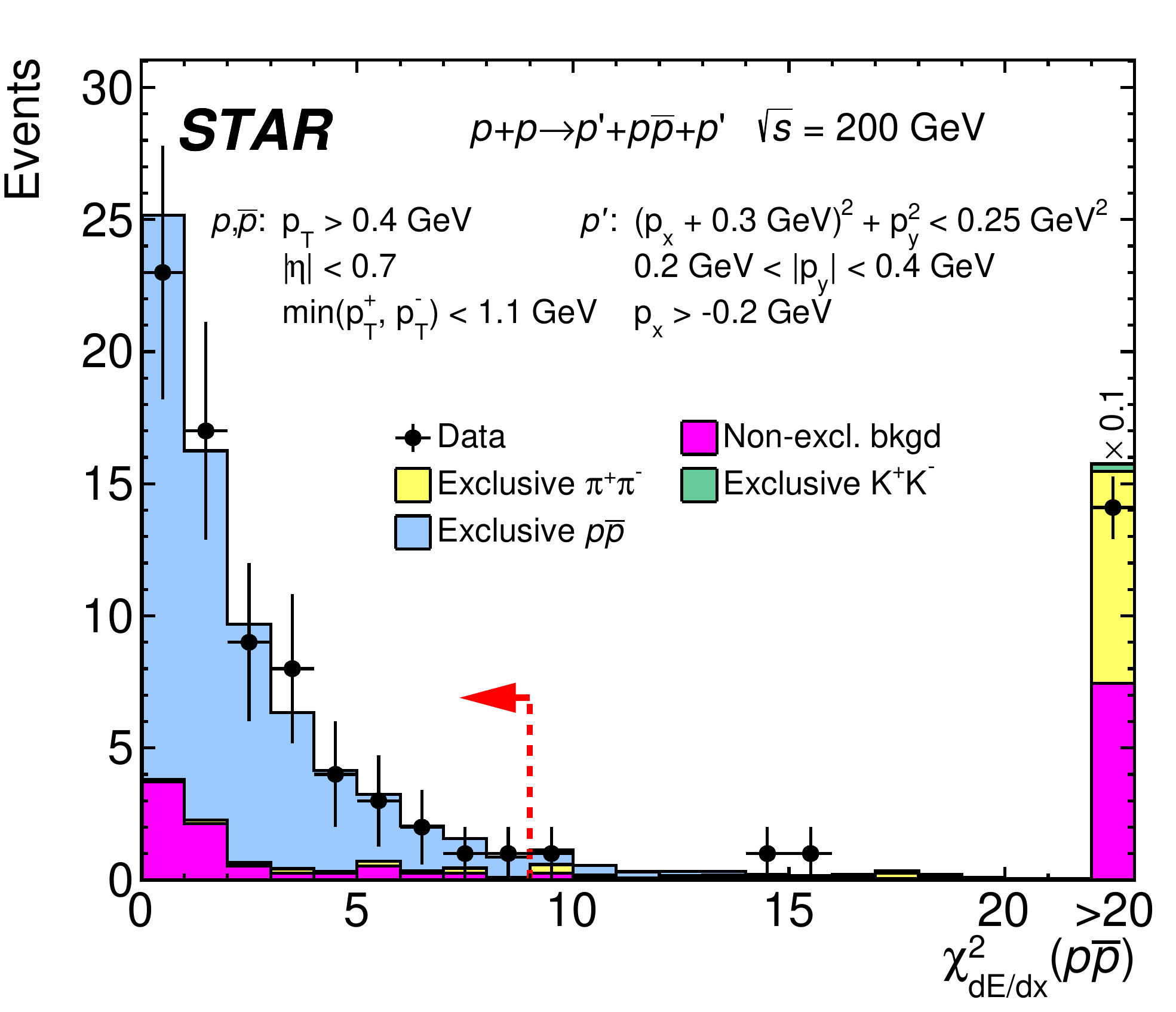}
\hfill
\includegraphics[width=.485\textwidth,page=1]{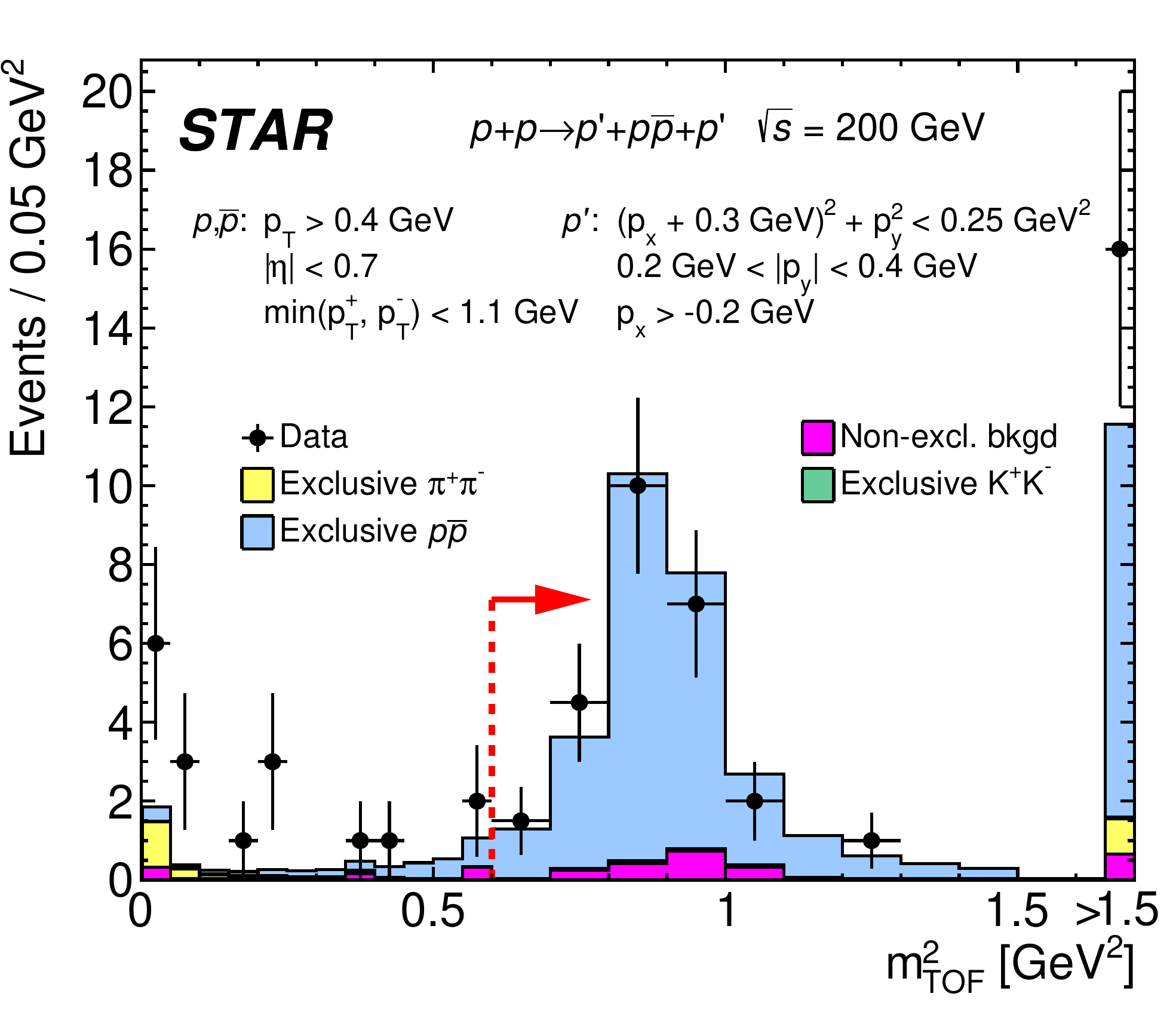}%
\caption{Distributions of $\chi^{2}_{dE/dx}$ (left column) and $m^{2}_\mathrm{\scriptscriptstyle TOF}$ (right column) for exclusive $\pi^+\pi^-$ (top), $K^+K^-$ (middle) and $p\bar{p}$ (bottom) candidates after final event selection. The dashed red line and arrow indicate the value of the cut imposed on the plotted quantity to select exclusive pairs of a given particle species. Yellow, blue and green histograms correspond to the fast exclusive MC simulation while magenta shows the estimated amount of non-exclusive background in the data.}
\label{pid_plots}
\end{figure}

\begin{figure}[tbp]
\centering
\includegraphics[width=.485\textwidth,page=1]{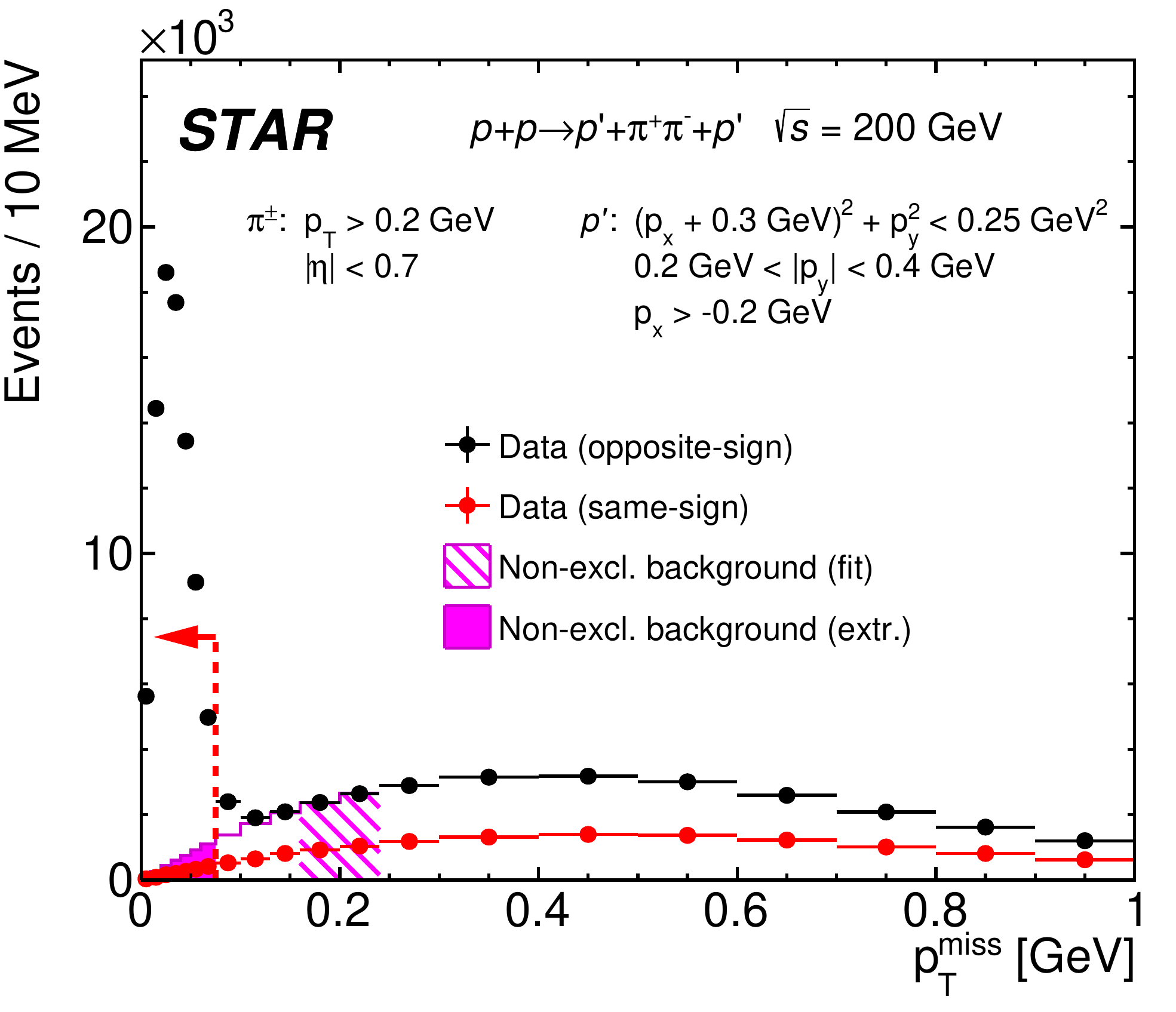}
\hfill
\includegraphics[width=.485\textwidth,page=1]{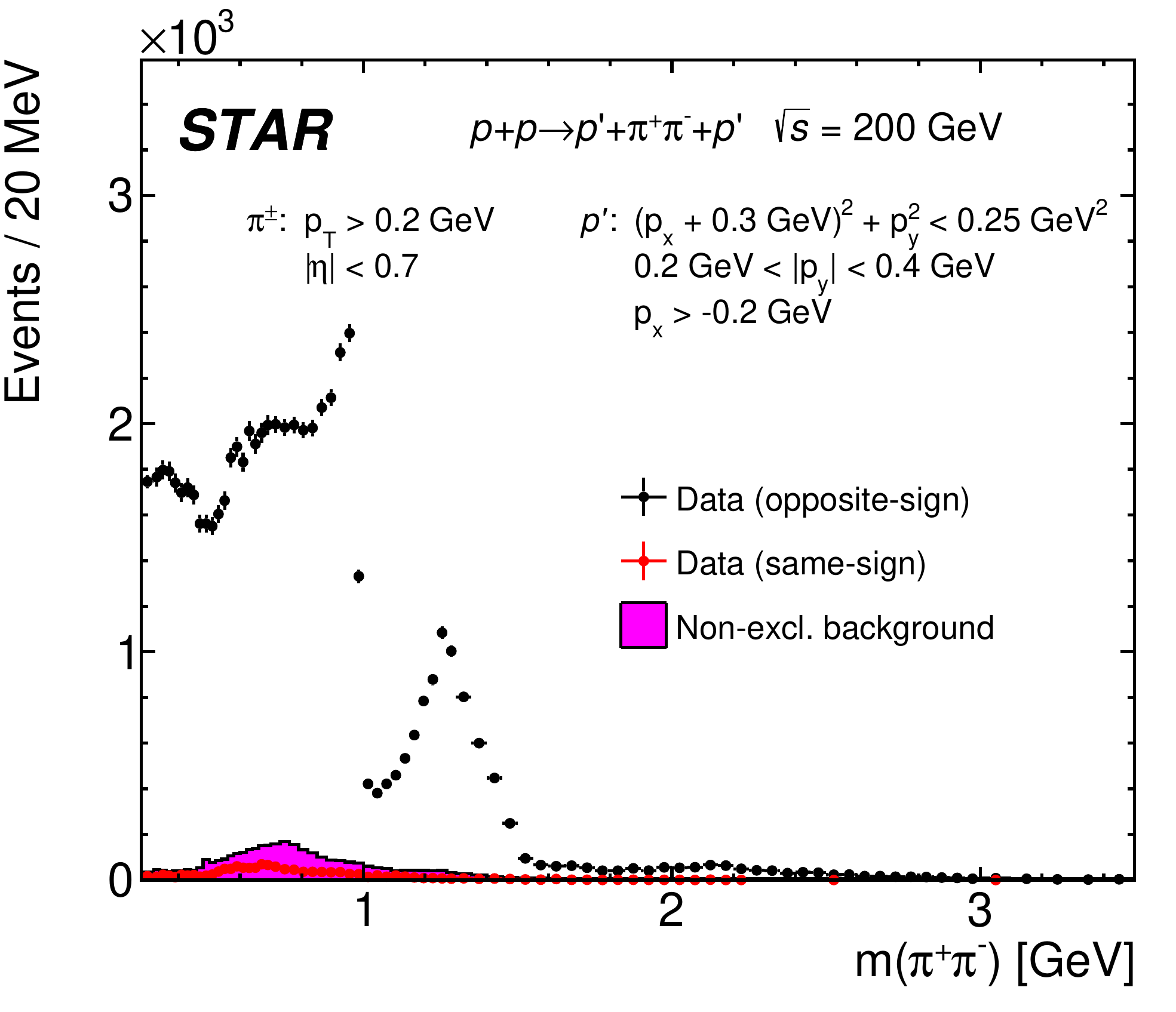}
\newline
\includegraphics[width=.485\textwidth,page=1]{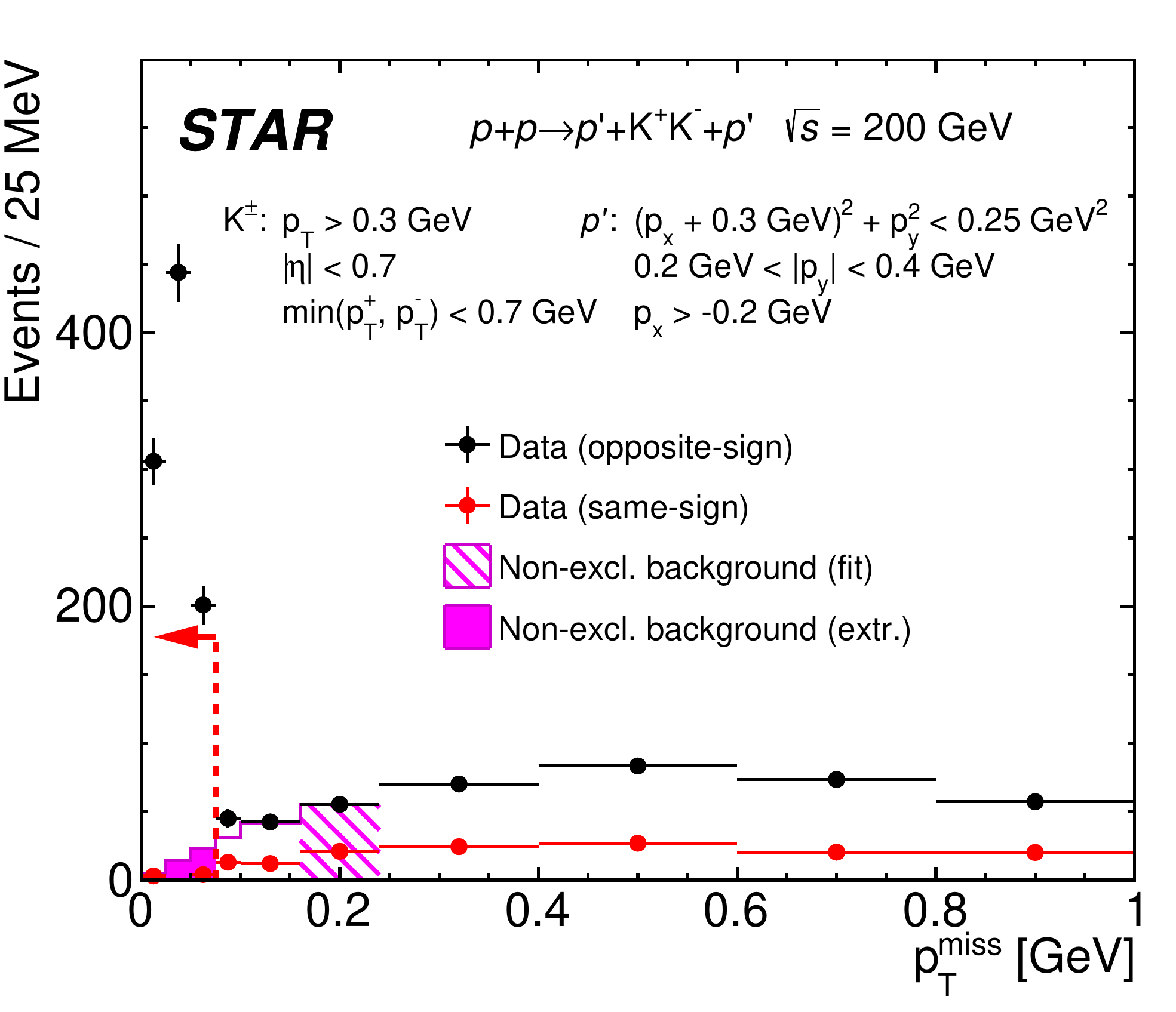}
\hfill
\includegraphics[width=.485\textwidth,page=1]{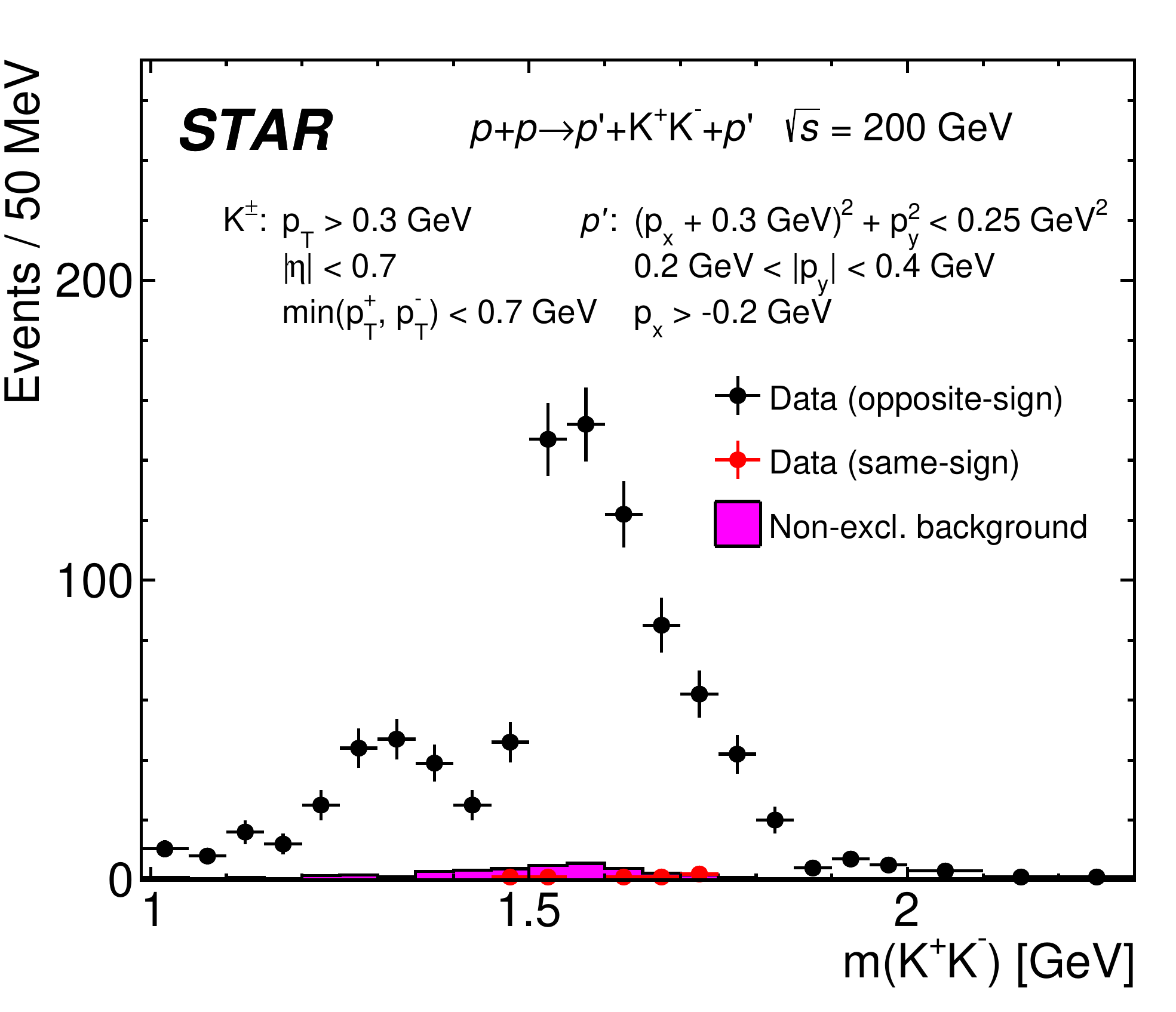}
\newline
\includegraphics[width=.485\textwidth,page=1]{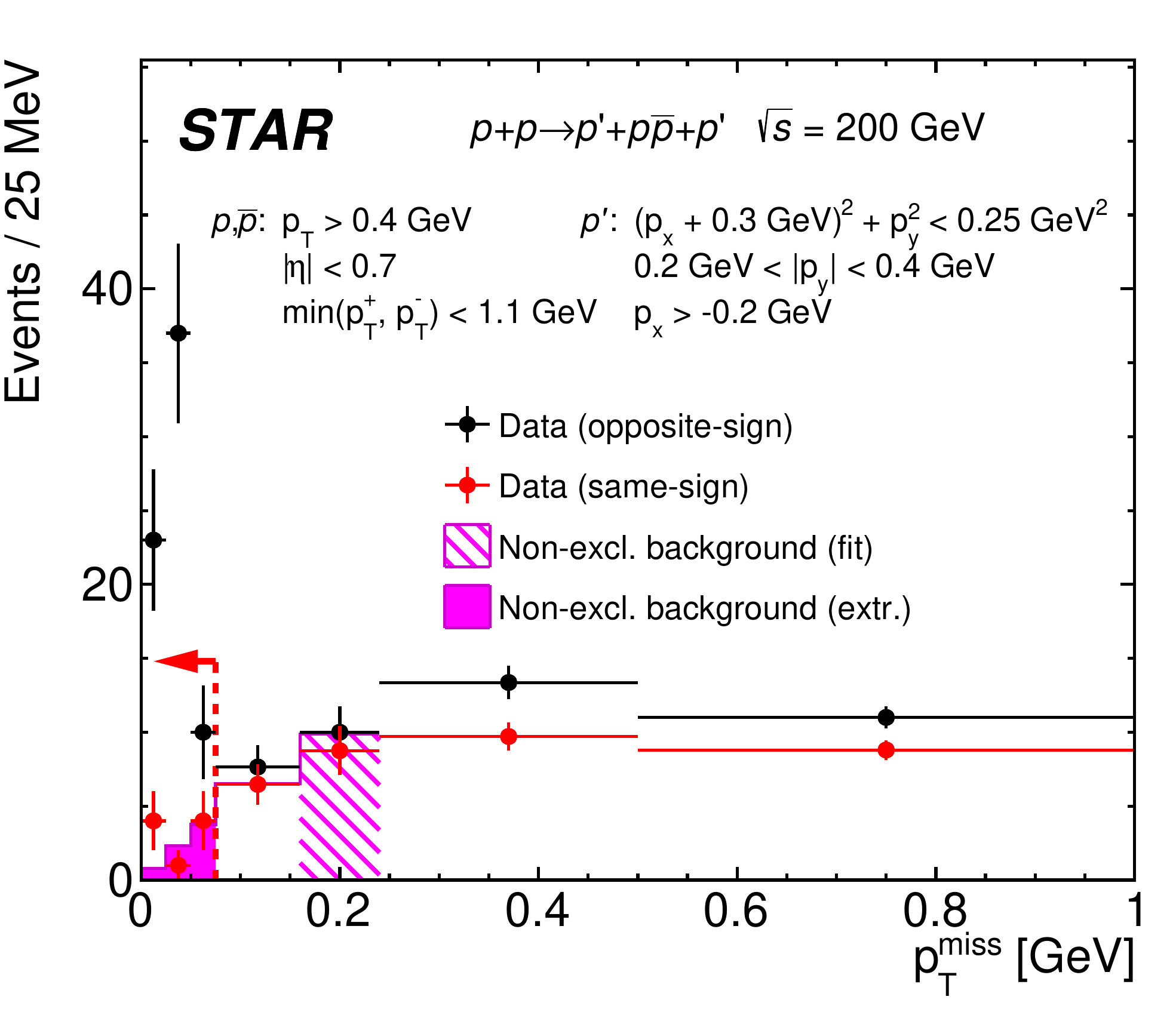}
\hfill
\includegraphics[width=.485\textwidth,page=1]{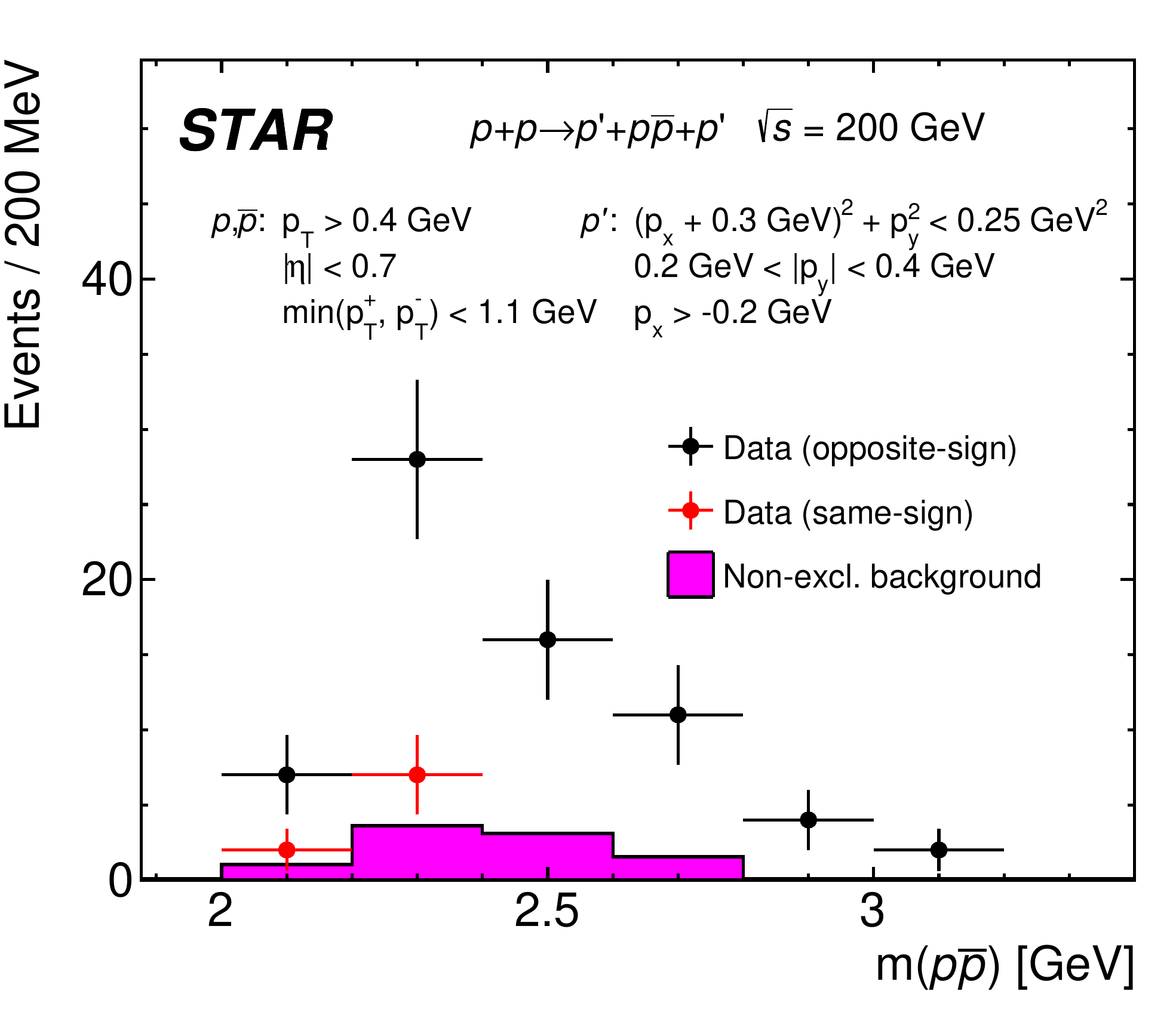}\\%
\vspace*{-4pt}
\caption{Uncorrected distributions of the CEP event candidates' missing transverse momentum $p_\mathrm{T}^\mathrm{\scriptstyle miss}$ (left column) and invariant mass of the charged particle pairs produced in the final state (right column) for $\pi^+\pi^-$ (top), $K^+K^-$ (middle) and $p\bar{p}$ (bottom) pairs. Invariant mass distributions are obtained for the signal dominated regions marked with the red arrows on the $p_\mathrm{T}^\mathrm{\scriptstyle miss}$ plots. Distributions for opposite-sign and same-sign particle pairs are shown as black and red symbols, respectively. The vertical error bars represent statistical uncertainties. The horizontal bars represent bin sizes. Solid magenta histograms correspond to the estimated non-exclusive background, determined differentially from the number of counts in the hatched range $0.16\,\text{GeV}<p_\mathrm{T}^\mathrm{\scriptstyle miss}<0.24\,\text{GeV}$, and extrapolated to the signal region indicated with dashed red line and arrow.}
\label{control_plots_1}\vspace*{-15pt}
\end{figure}

The CEP events were triggered by requiring signals in at least one RP station on each side of the IP, and at least 2 hits in the TOF to ensure the presence of at least two in-time tracks in the TPC. In addition, a lack of activity in both the small BBC tiles and the ZDC detectors is required to ensure the double gap in pseudorapidity topology characteristic of CEP events. In order to reduce pile-up events, or events involving proton dissociation, a veto is imposed on events containing signals in both upper and lower RP stations on the same side of the IP. 560 million CEP event candidates were triggered in total, corresponding to 14.2~pb$^{-1}$ of integrated luminosity. An average trigger prescale of 5 was used during the entire data-taking period due to limited data acquisition bandwidth.\\
\indent
In the offline analysis, the protons tagged in the RPs are further required to have transverse momenta ($p_x$, $p_y$) in the fiducial region defined as
\begin{equation}
(p_x+0.3\,\mbox{GeV})^2 + p_y^2 < 0.25\,\mbox{GeV}^2\;\;\; \& \;\;\; 0.2\,\mbox{GeV} < |p_y| < 0.4\,\mbox{GeV} \;\;\; \& \;\;\; p_x>-0.2\,\mbox{GeV}.\;
\label{fp_fiducial}
\end{equation}
\begin{figure}[t]
\centering
\includegraphics[width=.465\textwidth]{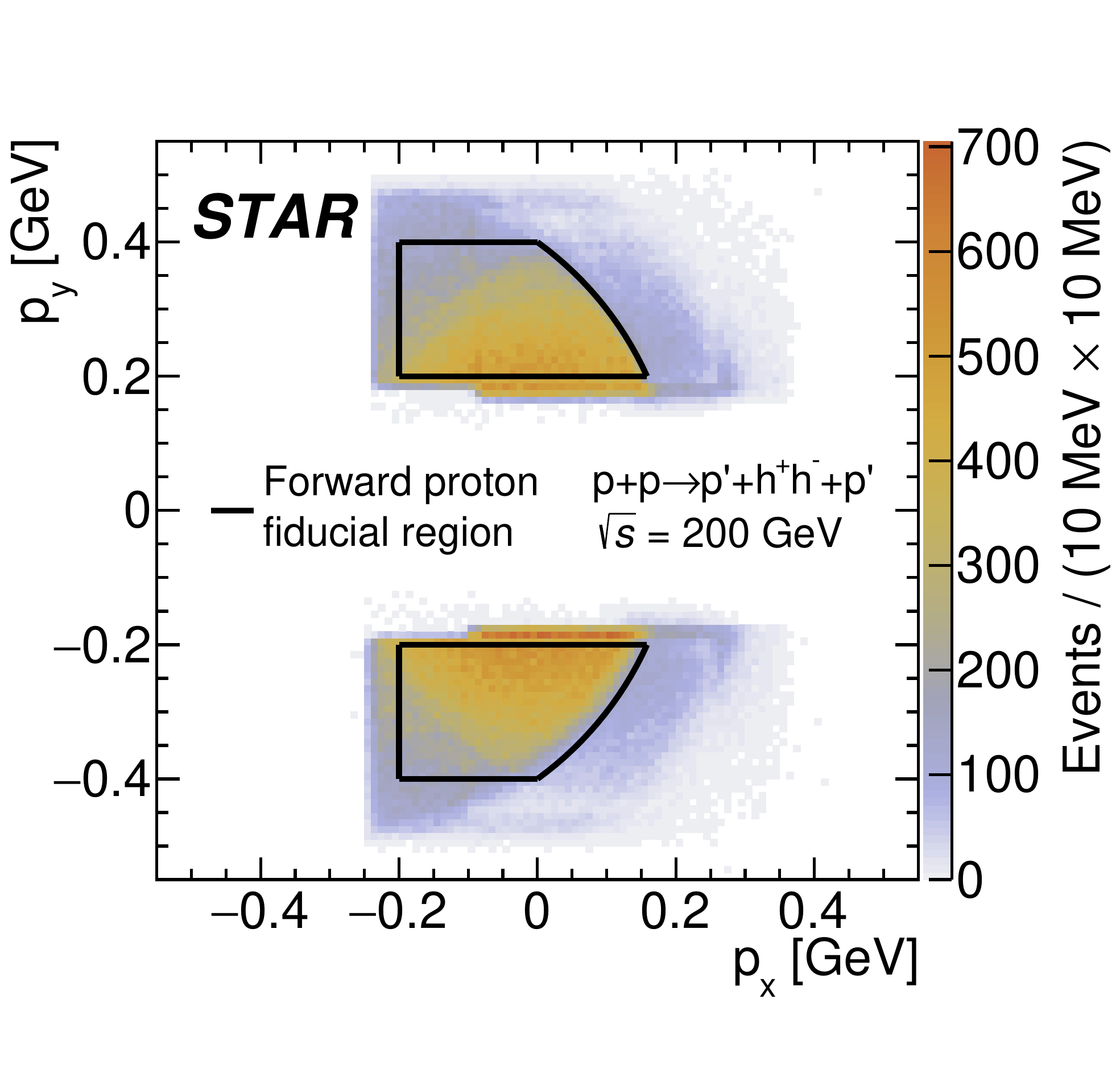}
\includegraphics[width=.523\textwidth]{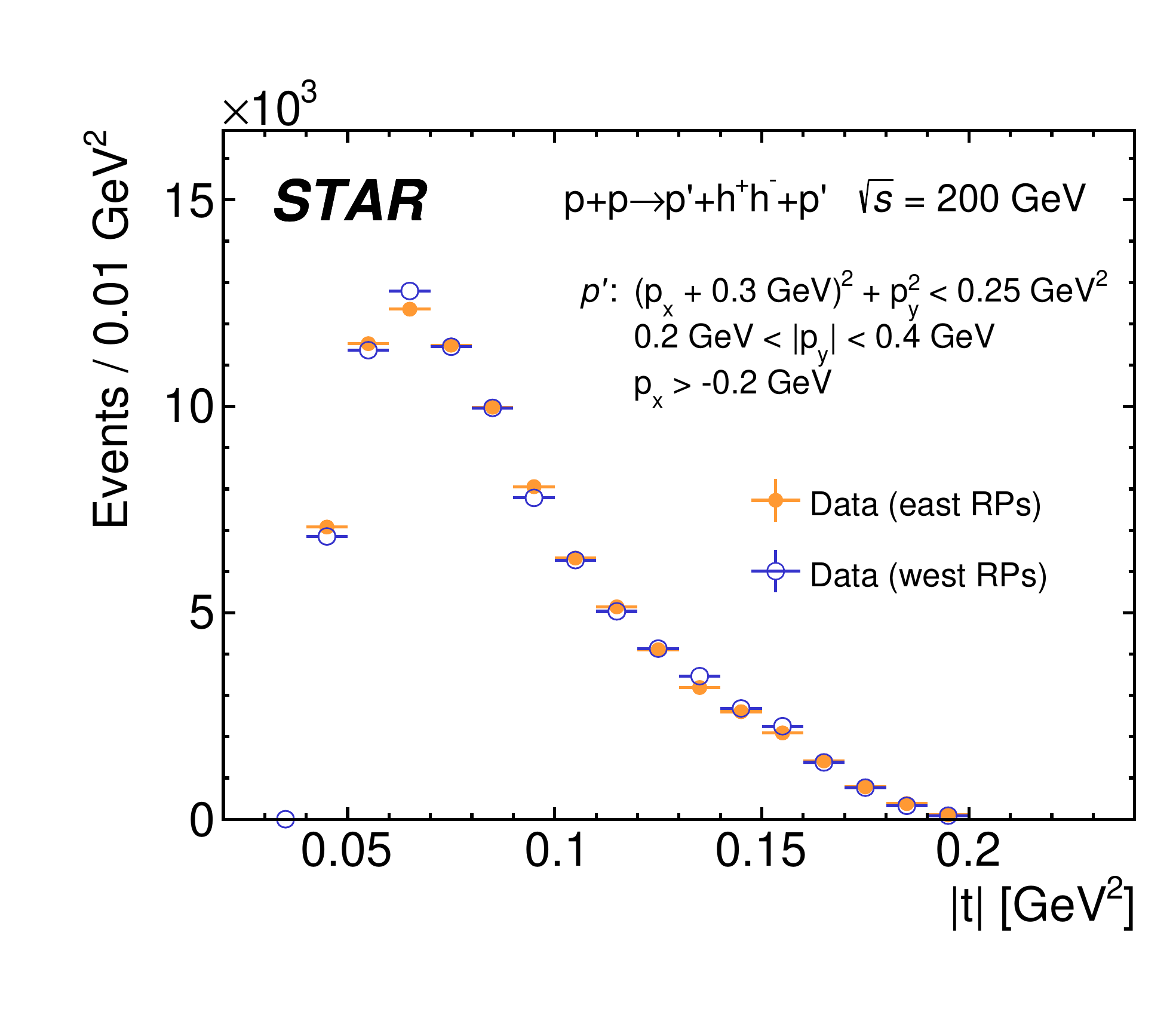}
\vspace*{-29pt}
\caption{(left) Combined distributions of diffractively-scattered protons' momenta $p_y$ vs. $p_x$ reconstructed with the East and West RP stations. The kinematic region used in the measurement is enclosed with the black line. (right) Distributions of measured four-momentum transfers at the proton vertices are shown for the East and West stations with yellow and blue colours, respectively.}
\label{rp_hits}
\end{figure}
\noindent
This fiducial region is chosen to achieve high geometrical acceptance and track reconstruction efficiency and also to minimise systematic uncertainties.
Figure~\ref{rp_hits} (left) shows the combined distributions of the momenta, $p_y$ vs. $p_x$, of the diffractively scattered protons in exclusive $h^{+}h^{-}$ events reconstructed with the East and West RP stations. The kinematic region used in the measurement, defined in Eq.~\eqref{fp_fiducial}, is enclosed with the black line. Figure~\ref{rp_hits} (right) shows the distributions of measured four-momentum transfers at the proton vertices separately for the East and West stations.\\
\indent
In this analysis, the CEP events must consist of only one pair of oppositely charged mid-rapidity particles, besides the two forward-scattered protons. Therefore, only events with exactly two opposite-sign TPC tracks, each matched with hits in the TOF and originating from a common vertex, are selected. These tracks have at least 25 hits (out of a possible 45). All tracks are required to be within the fiducial region defined by $p_\mathrm{T}^\mathrm{\scriptstyle track}>0.2$ GeV and $|\eta^\mathrm{\scriptstyle track}|<0.7$. 
The $z$-position of the event vertex obtained from the TPC tracks is limited to $|z_\mathrm{\scriptstyle vtx}|<80$ cm. The above limits on $\eta^\mathrm{\scriptstyle track}$ and $z_\mathrm{\scriptstyle vtx}$ 
were chosen to ensure high geometrical acceptance in the entire fiducial phase space. In addition, it is required that the $z$-position of the vertex obtained from the time difference of the signals from the forward protons in the RPs agrees with the TPC vertex within 36~cm, corresponding to three-and-a-half standard deviations.
To further suppress the residual backgrounds, a veto is imposed on signals in the large BBC tiles ($2.1<|\eta|<3.3$), as well as on events with more than one additional TOF hit not matching either of the two TPC tracks. This mainly removes higher-multiplicity events where some particles are either not reconstructed in the TPC or produced outside the TPC acceptance.\\
\indent
PID involves a few steps. First, the $p\bar{p}$ hypothesis is checked:
\protect \begin{equation}\label{eq:pidPPbar}\chi^{2}_{dE/dx}(p\bar{p})<9\;\;\; \& \;\;\; \chi^{2}_{dE/dx}(\pi^{+}\pi^{-})>9\;\;\; \& \;\;\; \chi^{2}_{dE/dx}(K^{+}K^{-})>9\;\;\; \& \;\;\; m^{2}_\mathrm{\scriptscriptstyle TOF} > 0.6~\mbox{GeV}^{2}.\end{equation}
\noindent
If the conditions~\eqref{eq:pidPPbar} are satisfied, the pair is assumed to be $p\bar{p}$. If not, the pair is checked for compatibility with the $K^{+}K^{-}$ hypothesis: \protect \begin{equation}\label{eq:pidKK}\chi^{2}_{dE/dx}(K^{+}K^{-})<9\;\;\; \& \;\;\; \chi^{2}_{dE/dx}(\pi^{+}\pi^{-})>9\;\;\; \& \;\;\; \chi^{2}_{dE/dx}(p\bar{p})>9\;\;\; \& \;\;\; m^{2}_\mathrm{\scriptscriptstyle TOF} > 0.15~\mbox{GeV}^{2}.\end{equation}
If the pair is not compatible with either the $p\bar{p}$ or the  $K^{+}K^{-}$ hypothesis, it is assumed to be a $\pi^{+}\pi^{-}$ pair if
\protect \begin{equation}\label{eq:pidPiPi}
\chi^{2}_{dE/dx}(\pi^{+}\pi^{-})<12.
\end{equation}
\indent
For pairs identified as kaons or protons, a more restrictive cut on each track's $p_\mathrm{T}$ is imposed: $p_\mathrm{T}>0.3$ GeV for $K^\pm$ and $p_\mathrm{T}>0.4$ GeV for $p(\bar{p})$. In addition,  it is required that the lower-$p_\mathrm{T}$ track in the pair has $p_\mathrm{T} < 0.7$~GeV ($K^+K^-$) or $p_\mathrm{T} < 1.1$~GeV ($p\bar{p}$). These additional cuts are intended to constrain the fiducial range of high track reconstruction efficiency (lower cut) and high pair identification efficiency (upper cut). 
The criteria used for PID given by Eqs.~\eqref{eq:pidPPbar}, \eqref{eq:pidKK} and~\eqref{eq:pidPiPi}, and the $\mathrm{min}(p^+_T,p^-_T)$ cut discussed above, were chosen to suppress exclusive background below 1\% for $\pi^+\pi^-$ and $p\bar{p}$ and below $3\%$ for $K^+K^-$.
The distributions of $\chi^{2}_{dE/dx}$ and $m^{2}_\mathrm{\scriptscriptstyle TOF}$ for all studied particle species are shown in Fig.~\protect\ref{pid_plots}, together with the fast MC predictions.
The $\chi^{2}_{dE/dx}$ distributions are shown after the corresponding $m^{2}_\mathrm{\scriptscriptstyle TOF}$ cut listed in Eq.~\eqref{eq:pidPPbar} and Eq.~\eqref{eq:pidKK}. Similarly, the $m^{2}_\mathrm{\scriptscriptstyle TOF}$ distributions are shown after the corresponding $\chi^{2}_{dE/dx}$ cut.\\
\indent
Finally, the missing transverse momentum in the event, $p_\mathrm{T}^\mathrm{\scriptstyle miss}$, obtained from the transverse momenta of the protons tagged in the RPs and the tracks of the centrally produced pair, is required to be less than 75~MeV to suppress the non-exclusive background. Figure~\ref{control_plots_1}~(left column) shows the  $p_\mathrm{T}^\mathrm{\scriptstyle miss}$ distributions for all studied particle species together with the  $p_\mathrm{T}^\mathrm{\scriptstyle miss}$ distributions for like-sign control sample. 
After all the above selection cuts, the approximate numbers of CEP event candidates are  85600 for $\pi^+\pi^-$ pairs, 930 for $K^+K^-$ pairs and 70 for $p\bar{p}$ pairs in the final state.\\
%
%
%
\indent
Uncorrected invariant mass distributions of the $\pi^+\pi^-$, $K^+K^-$ and $p\bar{p}$ pairs after the final selection cuts are shown in Fig. \ref{control_plots_1} (right column). The same-sign control sample is also shown, obtained with exactly the same event selection cuts as the nominal sample except for the requirement that the two centrally produced tracks should have opposite electric charges. Instead it is required that the charges of the tracks are the same. The same-sign control sample is at the level of a few percent of the final sample.\\

\FloatBarrier

\section{Kinematic variables and fiducial region}
\label{fiducial_region}
The measurements are done differentially in several kinematic variables, which include:
\begin{itemize}
\item the invariant mass of the central state, $m(h^+h^-)$, with resolution rising approximately linearly with increasing invariant mass, starting from about 2~MeV at 0.3~GeV and reaching $20-30$~MeV at 3~GeV and above,
\item the rapidity of the central pair, $y(h^+h^-)$, with relatively constant resolution of about $0.01$~unit,
\item the difference of azimuthal angles, $\Delta\upvarphi$, of the forward-scattered protons with typical resolution of $5^\circ-7^\circ$,
\item the sum of the squares of the four-momentum transfers at the proton vertices, $|t_1+t_2|$, with resolution of $0.01-0.02$~GeV$^{2}$,
\item the cosine of polar angle ($\cos{\uptheta^\mathrm{CS}}$) and the azimuthal angle ($\upphi^\mathrm{CS}$) of positively charged central particle in the Collins-Soper frame~\cite{cs_frame}~\footnote{
 Collins-Soper frame is the centre-of-mass frame of the charged particles pair with the $z$-axis making equal angles with the beam protons momenta, which in addition defines the new $x-z$ plane. It can be reached from the laboratory frame (proton-proton c.m.s.) in two steps. First, boost along the $z$-axis to an intermediate frame in which the pair longitudinal momentum is equal to zero. In this frame the beam protons momenta remain parallel to the $z$-axis and the transverse momentum of the pair remains unchanged. Second, boost in the direction of the transverse momentum of the pair, to get to the pair c.m.s. frame.} 
 with typical resolutions of $0.005-0.01$ and $1^\circ-2^\circ$, respectively.
\end{itemize}
The differential cross sections  are obtained in the fiducial region defined by the kinematical cuts imposed on the forward-scattered protons given in Eq.~\eqref{fp_fiducial}, and by the cuts on the final state charged particles' pseudorapidities: $|\eta|<0.7$ and transverse momenta $p_\mathrm{T}>0.2$ GeV ($\pi^+\pi^-$), $p_\mathrm{T}>0.3$ GeV ($K^+K^-$) and  $p_\mathrm{T}>0.4$ GeV ($p\bar{p}$). In addition, in the case of $K^+K^-$ and $p\bar{p}$ pairs, the fiducial volume is restricted to the region with the lower $p_\mathrm{T}$ in the pair below 0.7~GeV or 1.1~GeV, respectively.
\newpage
\FloatBarrier
\section{Background estimation}
Background in the analysis arises from non-exclusive processes leading to correlated signals in the RP and TOF/TPC (`single source') and from coincidences of a signal in the RP with an uncorrelated signal in the TOF/TPC (`pile-up').
Other sources of background are exclusive processes in which the particle pair was misidentified.\\
\indent
The `single source' non-exclusive contribution is dominated by  Central Diffraction, $p+p\rightarrow p^\prime+h^+h^-+Y+p^\prime$, where $Y$ is any number of particles produced (but not measured) in addition to the measured $h^+h^-$ pair. In the `pile-up' background, the signal in the central detector almost always arises from an inelastic $pp$ collision while the RP signal occurs due to `pile-up' from real forward-going protons from elastic scattering, central diffraction, showering in single, double or non-diffractive events or beam-induced sources. 
All the above sources of background are estimated using a data-driven method. Both undetected particles in `single source' events and the random character of `pile-up' events lead to breaking the correlation between the central $h^+h^-$ pair and the forward protons and to a much flatter $p_\mathrm{T}^\mathrm{\scriptstyle miss}$ distribution. 
This can be seen in Fig.~\ref{control_plots_1} (left column), where the $p_\mathrm{T}^\mathrm{\scriptstyle miss}$ distribution starts to increase above 100 MeV. Background estimation is based on the extrapolation into the signal region of the second-degree polynomial function fitted to the signal-free region, as shown by the magenta histograms. The polynomial is constrained in the fit to vanish at $p_\mathrm{T}^\mathrm{\scriptstyle miss}=0$. This procedure is repeated differentially for all the kinematic variables presented later. As an example, the resulting background estimation is presented in Fig.~\ref{control_plots_1} (right column) differentially in $m(h^+h^-)$ by the magenta histograms. The shape of the background as a function of $p_\mathrm{T}^\mathrm{\scriptstyle miss}$ is confirmed by MC predictions and like-sign events (shown by red points), which are, by definition, background. On average, this background amounts to $5.3\%$ ($\pi^+\pi^-$), $5.4\%$ ($K^+K^-$) and $12\%$ ($p\bar{p}$).\\
\indent
The exclusive background was estimated based on the fast MC simulation. 
The resulting distributions of $\chi^2_{dE/dx}$ and $m^2_\mathrm{\scriptscriptstyle TOF}$ for all particle species are presented in Fig.~\ref{pid_plots}.
%
\FloatBarrier
\section{Corrections}
For all cross section calculations, bin sizes are chosen to correspond to about three times the detector resolution so that migrations between bins could be neglected.\\
\indent
Particles passing through the detector material lose some energy. To minimise biases from this effect, a correction procedure is applied during track-momentum reconstruction for both data and MC simulation based on the expected material budget for the given track.  
In this procedure, all tracks are assumed to be pions, therefore the reconstructed momenta of the remaining particle species exhibit some bias.
For tracks identified as kaons and protons, an additional energy loss correction was applied based on the single particle MC. The correction is up to 10 and 20 MeV for low-momentum kaons and protons, respectively.\\
\indent
Several corrections were implemented to account for the limited efficiency of the measurement. The RP and TOF trigger efficiencies were estimated from the unbiased data to be $\sim 100\%$ and 98\%, respectively. The RP trigger efficiency was evaluated as the probability that the trigger was set when a proton was reconstructed in the given station. For events with exactly two TPC tracks, each matched with hits in the TOF and originating from a common vertex, and at most one additional TOF hit, the TOF trigger efficiency was estimated as the fraction of events that passed the TOF trigger conditions. The average number of inelastic collisions per bunch crossing varies between 0.2 and 0.9 and leads to a sizeable probability that the exclusive signal overlaps with another process, which causes signal loss due to the trigger or offline selections. This probability is estimated based on zero-bias data for each run independently and parameterised as a function of instantaneous luminosity for each of the four possible combinations of RPs topology.
An event is rejected if the overlapping process produces a signal in the BBC, ZDC, TOF, or the RP station not belonging to the studied combination, since our selection criteria include vetos on these detectors. The overall veto efficiency varies between $40-80\%$. The efficiency of the $|z_\mathrm{\scriptstyle vtx}|<80$~cm selection cut was estimated for each RHIC fill independently based on the estimated values of the mean and standard deviation, assuming a normal distribution in the data. A typical vertex-cut efficiency is 88\%. \\
\indent
The above corrections were applied independent of all kinematic variables, affecting only the normalisation. Corrections described below were applied as functions of the relevant variables, affecting the shapes of all measured distributions.\\
\indent
Protons successfully reconstructed in a RP station may still produce secondaries in the dead material in the other RP branch, which cause the trigger to veto the event. The probability to pass the veto trigger was estimated using the embedded GenEx sample as a 3D function of proton momenta $p_x$, $p_y$ and $z_\mathrm{\scriptstyle vtx}$.
The same sample was used to study the proton reconstruction efficiency. Secondaries produced in the dead material or from 'pile-up' processes introduce inefficiencies in the reconstruction procedure. This inefficiency was also calculated using a data-driven tag-and-probe method using elastic scattering data, where one of the protons serves as a tag to probe the reconstruction efficiency of the second proton. The joint efficiency of the proton reconstruction and the trigger veto caused by the proton interaction with dead material is typically 98\%, but goes down to 60\%  in the $p_x,p_y$ region where the protons are expected to pass the RF shield present between the two RP stations.\\
\indent
TPC and TOF efficiencies were calculated as functions of particle $p_T, \eta$ and $z_\mathrm{\scriptstyle vtx}$. 
The joint acceptance and reconstruction efficiency of a track in the TPC was measured separately for $\pi^+,\pi^-,K^+,K^-,p$ and $\bar{p}$ using the single particle MC embedded into zero-bias trigger data taken simultaneously with physics triggers.
TPC inefficiencies are caused by empty spaces between the sectors, 
fiducial cuts on the positions of space points, the influence of a high density of off-time tracks, the interactions of particles in dead material in front of the TPC, dead TPC modules, and natural decays of pions and kaons before or inside the TPC.
The TPC efficiency increases with $p_T$. The efficiency at the lowest transverse momentum used in the analysis is 70\% for pions, 40\% for kaons and 75\% for protons. Above 1~GeV, the efficiency plateaus at 80\% for pions, 70\% for kaons and 85\% for protons. The efficiency is roughly  independent of particle charge, except that the efficiency for anti-protons is around 2\% smaller than for protons.
The TPC efficiencies depend on the track selection criteria. To check the sensitivity of the results on the track selection, the TPC efficiencies were calculated with looser and tighter matching criteria of the TPC tracks with a vertex and also with an increased (lowered) required number of associated hits 28 (20). These changes led to $\pm 4$\% changes in the efficiencies. \\
\indent
The combined TOF acceptance, hit-reconstruction efficiency and matching efficiency with TPC tracks was measured separately for $\pi^+,\pi^-,K^+,K^-,p$ and $\bar{p}$ using single particle  MC embedded into zero-bias trigger data taken simultaneously with physics triggers.
For high statistics exclusive $\pi^+\pi^-$ production, the TOF efficiency was also measured using a data-driven tag-and-probe method where one of the pions matched with the TOF serves as a tag and the efficiency of the TOF was measured for the second pion. The differences between the data-driven and MC-based efficiencies were added to the MC-based efficiencies, assuming that they are independent of $z_\mathrm{vtx}$. The same procedure was applied to kaons and protons. The average $p_\mathrm{T}$- and $\eta$-dependent correction to the TOF efficiency is $3\%$. Finally, the TOF efficiency was measured directly by selecting in-time TPC tracks using an independent data sample with the HFT signal recorded. Reconstruction of TPC tracks containing hits in silicon layers of the HFT guarantees that tracks are in-time with the TOF hits. HFT matched tracks, however, have limited coverage of $z_\mathrm{vtx}$, with $|z_\mathrm{vtx}|<20$~cm. The average additional correction is 1\% for pions, 3\% for kaons and 2\% for protons.\\
\indent
A small fraction of signal events are rejected by the  $p_\mathrm{T}^\mathrm{\scriptstyle miss}<75$~MeV requirement. The leakage is caused by the finite resolution of particle momenta which, in the case of a particle measured in the TPC, depends on the momentum. The efficiency of this cut was measured as a function of central particle momenta using a fast phase space MC simulation. This efficiency is 97\% if both tracks have transverse momentum less than 0.4~GeV, and decreases to 89\% for tracks with transverse momentum above 1.5 GeV.\\
\indent
The PID efficiency, defined as the probability that the particle pair $h^+h^-$ passes the relevant PID selection criteria given by Eqs.~\eqref{eq:pidPPbar}, \eqref{eq:pidKK} and \eqref{eq:pidPiPi}, was calculated as a function of the central particles' momenta using the fast phase space MC simulation. The PID efficiency for $\pi^+\pi^-$ pairs is almost 100\% in the whole fiducial region. For $K^+K^-$ and $p\bar{p}$ it is also close to 100\% if the lower of the two transverse momenta in the pair, $p_\mathrm{T}^\mathrm{\scriptstyle min}$, is less than 0.6 GeV for $K^+K^-$ or less than 1~GeV for $p\bar{p}$. At larger values of $p_\mathrm{T}^\mathrm{\scriptstyle min}$, efficiencies for $K^+K^-$ and $p\bar{p}$ identification decrease significantly due to the $\chi^{2}_{dE/dx}(\pi^{+}\pi^{-})>9$ requirement 
used to limit 
misidentified $\pi^+\pi^-$ pairs in the $K^+K^-$ and $p\bar{p}$ samples.\\
\indent
The tracking efficiencies provide corrections for true particles inside the fiducial volume to be reconstructed in the TPC or RP detectors. Additional corrections were applied to account for true particles inside the fiducial volume that are reconstructed outside this volume, and true particles outside the fiducial region that are reconstructed inside this volume. Such migration is caused by finite detector resolutions and the intrinsic smearing of the forward proton kinematics due to the RHIC angular beam divergence. It is also possible that the presence of tracks not associated with the true particle causes an incomplete exclusive event to pass all the selection cuts. Such fake tracks may come from  interactions of true particles with material in the detector or from additional pile-up processes. 
Correction factors for migrations through the boundary of the fiducial region, and for fake tracks, were estimated from GenEx and single particle samples. Joint correction factors for the migrations are generally very small, but can be up to 5\% close to the edges of the fiducial region for central particles and up to 30\% for forward protons. Corrections for fake particles only weakly depend on transverse momentum and are below 2\% for central particles and up to 5\% for forward protons.
\FloatBarrier
\section{Systematic uncertainties}
\label{sec:systematics}
Several sources of possible systematic uncertainties have been considered in this analysis.
The largest contributions to the systematic uncertainty arise from the modelling of the RP system and the beam-line elements, detector alignment and the embedding technique.  The overall uncertainty on measurement efficiencies related to the RP system is typically 6\%, but up to 30\% for $|t_1+t_2|>0.3$ GeV$^2$. This uncertainty is derived from the difference between efficiencies obtained from the MC simulation and from the data-driven tag-and-probe method using elastic scattering data.\\
\indent
The uncertainties related to the TPC efficiency are dominated by modelling of the disturbing activity in the detector caused by the high density of off-time tracks. An uncertainty of 1\% was estimated by studying the  consistency of the corrections obtained from the embedding technique for different rates of off-time tracks. 
An additional 1\% uncertainty arises from the extrapolation of the embedding result, obtained from only a subset of data sample, to the full sample. 
Finally, a 0.5\% uncertainty related to the amount of inactive material between the primary vertex and the STAR TPC was estimated based on the comparison of rates of secondary vertices between data and simulation. In addition, we observed up to $\pm 1.5$\% changes in the cross sections by applying looser and tighter TPC track selection criteria and correcting using the TPC efficiency obtained for a given set of selection cuts. We treat these deviations as an additional source of systematic uncertainty.
A typical TPC-related total systematic uncertainty on the cross section is $4\%$ for $\pi^+\pi^-$ and $p\bar{p}$ and $6\%-7\%$ for $K^+K^-$.\\
\indent
The TOF-related uncertainties were estimated as the difference between results obtained with simulation and data-driven tag-and-probe methods using exclusive events and those obtained with the direct method using an independent sample
of HFT-tagged tracks. A typical total TOF-related systematic uncertainty on the cross section is 3\% for $\pi^+\pi^-$, 5\% for $p\bar{p}$ and 10\% for $K^+K^-$.\\
%
%
\begin{figure}
\centering
\includegraphics[width=.48\textwidth,page=1]{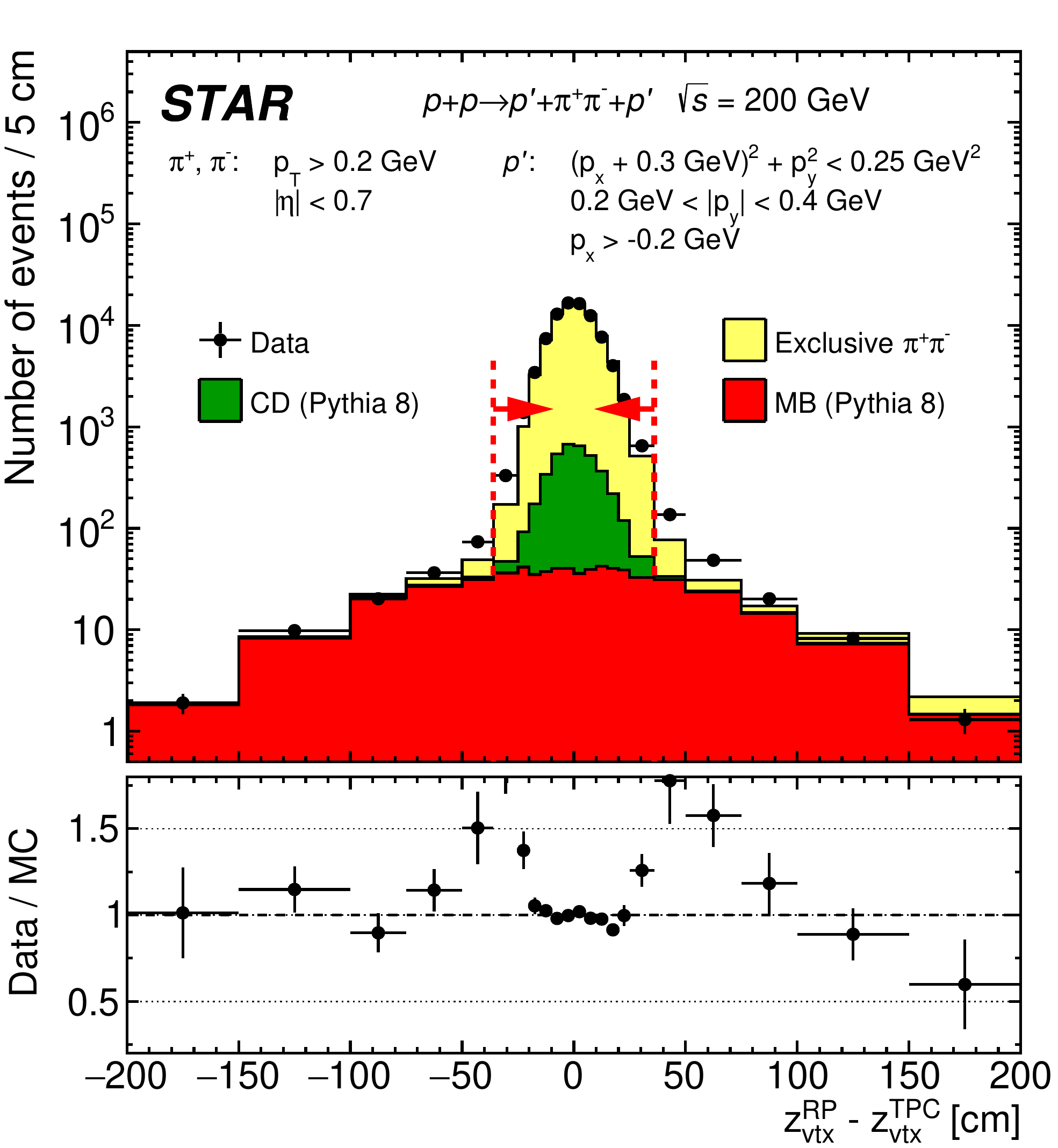}~~
\includegraphics[width=.48\textwidth,page=1]{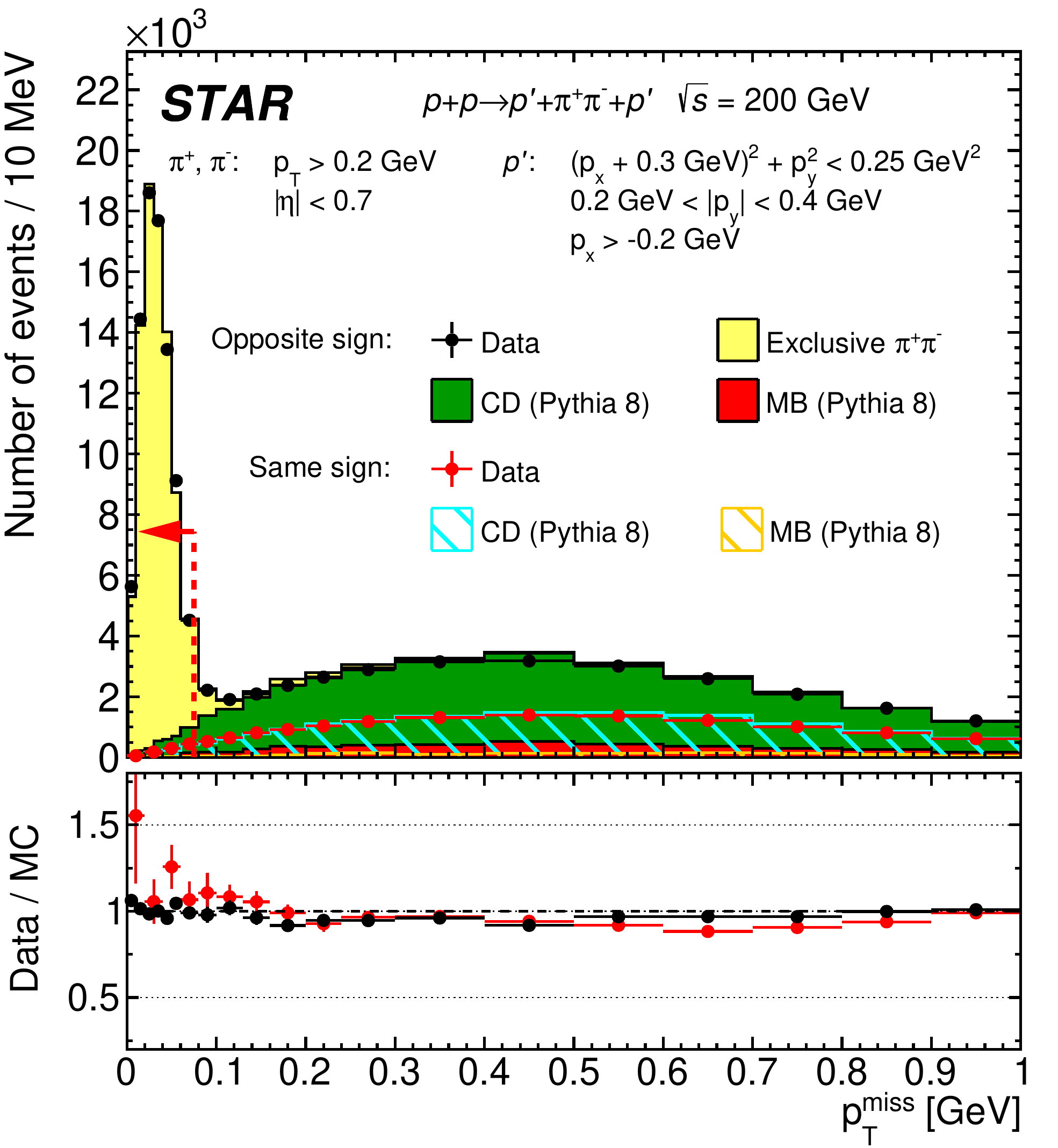}\vspace*{-6pt}
\caption{(left) Comparison of $\Delta z_{\text{vtx}}$ for CEP $\pi^{+}\pi^{-}$ events between data (points) and MC~(stacked colour histograms) after offline selection excluding the cut on agreement between longitudinal vertex position measured in RPs and TPC, marked with dashed red lines and arrows. 
(right) Comparison of $p_\mathrm{T}^{\text{miss}}$ for CEP $\pi^{+}\pi^{-}$ event candidates between data and MC after offline selection excluding the total transverse momentum cut, marked with dashed red line and arrow. In addition to the signal channel (opposite-sign particles), the control background channel (same-sign particles) is also shown in the plots. Data are represented by black (opposite-sign) or red (same-sign) points, while stacked MC predictions are drawn as filled (opposite-sign) or hatched (same-sign) histograms of different colours. Vertical error bars represent statistical uncertainties and horizontal bars represent bin sizes.}
\label{deltaZVtx_missingPt_dataVsMC}\vspace*{-16pt}
\end{figure}
\begin{figure}
\centering
\includegraphics[width=.48\textwidth,page=1]{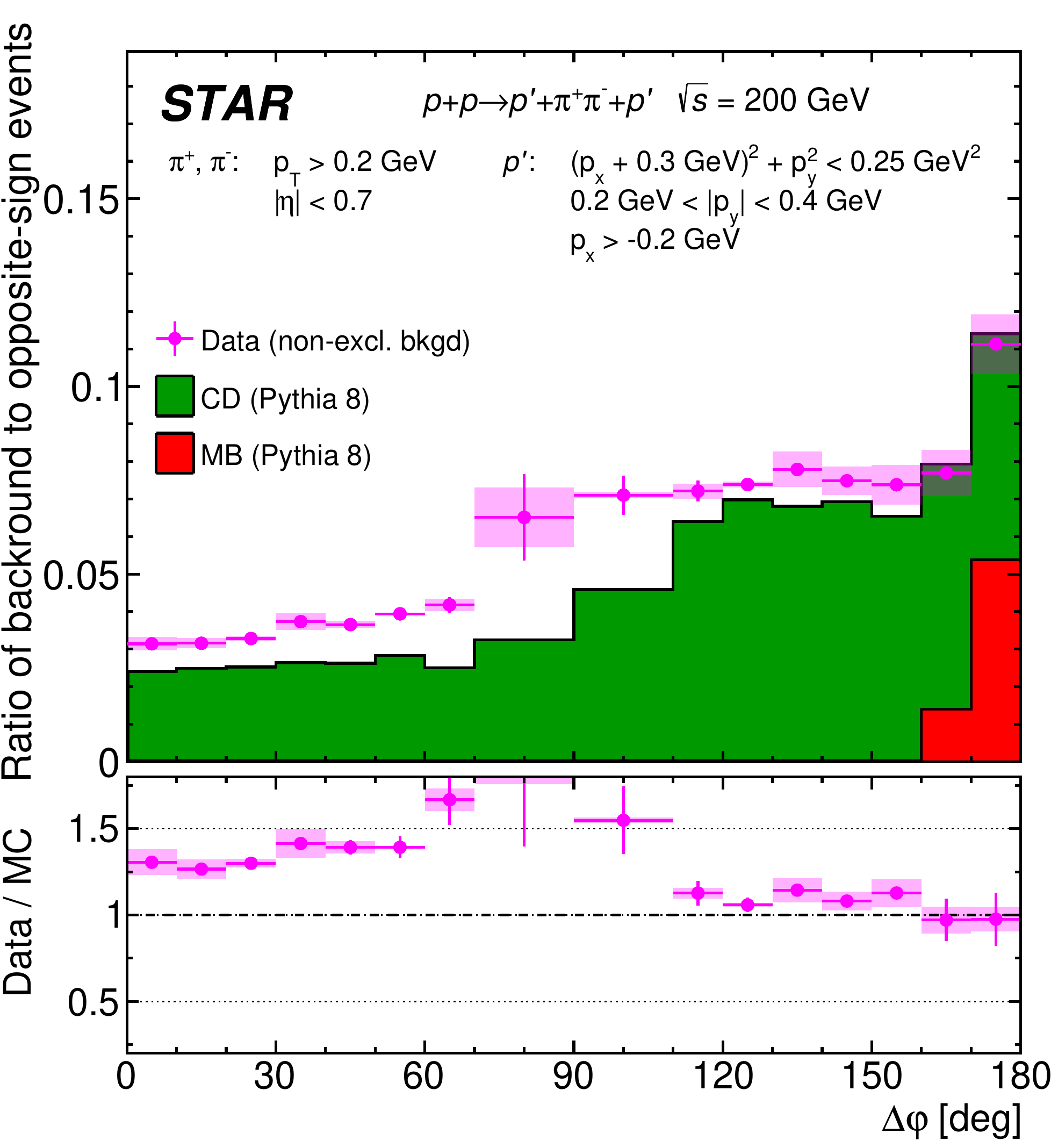}~~
\includegraphics[width=.48\textwidth,page=1]{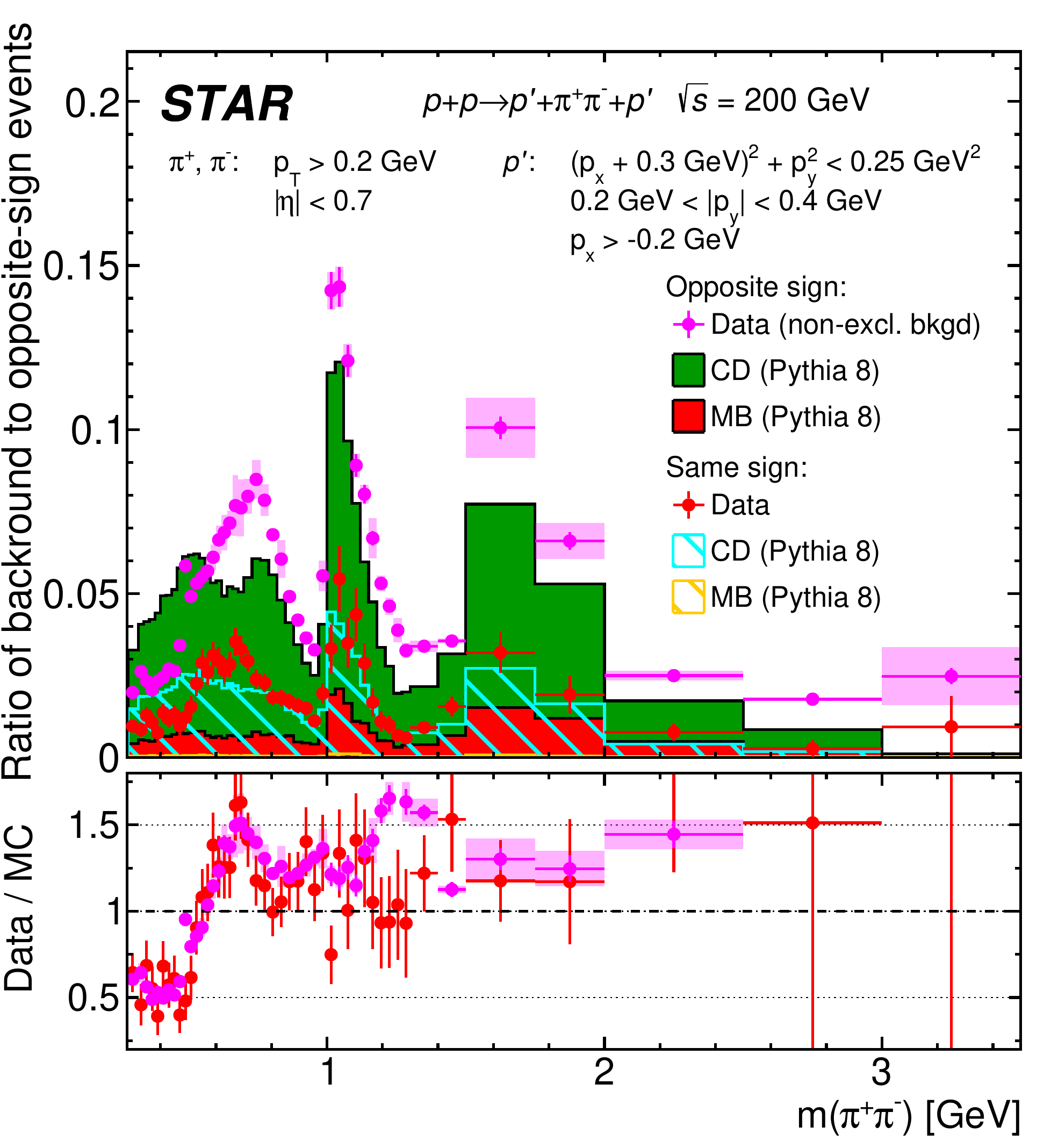}\vspace*{-6pt}
\caption{Comparison of the ratio of non-exclusive background extracted from the data (points) and predicted by MC (filled histograms) to all opposite-sign events in the final CEP $\pi^{+}\pi^{-}$, shown as a~function of $\Delta\upvarphi$ (left) and $m(\pi\pi)$ (right). In both plots, the  vertical error bars represent statistical uncertainties and horizontal bars represent bin sizes. Data points showing opposite-sign non-exclusive background are accompanied by a shaded area denoting the estimated systematic uncertainty related to the background determination method. Same-sign control events are shown in the right plot, marked with red points for the data, and drawn as stacked hatched histograms for MC predictions. The bottom panels show the ratio of the data to corresponding MC predictions.}
\label{invMass_deltaPhi_bkgd_dataVsMC}\vspace*{-13pt}
\end{figure}
\indent
The only sources of systematic uncertainties which may vary significantly as a function of $m(h^+h^-)$ are non-exclusive and pile-up backgrounds.
The quality of the description of these two sources of backgrounds is investigated using MC samples of Central Diffraction and Minimum Bias embedded into zero-bias collision data, using control samples enhanced in the background. The normalisation of both background sources for $\pi^+\pi^-$ CEP are tuned to match the data in the signal-free regions of the control samples.
The control sample for pile-up background normalisation was obtained by imposing
all the standard selection cuts, except the requirement of consistency of the $z$-position of the vertex obtained from time difference of the signals from the forward protons in the RPs with the TPC vertex, $\Delta z_\mathrm{\scriptscriptstyle vtx}$. 
The distribution of $\Delta z_\mathrm{\scriptscriptstyle vtx}$ is shown in Fig.~\ref{deltaZVtx_missingPt_dataVsMC} (left). The normalisation of the pile-up background shown by the red histogram is tuned to describe the data in the signal-free region of large difference between estimates of $z$-vertex. In the  control sample for non-exclusive background normalisation, only the  $p_\mathrm{T}^\mathrm{\scriptstyle miss}<75$ MeV cut was not used. The resulting $p_\mathrm{T}^\mathrm{\scriptstyle miss}$ distribution is shown in Fig.~\ref{deltaZVtx_missingPt_dataVsMC} (right). 
The normalisation of the non-exclusive background shown by the green histogram is tuned to describe the data in the signal-free region of large values of $p_\mathrm{T}^\mathrm{\scriptstyle miss}$. After tuning, the description of the data in both control regions is very good. The good agreement between data and simulation shown in Fig.~\ref{deltaZVtx_missingPt_dataVsMC} (right) is achieved only by removing the final states consisting of $\pi^+ + \pi^- +$ neutrals (mainly $\pi^+ + \pi^- +\pi^0+\pi^0$) from the PYTHIA8 CD prediction.  
Removal of $\pi^+ + \pi^- +$ neutrals not only makes the shape of the $p_\mathrm{T}^\mathrm{\scriptstyle miss}$ distribution compatible with data, but also correctly predicts the ratio of same-sign to opposite-sign pairs outside the signal region. Otherwise, this ratio is underestimated by 50\%. The background in the signal sample is estimated by summing the predictions from both samples after applying all the selections cuts. Contributions of the remaining background in the data sample are shown in Fig.~\ref{invMass_deltaPhi_bkgd_dataVsMC} together with estimations of the background using the (nominal) fully data-driven method, shown by magenta circles. For $\Delta\upvarphi$ (Fig.~\ref{invMass_deltaPhi_bkgd_dataVsMC} (left)), the background contributions from the nominal and alternative methods agree well. Both methods show that the pile-up background is mainly located close to $\Delta\upvarphi = 180^\circ$, as expected for pile-up from elastic scattering  events. For $m(\pi\pi)$ (Fig.~\ref{invMass_deltaPhi_bkgd_dataVsMC} (right)), the background contributions using the nominal and alternative methods disagree by up to 50\% 
depending on the mass region. The enhancement of the data-driven estimate over the MC prediction around the $f_2(1270)$ and $\rho^0$ mass regions may be caused by imperfect modelling of the resonant states in the hadronisation model used in the PYTHIA generator. This discrepancy was not used as a systematic uncertainty on the  background estimation. The third possible method for background estimation is another data-driven method using the same-sign control sample normalised to the opposite-sign sample in the signal-free region of  $p_\mathrm{T}^\mathrm{\scriptstyle miss}>0.15$ GeV. The unscaled same-sign contribution is shown by red circles in Figs.~\ref{deltaZVtx_missingPt_dataVsMC} and~\ref{invMass_deltaPhi_bkgd_dataVsMC},  while the distributions predicted by PYTHIA 8 CD are shown by the blue hatched histogram. This method, by definition, accounts only for the combinatorial phase space background and cannot describe possible contributions from non-exclusive resonance production. This is seen in Fig.~\ref{invMass_deltaPhi_bkgd_dataVsMC} (right), where the lack of $\rho^0$ and $f_2(1270)$ contributions is clearly seen. This method was also not used to evaluate systematic uncertainty. The systematic uncertainty related to background subtraction was estimated by replacing the polynomial function describing the shape of the background distribution with a histogram template obtained from the  same-sign events normalised to opposite-sign events in the signal-free region of $p_\mathrm{T}^\mathrm{\scriptstyle miss}>0.15$ GeV. Agreement was found at the level of 10\%, which is used as the systematic uncertainty. This contributes up to 1\% uncertainty on the cross section.\\
\indent
The relative luminosity at STAR is determined from the coincidence rate in ZDC detectors in both beam directions. Absolute calibration is given by a special Van der Meer~\cite{van_der_meer} scan. For the precise measurement of the total and elastic cross sections~\cite{elastic_paper}, three dedicated Van der Meer scans were performed during a single RHIC fill to minimise systematic uncertainty on the luminosity measurement. The luminosity uncertainty was estimated to be 4\%. To account for possible fill-by-fill dependence in the luminosity measurement, an additional 4\% uncertainty was assigned to the luminosity. It was determined by comparing variations of the effective cross sections for elastic scattering relative to the measurement done solely based on data collected during the fill with the Van der Meer scans.
The overall luminosity uncertainty of 6\% was estimated by the quadratic sum of uncertainties from these two sources.\\
\indent
Other systematic uncertainties considered include those due to the vertex reconstruction efficiency, selection cuts, and the trigger efficiency. None of these produce uncertainties beyond the 2\% level. \\
\indent
The total systematic uncertainties are obtained by adding 
contributions from all sources in quadrature. They amount typically to between 10\% and 20\%, except at the extremes of the measurement range in $|t_1+t_2|$, and are highly correlated between bins. Table~\ref{tab:xSecSyst}~shows the systematic uncertainties, decomposed into their major components, of cross sections for CEP $\pi^{+}\pi^{-}$, $K^{+}K^{-}$ and $p\bar{p}$ pairs integrated over the fiducial region.
\begin{table}\centering
\begin{tabular}{l|ccccc|c}
 \multicolumn{7}{c}{  $\bm{ \delta_{\text{\bf{syst}}}/\sigma_{\text{\bf{fid}}}~[\text{\bf{\%}}]}$   } \\
 ~ & \bf{TOF} & \bf{TPC} & \bf{RP} & \bf{Other} & \bf{Lumi.} & \bf{Total} \\ \hline\hline
 
$\bm{\pi^{+}\pi^{-}}$ & $\prescript{3.0}{-2.8}{~~}$ & $\prescript{3.8}{-3.6}{~~}$ & $\prescript{5.8}{-5.1}{~~}$ & $\prescript{2.9}{-2.7}{~~}$ & $\prescript{6.4}{-5.7}{~}$ & $\prescript{10.3}{-9.3}{~}$ \\ $\bm{K^{+}K^{-}}$ & $\prescript{10.1}{-8.8}{~~}$ & $\prescript{6.7}{-6.3}{~~}$ & $\prescript{6.0}{-5.3}{~~}$ & $\prescript{5.1}{-5.0}{~~}$ & $\prescript{6.4}{-5.7}{~}$ & $\prescript{15.8}{-14.2}{~~}$ \\ $\bm{p\bar{p}}$ & $\prescript{5.6}{-5.1}{~~}$ & $\prescript{4.1}{-3.9}{~~}$ & $\prescript{6.3}{-5.6}{~~}$ & $\prescript{10.0}{-9.8}{~~}$ & $\prescript{6.4}{-5.7}{~}$ & $\prescript{15.1}{-14.2}{~~}$ \\
 
\end{tabular}
\caption{Typical fractional systematic uncertainties of the integrated fiducial cross sections for CEP of $\pi^{+}\pi^{-}$, $K^{+}K^{-}$ and $p\bar{p}$ pairs, decomposed into their major components. 
}
\label{tab:xSecSyst}\vspace*{-5pt}
\end{table}
\FloatBarrier
\section{Results}
\label{sec:results}
\subsection{Cross sections in fiducial region}
\begin{figure}
\centering
\includegraphics[width=.7\textwidth,page=1]{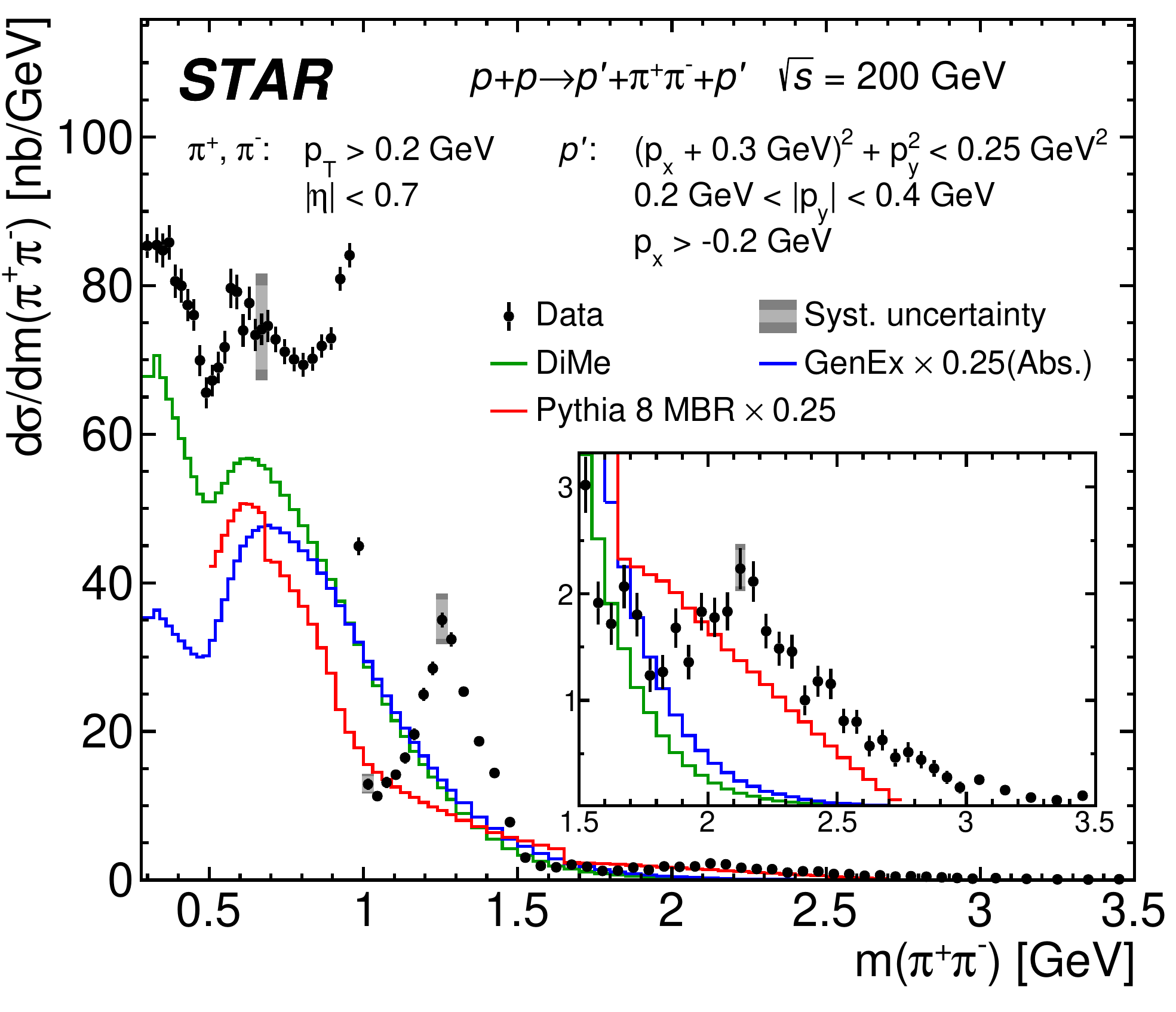}\vspace*{-14pt}
\caption{Differential cross section for CEP of $\pi^+\pi^-$ pairs as a function of the invariant mass of the pair in the fiducial region explained in the plot. Data are shown as solid points with error bars representing the statistical uncertainties. The typical systematic uncertainties are shown as dark/light gray boxes (with/without luminosity uncertainty included, respectively), for only a few data points as they are almost fully correlated between neighboring bins. Predictions from three MC models, GenEx, DiMe and MBR, are shown as histograms.}\vspace*{-14pt}
\label{results_01}
\end{figure}

\begin{figure}
\centering
\includegraphics[width=.48\textwidth,page=1]{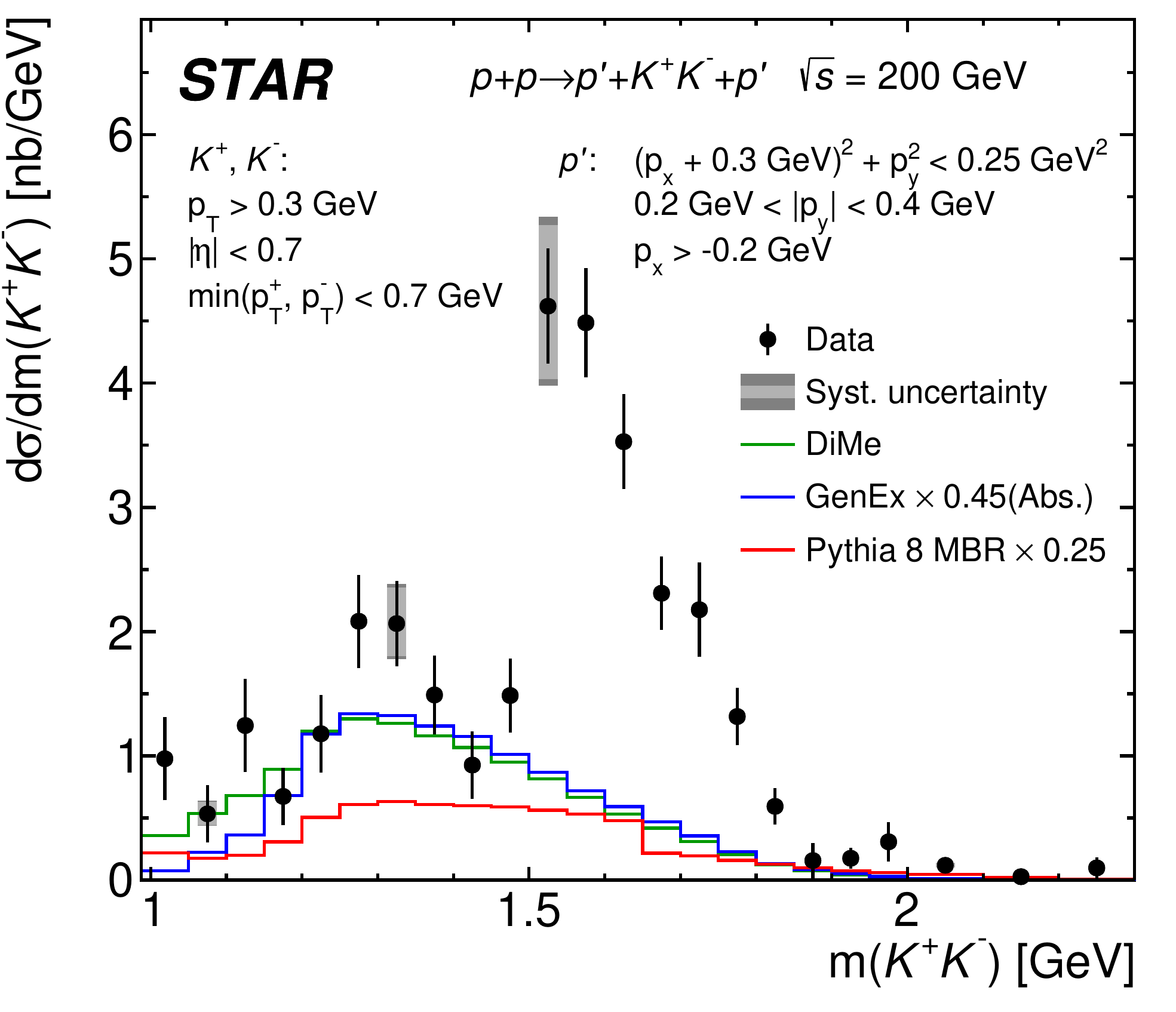}
\includegraphics[width=.48\textwidth,page=1]{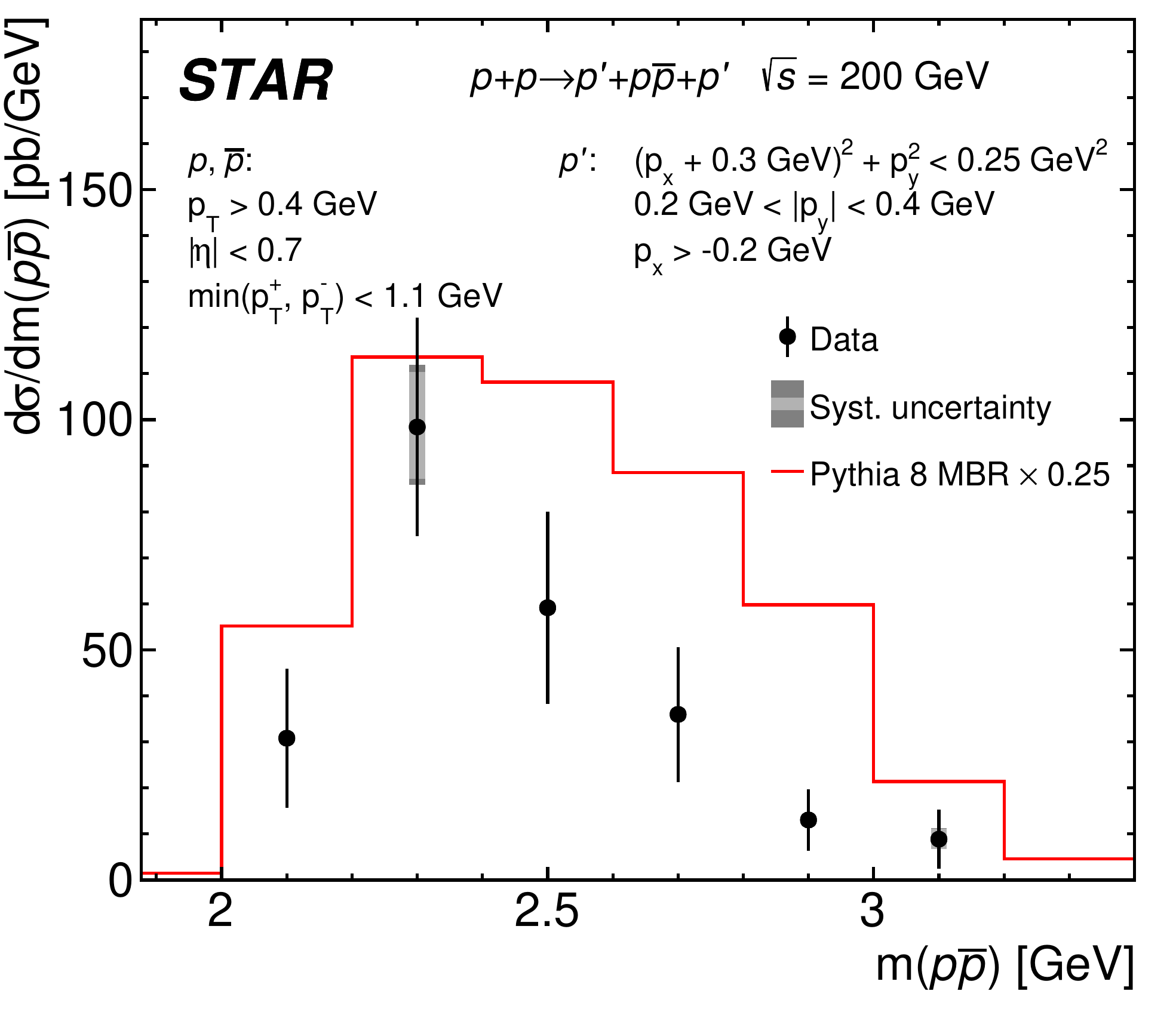}\vspace*{-10pt}
\caption{Differential cross sections for CEP of $K^+K^-$ (left) and $p\bar{p}$ (right) pairs as a function of the invariant mass of the pair in the fiducial region explained in the plots. \mbox{Data are shown as} solid points with error bars representing the statistical uncertainties. The typical systematic uncertainties are shown as gray boxes for only a few data points as they are almost fully correlated between neighboring bins. Predictions from three MC models, GenEx, DiMe and MBR, are shown as histograms. 
}\vspace*{-8pt}
\label{results_02}
\end{figure}

\begin{figure}[b]
\centering
\includegraphics[width=.32\textwidth,page=1]{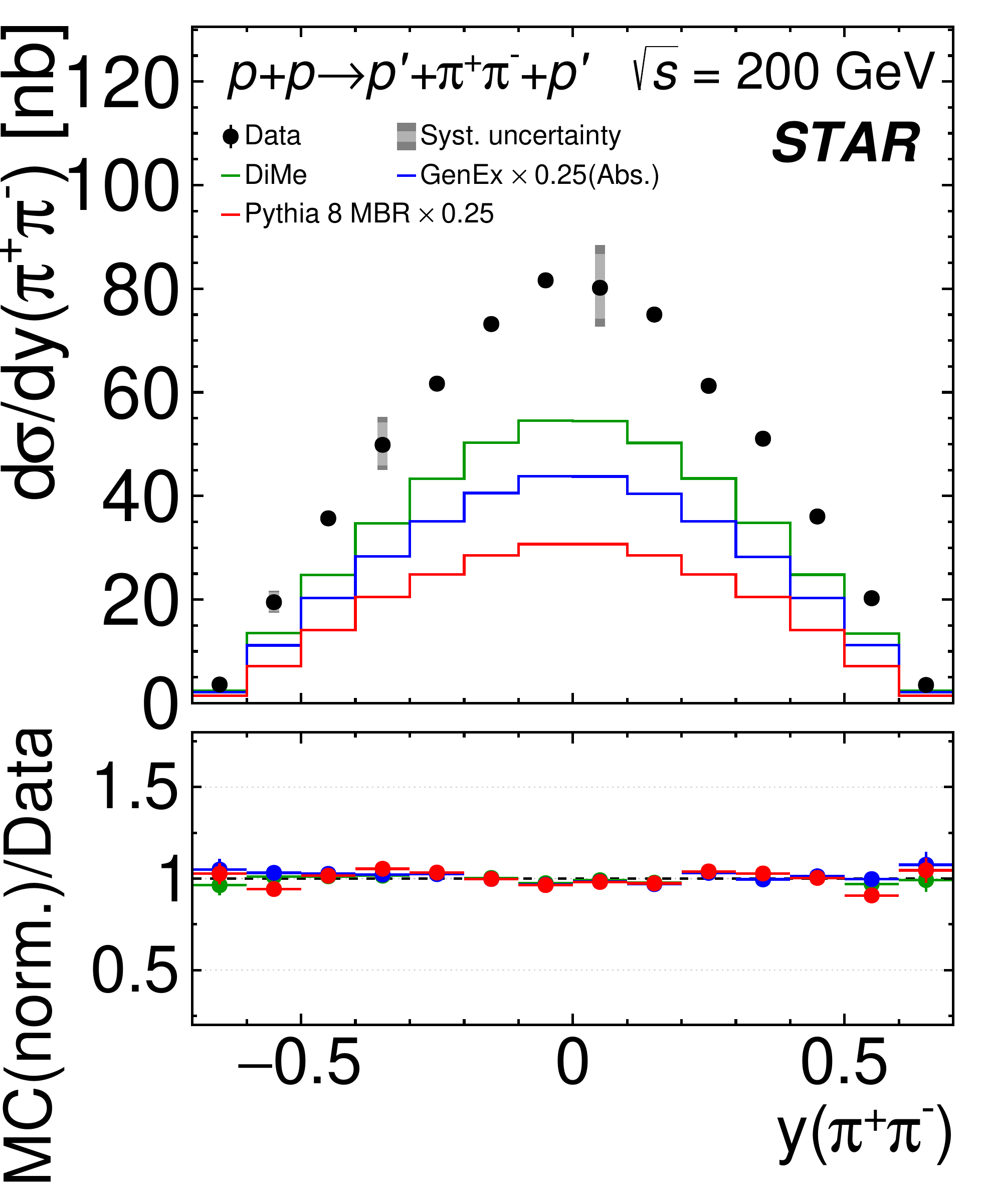}
\hfill
\includegraphics[width=.32\textwidth,page=1]{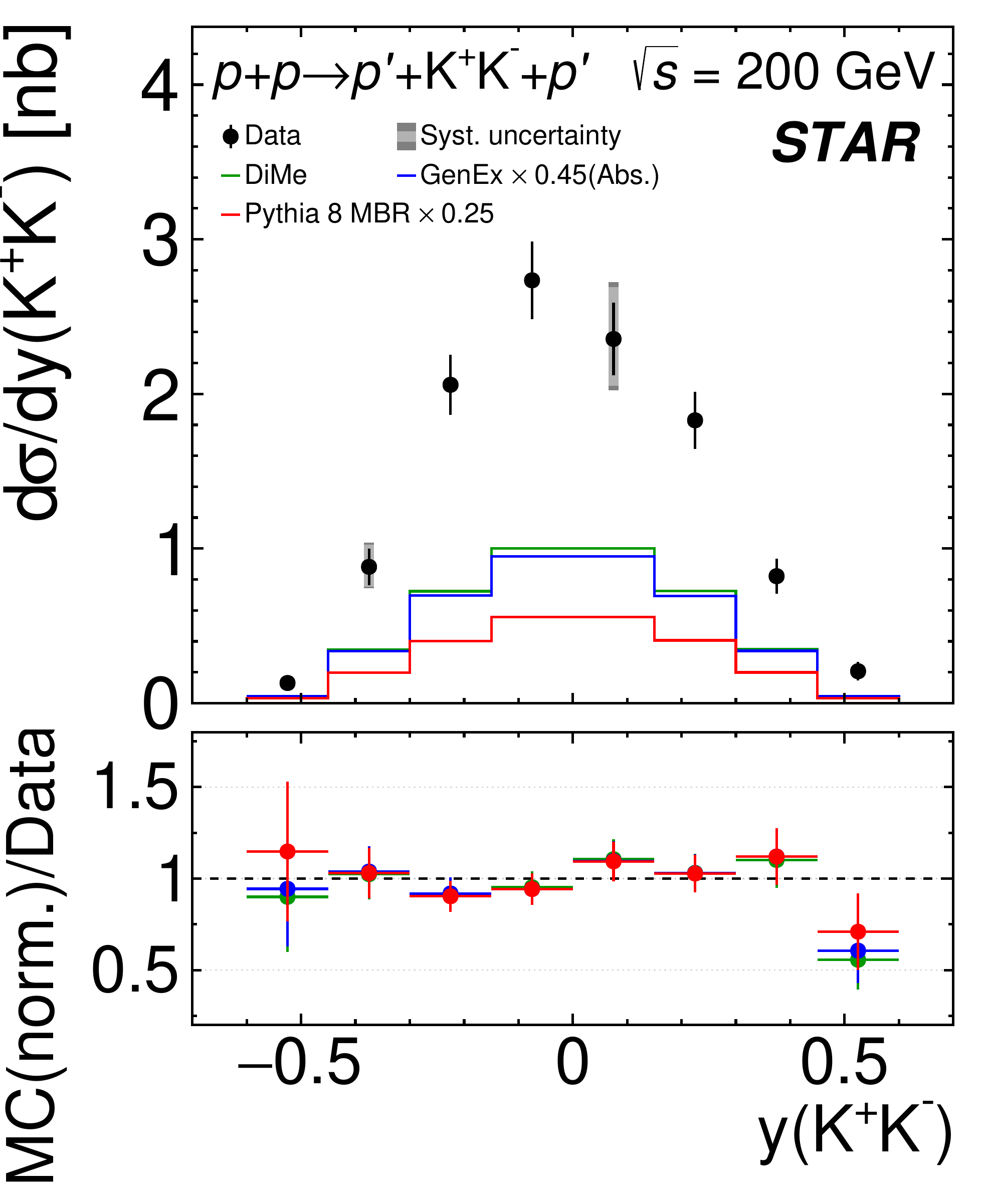}
\hfill
\includegraphics[width=.32\textwidth,page=1]{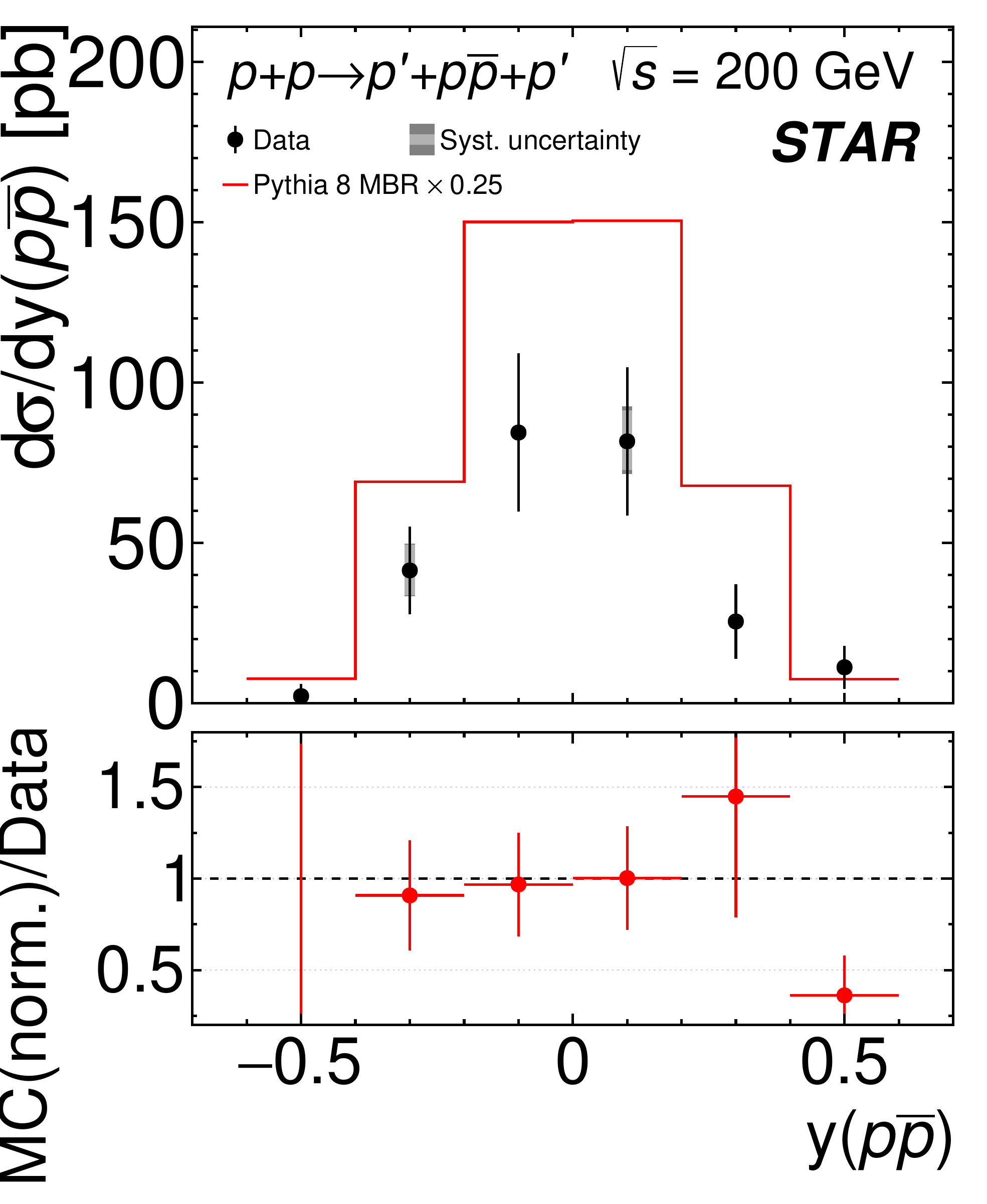}\vspace*{-6pt}
\caption{Differential cross sections for CEP of charged particle pairs $\pi^+\pi^-$ (left), $K^+K^-$ (middle) and $p\bar{p}$ (right) as a function of the pair rapidity measured in the fiducial region explained in the Sec.~\ref{sec:analysis}. Data are shown as solid points with error bars representing the statistical uncertainties. The typical systematic uncertainties are shown as gray boxes for only a few data points as they are almost fully correlated between neighboring bins. Predictions from three MC models, GenEx, DiMe and MBR, are shown as histograms. In the lower panels, the ratios between the MC predictions (scaled to the data for better shape comparison) and the data are shown.}\vspace*{-8pt}
\label{results_1}
\end{figure}

All results presented in this subsection are obtained in the fiducial region defined in Section~\ref{fiducial_region}.
%
In Figs.~\ref{results_01}, \ref{results_02} and \ref{results_1}, the differential fiducial cross sections related to the variables characterizing the centrally-produced hadron pairs are presented.
Figure~\ref{results_01} shows the differential cross section for CEP of $\pi^+\pi^-$ pairs as a function of the pair invariant mass.
There are several features of the distribution which need to be pointed out. The deep hole observed in the measured differential cross section $d\sigma/dm(\pi^+\pi^-)$ in the mass region $m(\pi^+\pi^-) < 0.6$~GeV is  mainly due to the fiducial cuts. At larger invariant masses,  resonance structures are seen in the data consistent with the $f_0(980)$ and $f_2(1270)$ mesons expected to be produced in the  Pomeron-Pomeron fusion process. At even higher invariant masses, another resonance is observed at $\sim 2.2$~GeV. The DiMe model roughly describes both the normalisation and the shape of the continuum production under the resonances, up to masses of about 1.9~GeV. In contrast, the GenEx model fails to describe the shape of the continuum production.
The MBR model prediction generally follows the shape of DiMe and GenEx predictions at masses below 1~GeV, but falls less rapidly with mass above 1~GeV. Notable are sharp drops of the predicted cross section at 0.7 GeV and 1.65 GeV. The former has been identified as a result of near-threshold-enhanced production of $\pi^++\pi^-$ + neutrals (mainly $2\pi^0$), which starts in PYTHIA8 around 0.7 GeV. It has already been demonstrated in Section~\ref{sec:systematics} that such events are extensively
overpopulated in PYTHIA8. The latter drop of the cross section at 1.65 GeV, present also in the prediction for $K^+K^-$, results from the fiducial cut on central particle pseudorapidities $|\eta|<0.7$ and peculiar correlation between the invariant mass and pseudorapidity of the final state
particles in PYTHIA8.

\begin{figure}[H]
\centering
\includegraphics[width=.49\textwidth,page=1]{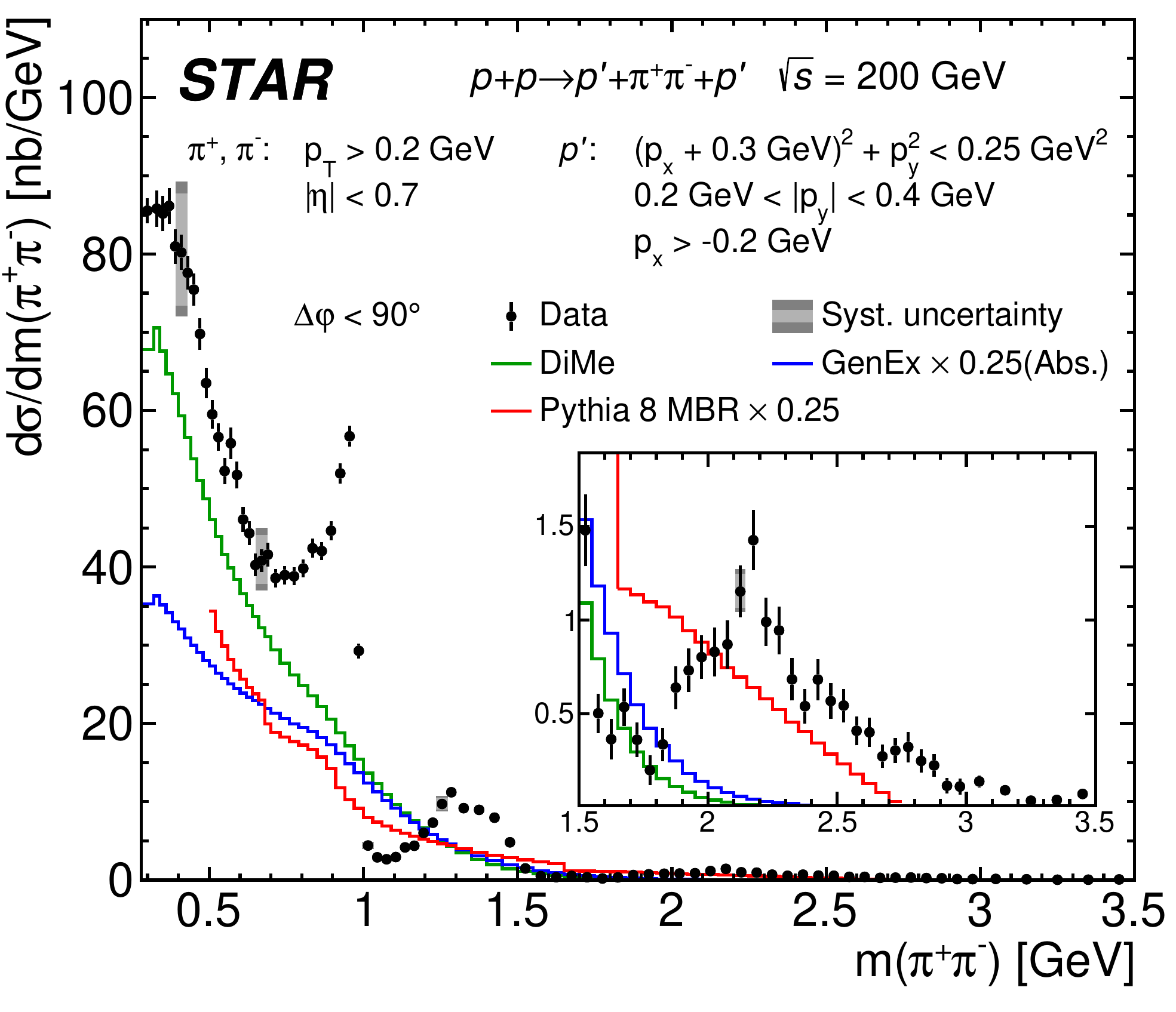}
\hfill
\includegraphics[width=.49\textwidth,page=1]{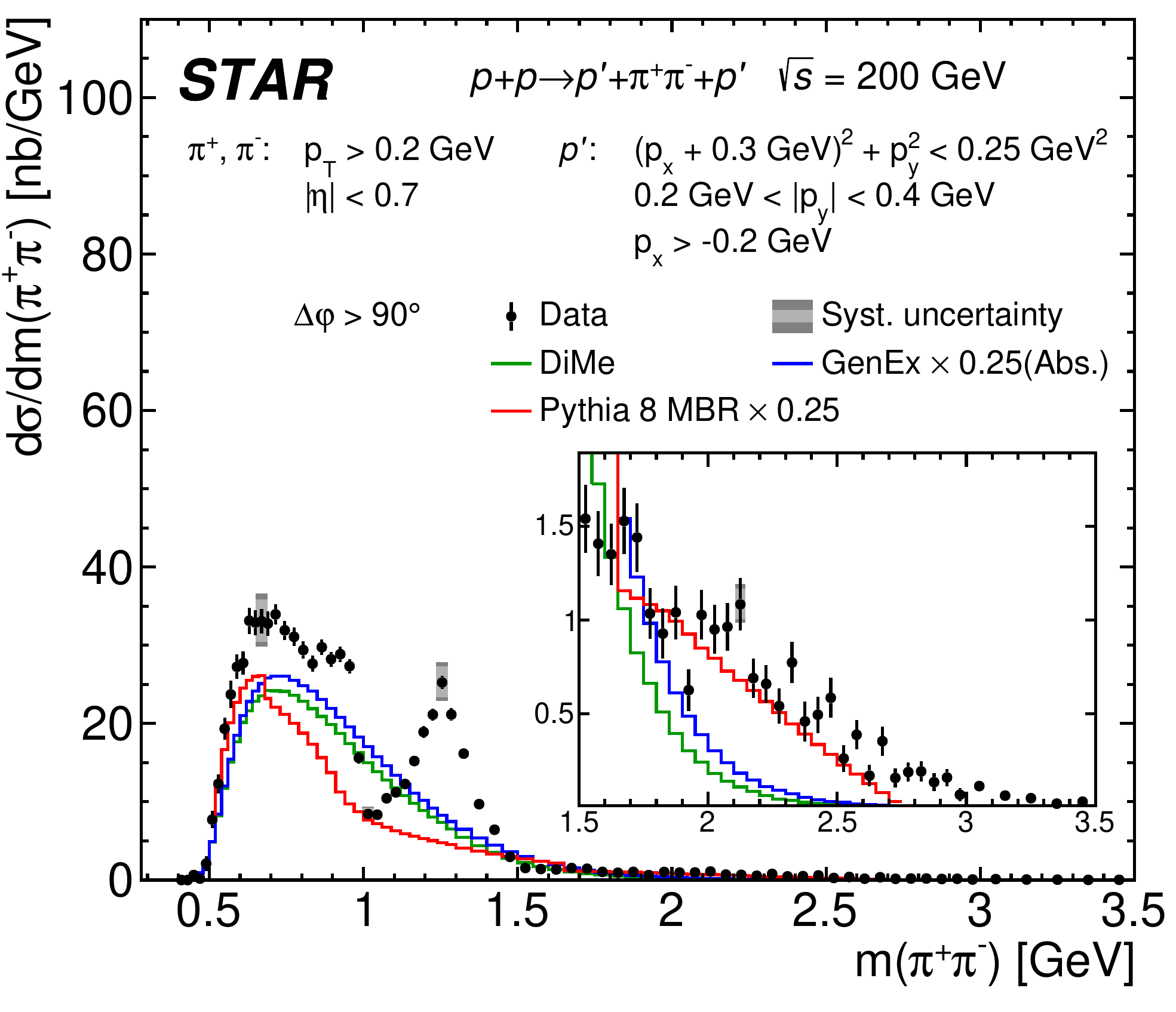}
\newline
\includegraphics[width=.49\textwidth,page=1]{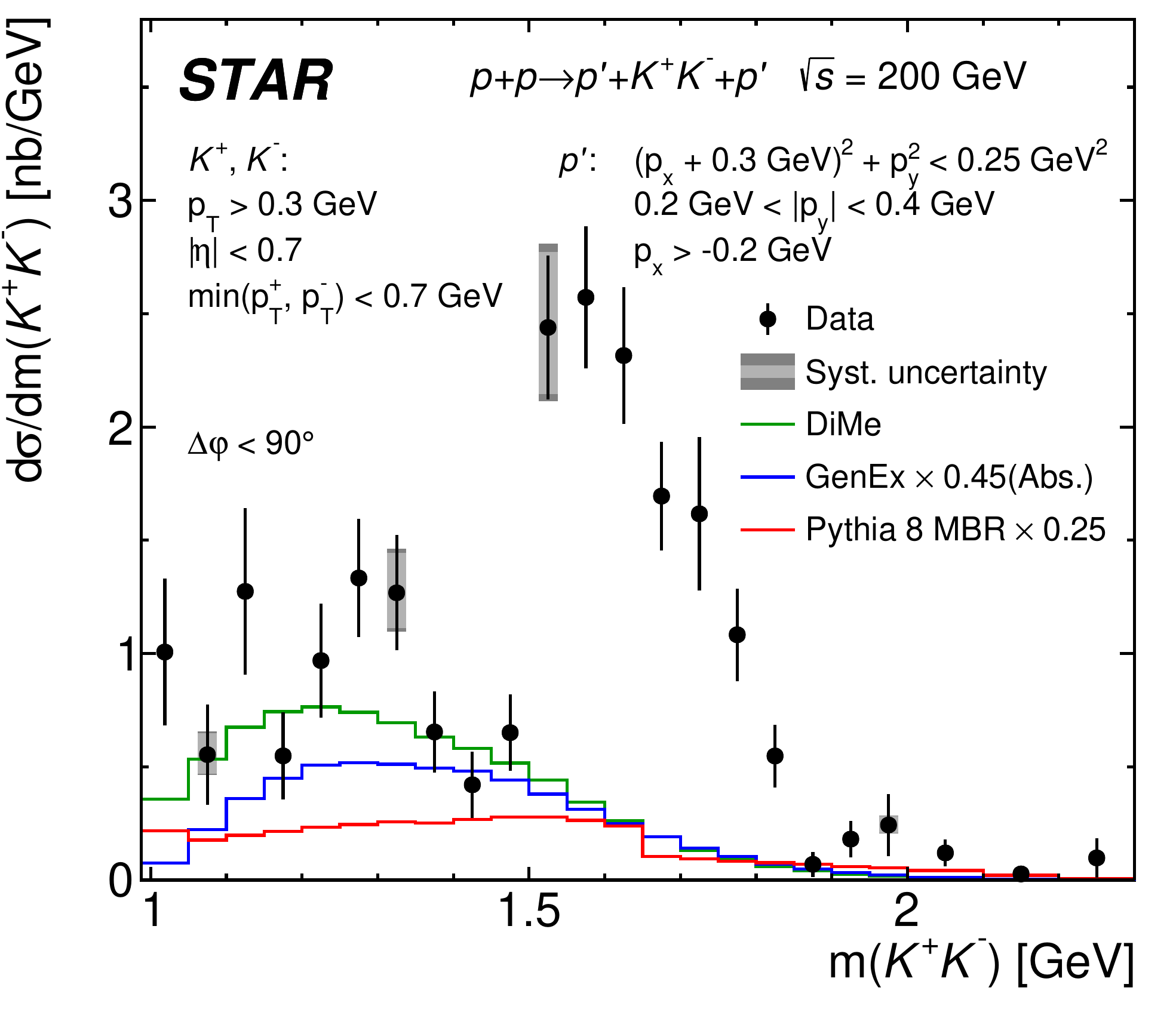}
\hfill
\includegraphics[width=.49\textwidth,page=1]{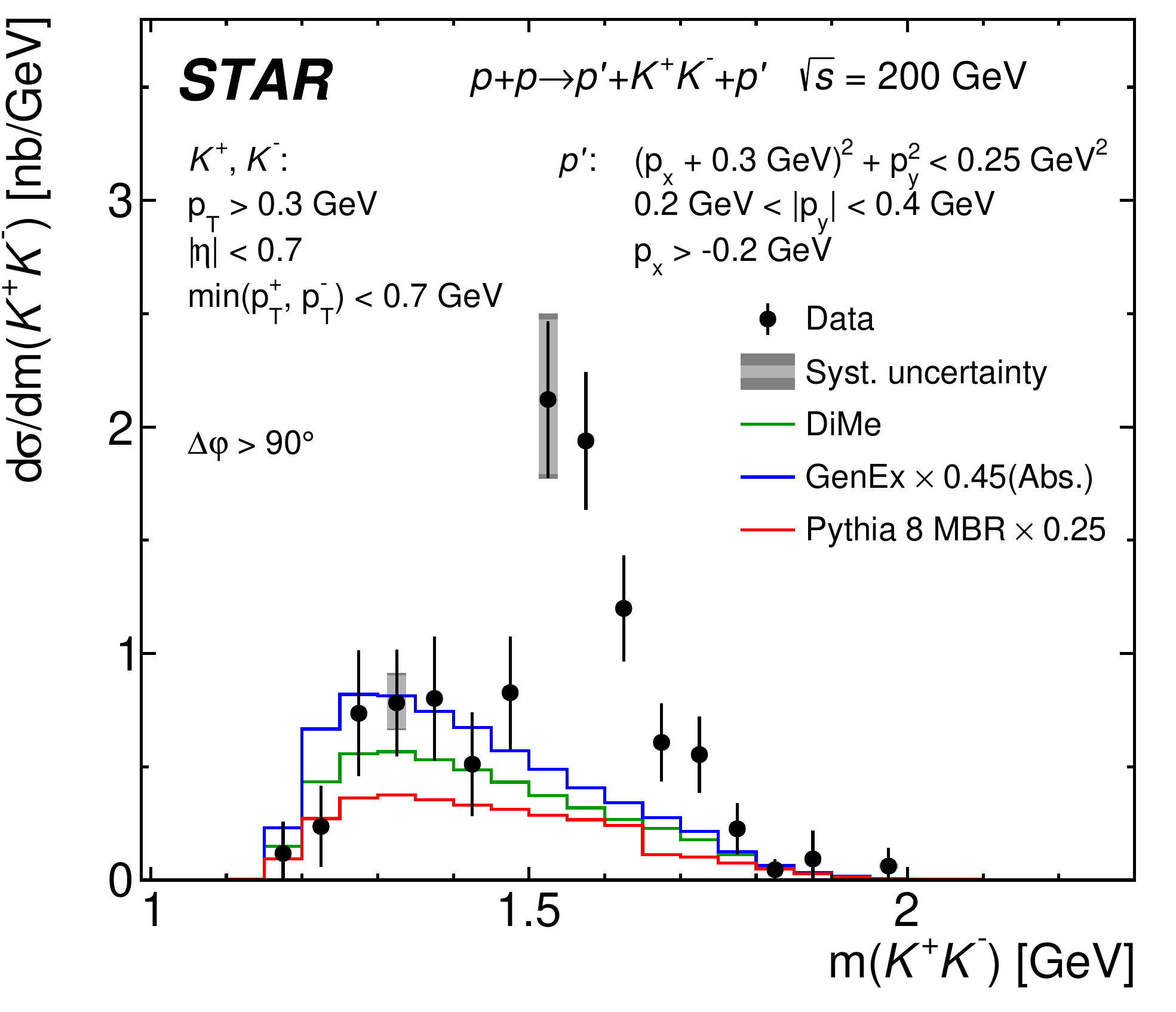}
\newline
\includegraphics[width=.49\textwidth,page=1]{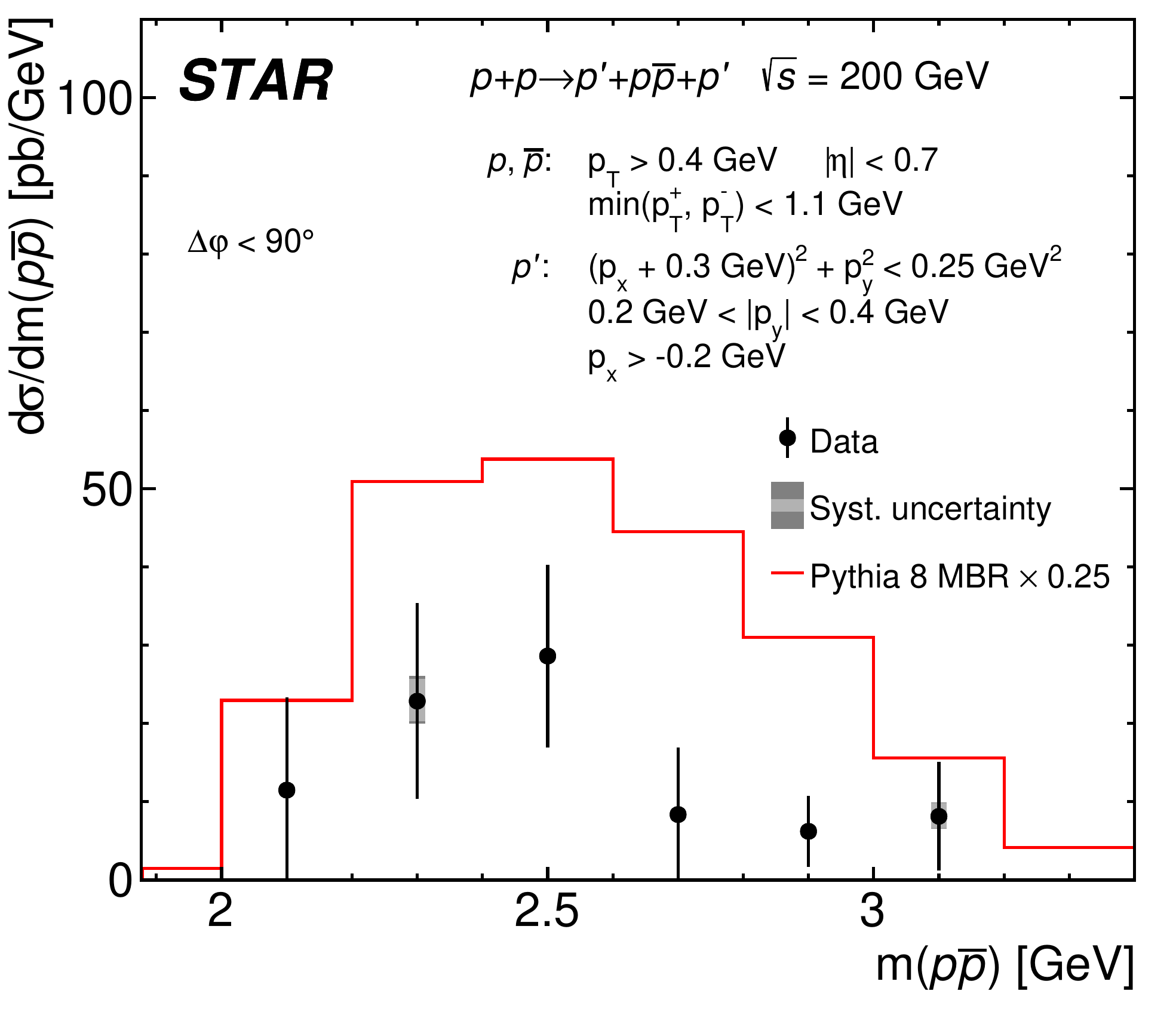}
\hfill
\includegraphics[width=.49\textwidth,page=1]{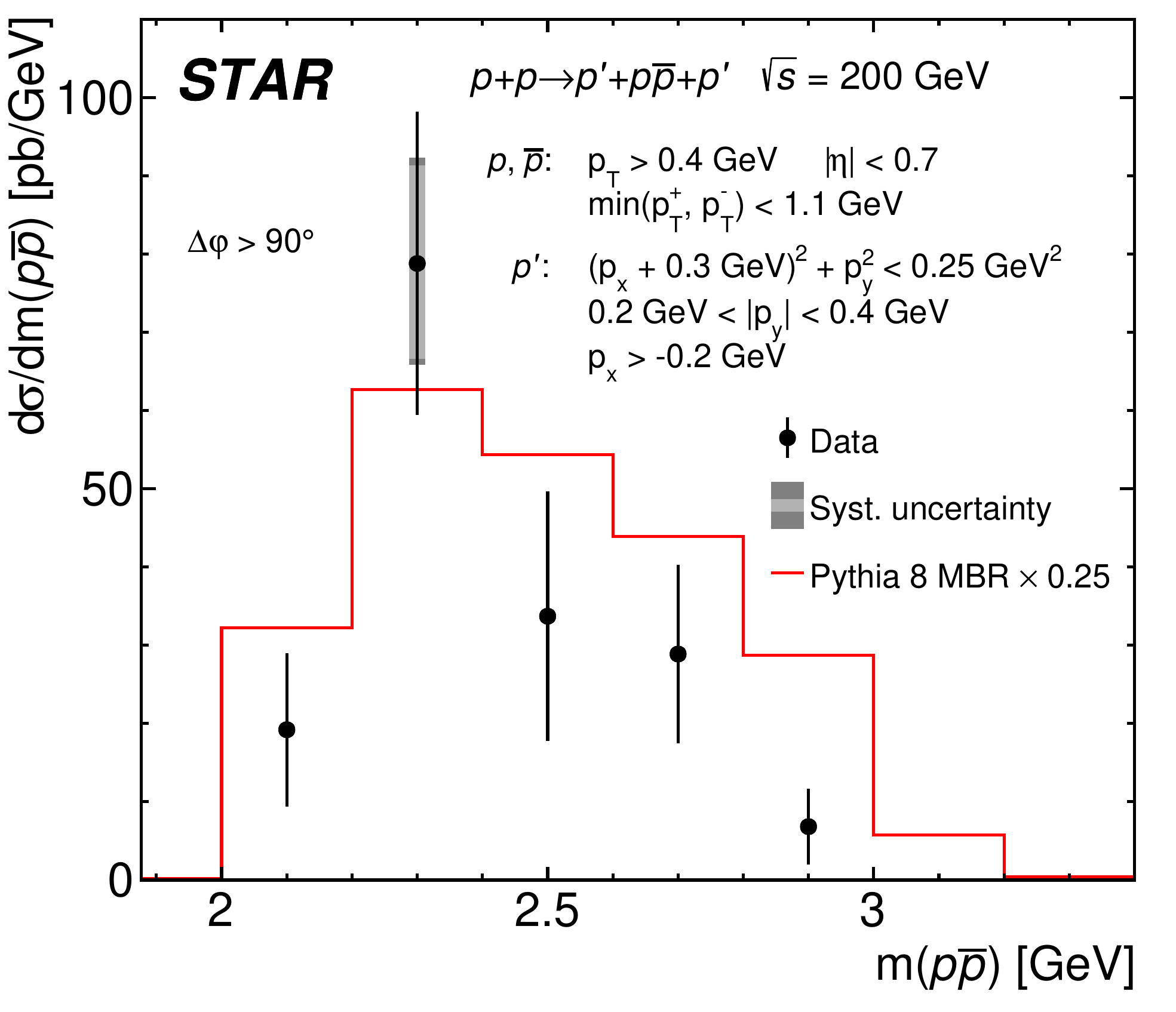}%
%
\caption{Differential cross sections for CEP of charged particle pairs $\pi^+\pi^-$ (top), $K^+K^-$ (middle) and $p\bar{p}$ (bottom) as a function of the invariant mass of the pair in two $\Delta\upvarphi$ regions, $\Delta\upvarphi<90^\circ$ (left column) and $\Delta\upvarphi>90^\circ$ (right column), measured in the fiducial region explained on the plots. Data are shown as solid points with error bars representing the statistical uncertainties. The typical systematic uncertainties are shown as gray boxes for only a few data points as they are almost fully correlated between neighboring bins. Predictions from three MC models, GenEx, DiMe and MBR, are shown as histograms.}
\label{results_3}
\end{figure}

\begin{figure}[H]%
\centering%
\includegraphics[width=.32\textwidth,page=1]{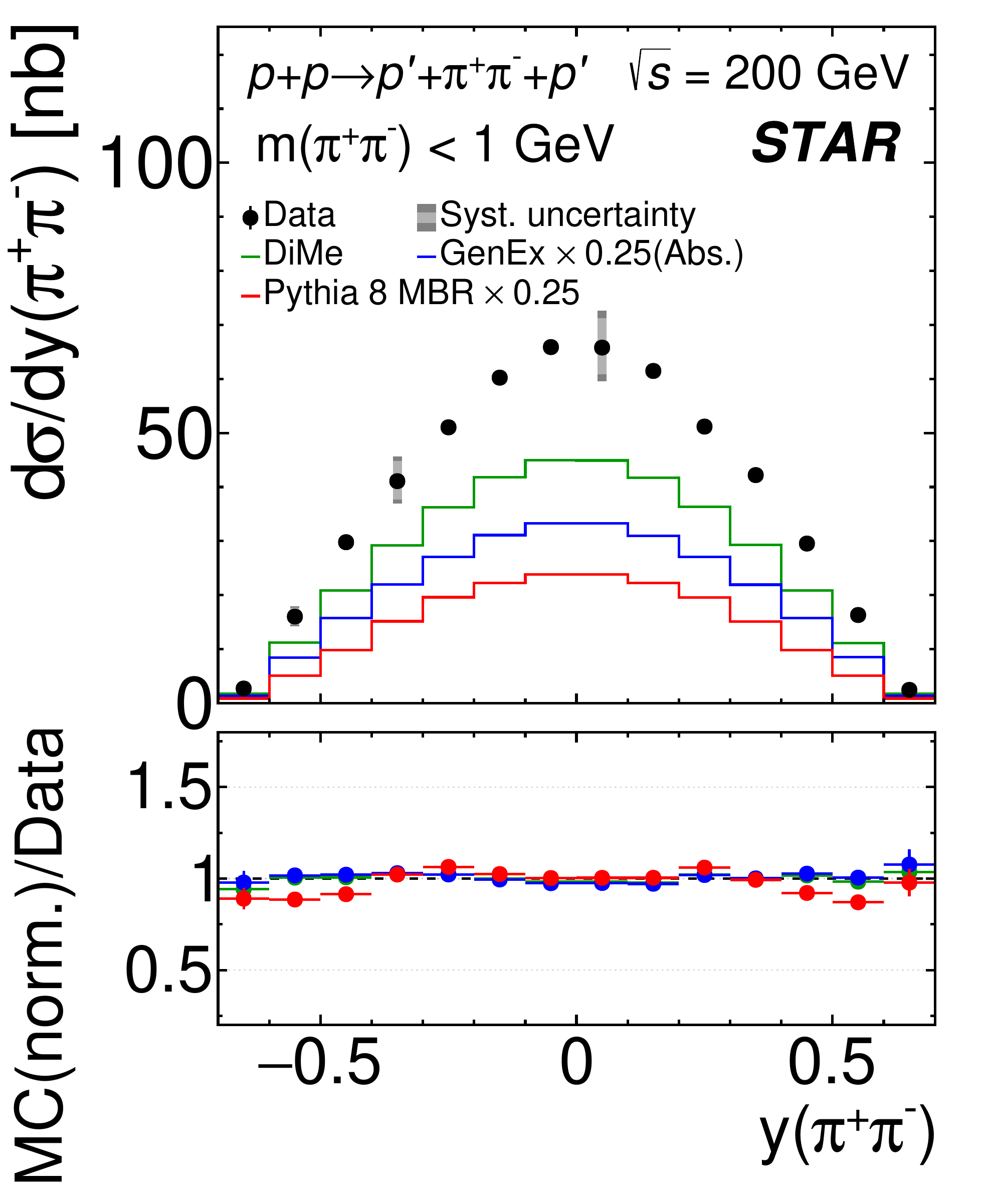}%
\hfill%
\includegraphics[width=.32\textwidth,page=1]{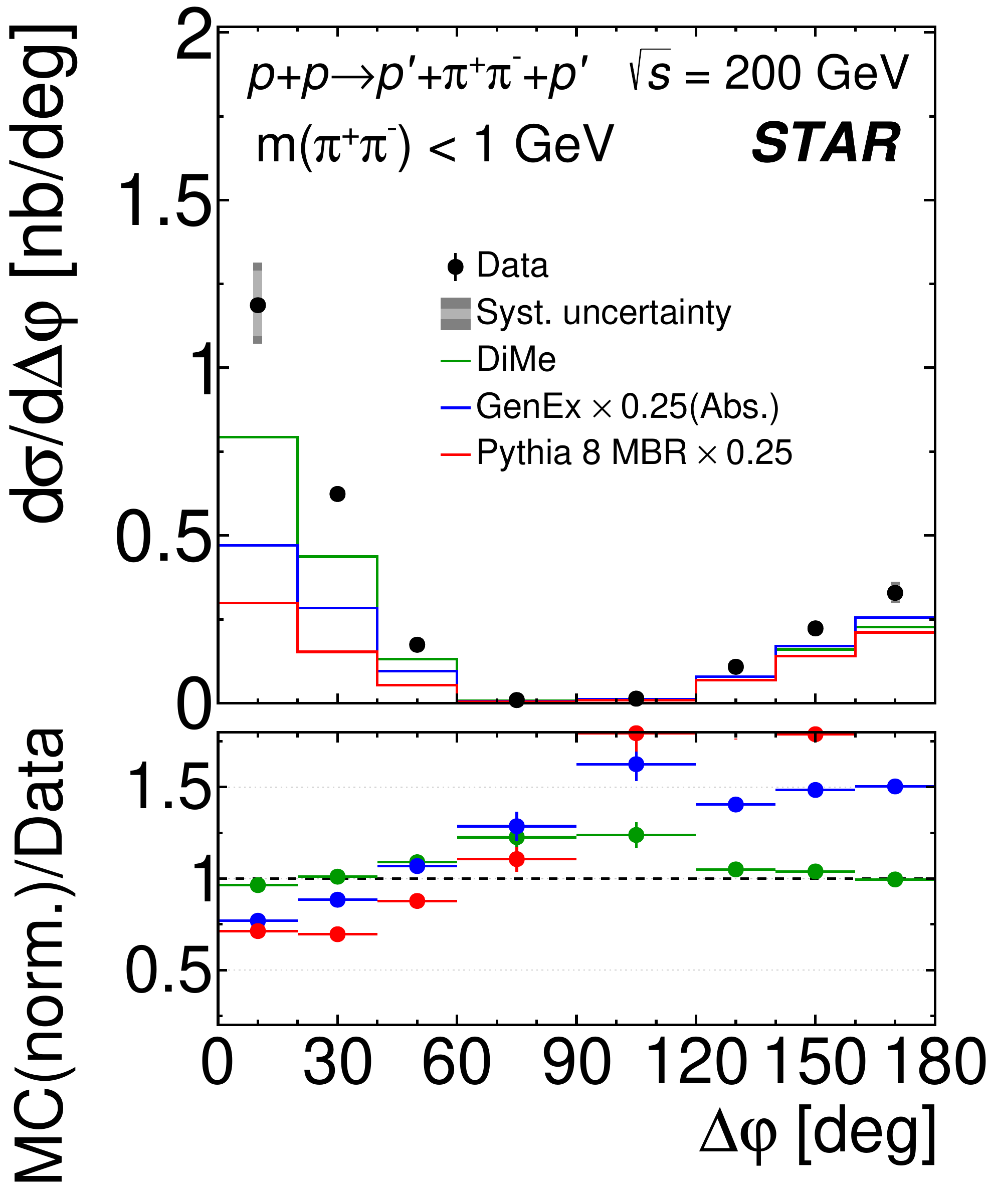}%
\hfill%
\includegraphics[width=.32\textwidth,page=1]{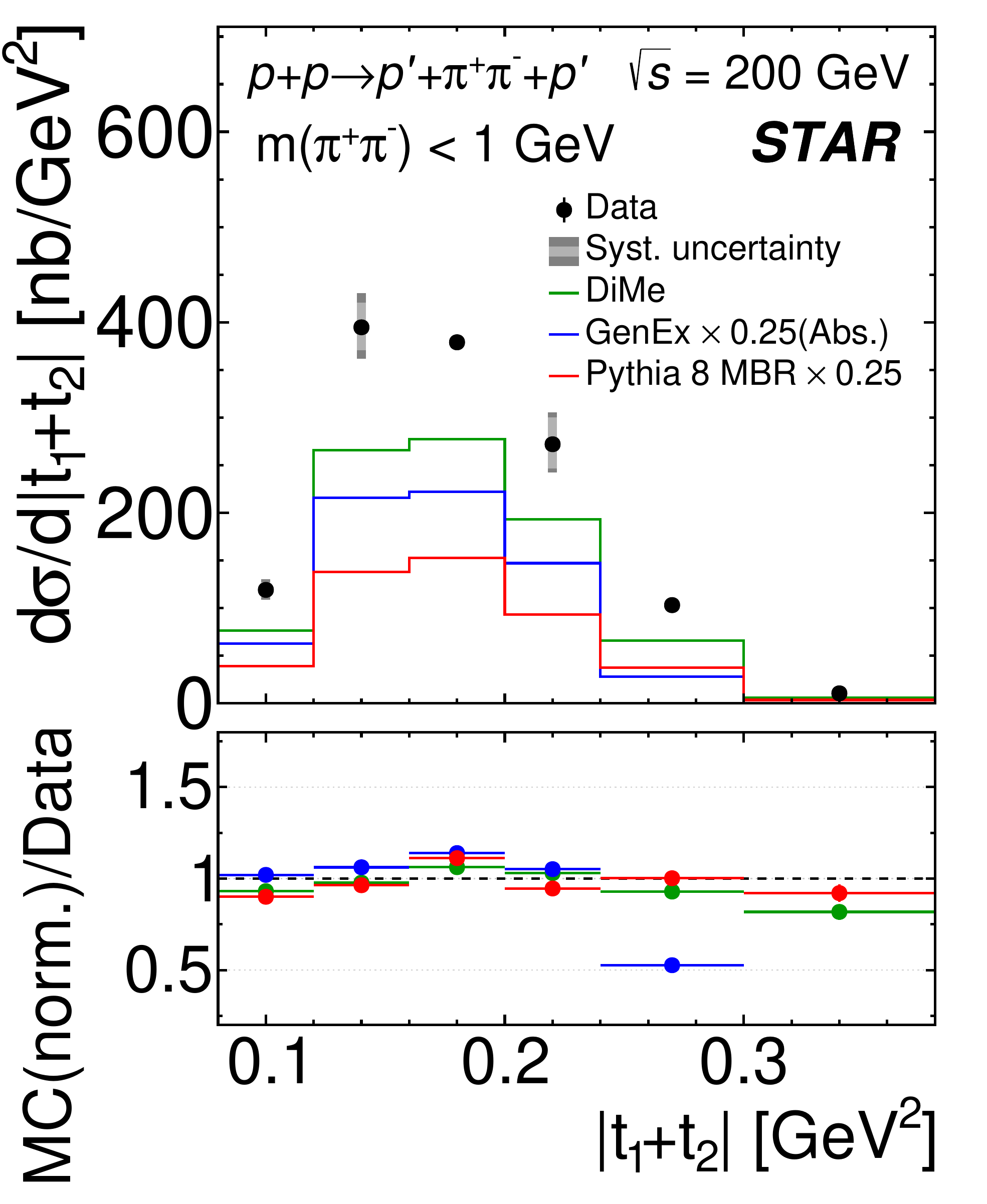}\\%
\includegraphics[width=.32\textwidth,page=1]{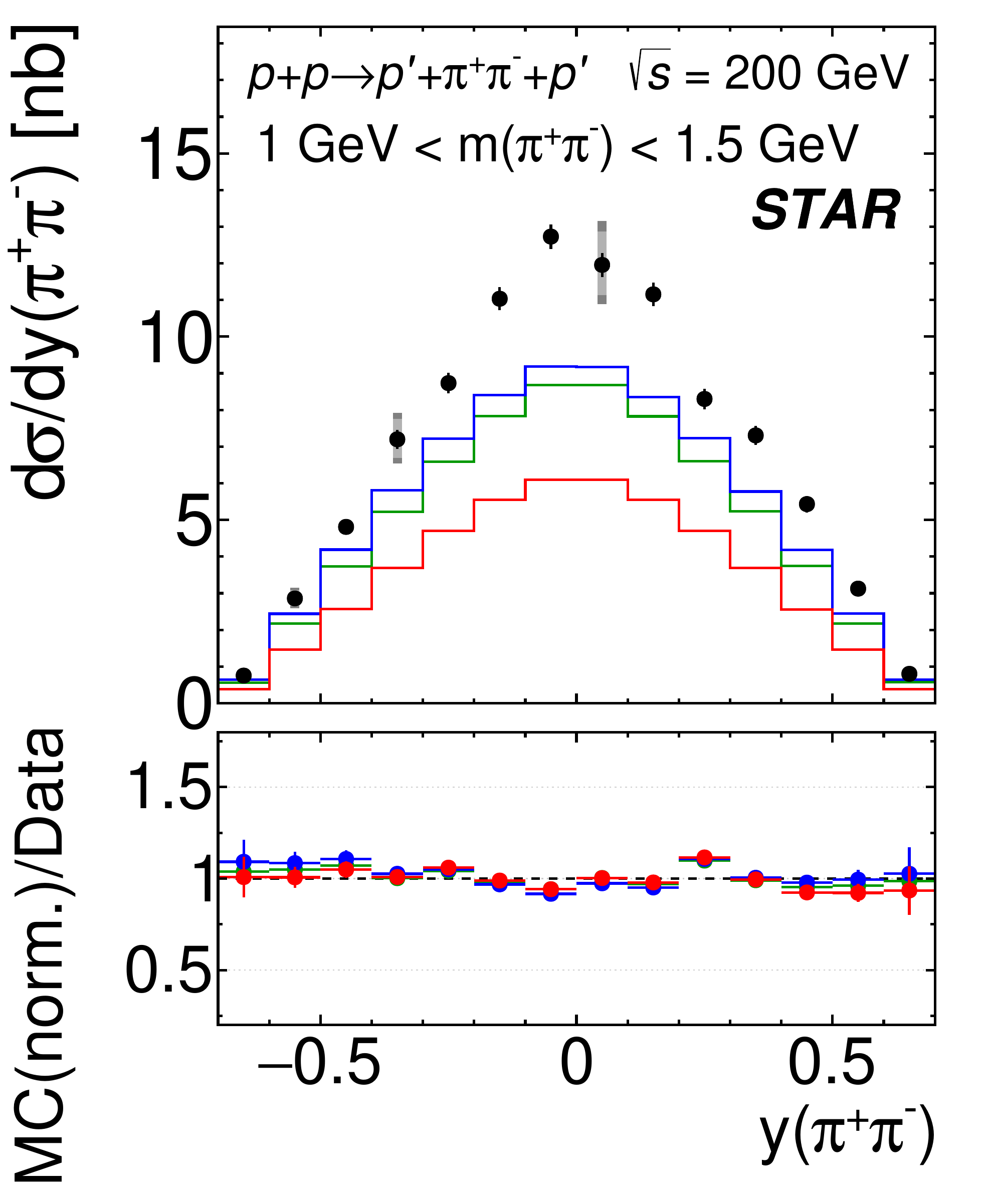}%
\hfill%
\includegraphics[width=.32\textwidth,page=1]{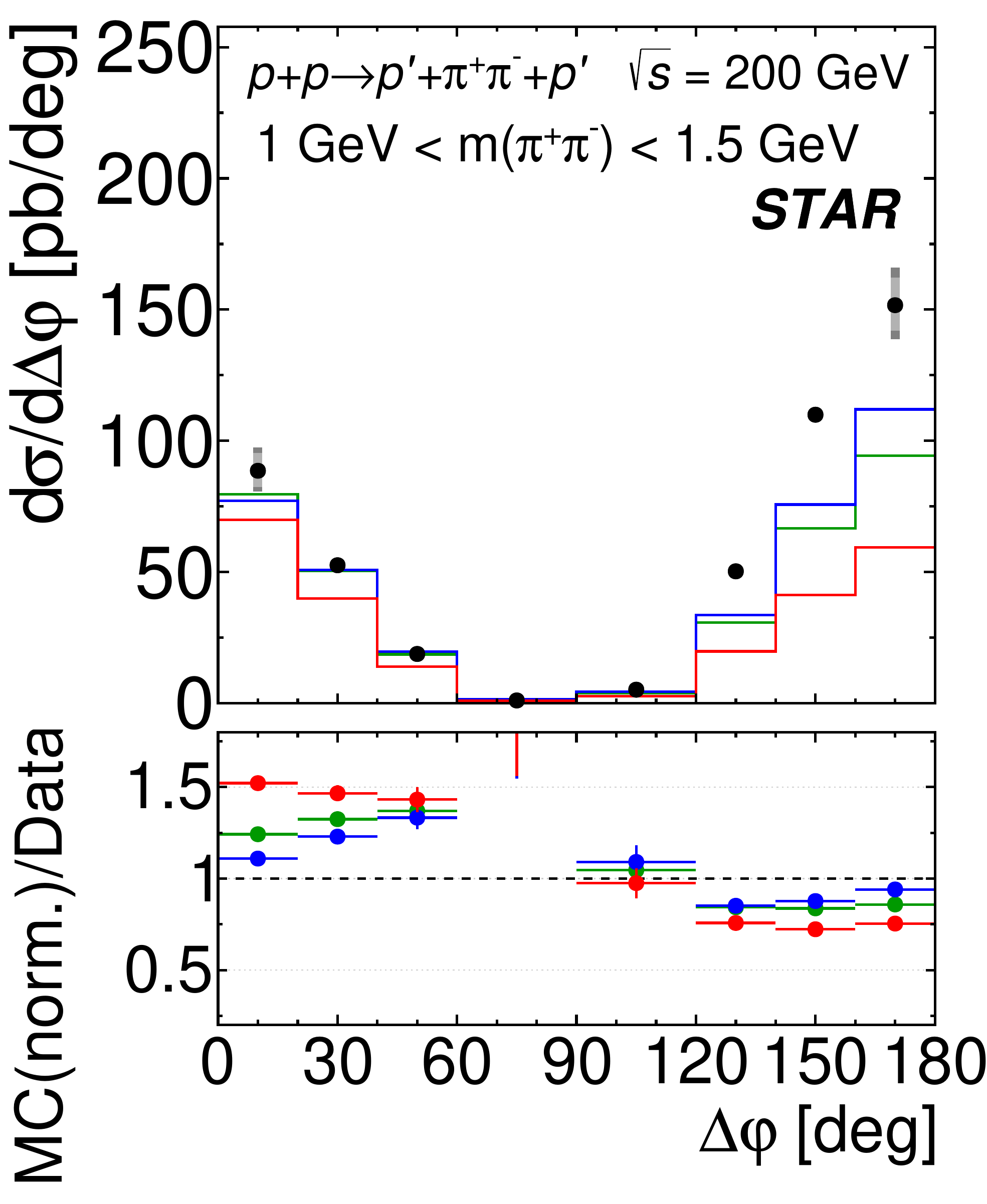}%
\hfill%
\includegraphics[width=.32\textwidth,page=1]{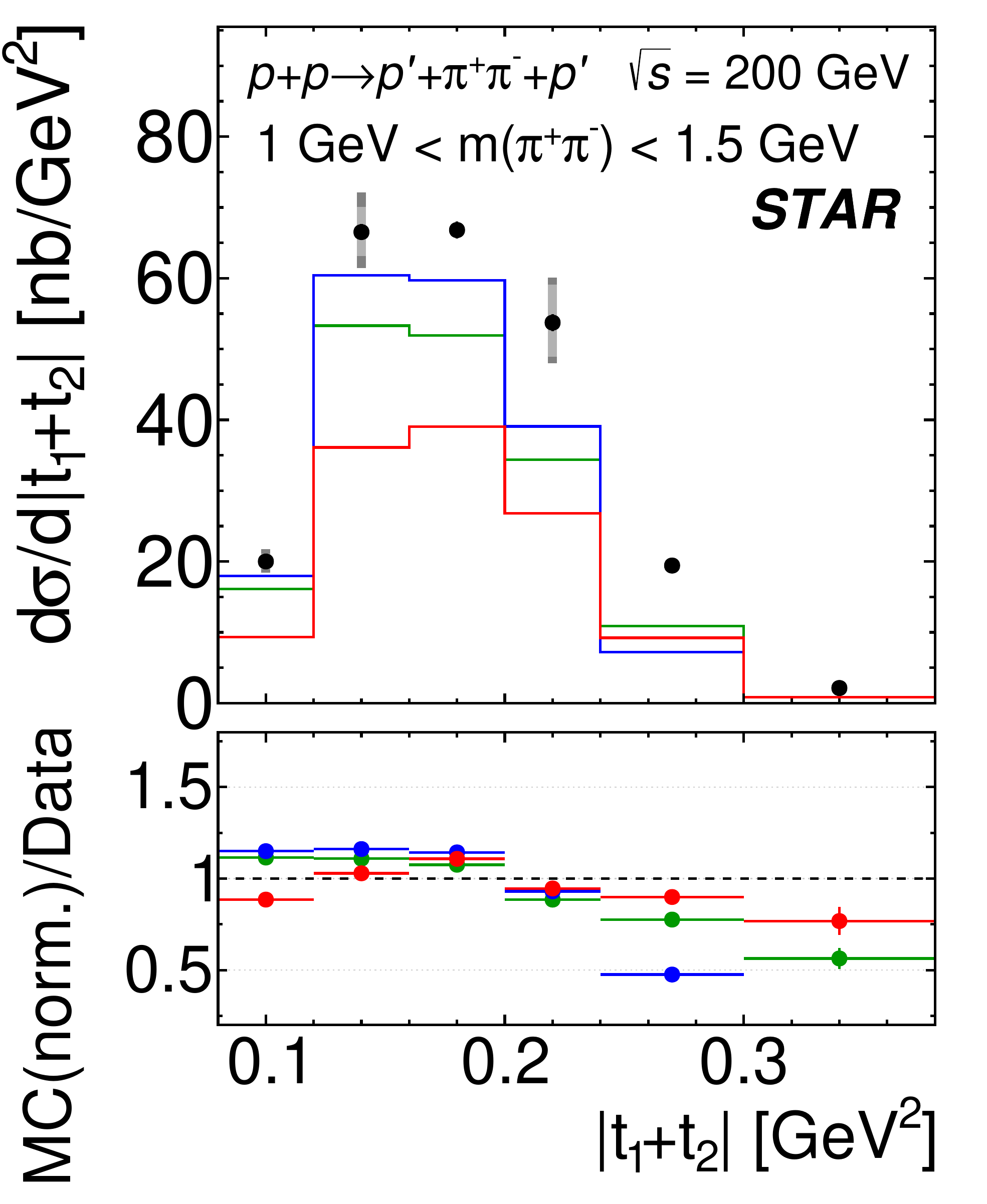}\\%
\includegraphics[width=.32\textwidth,page=1]{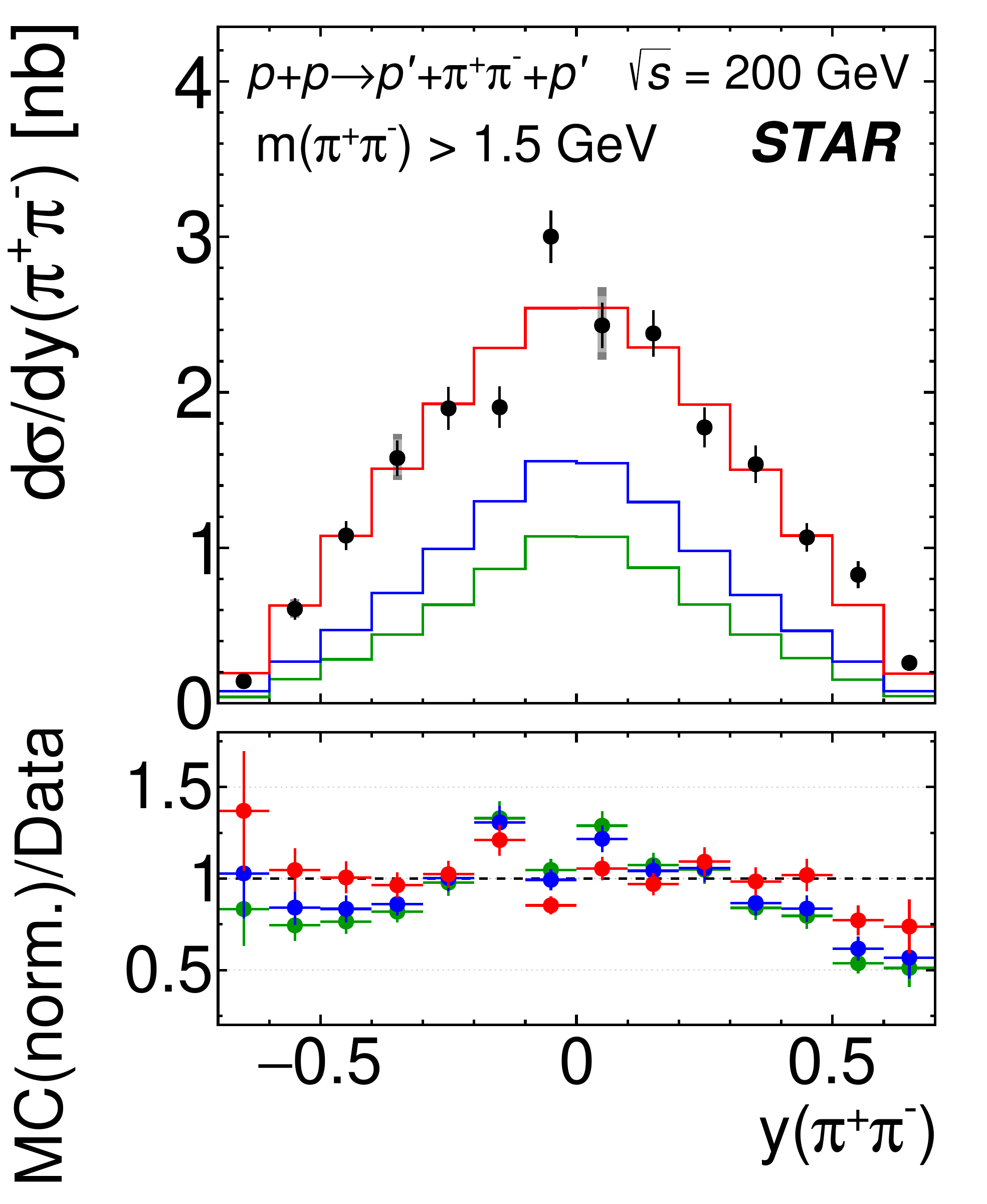}%
\hfill%
\includegraphics[width=.32\textwidth,page=1]{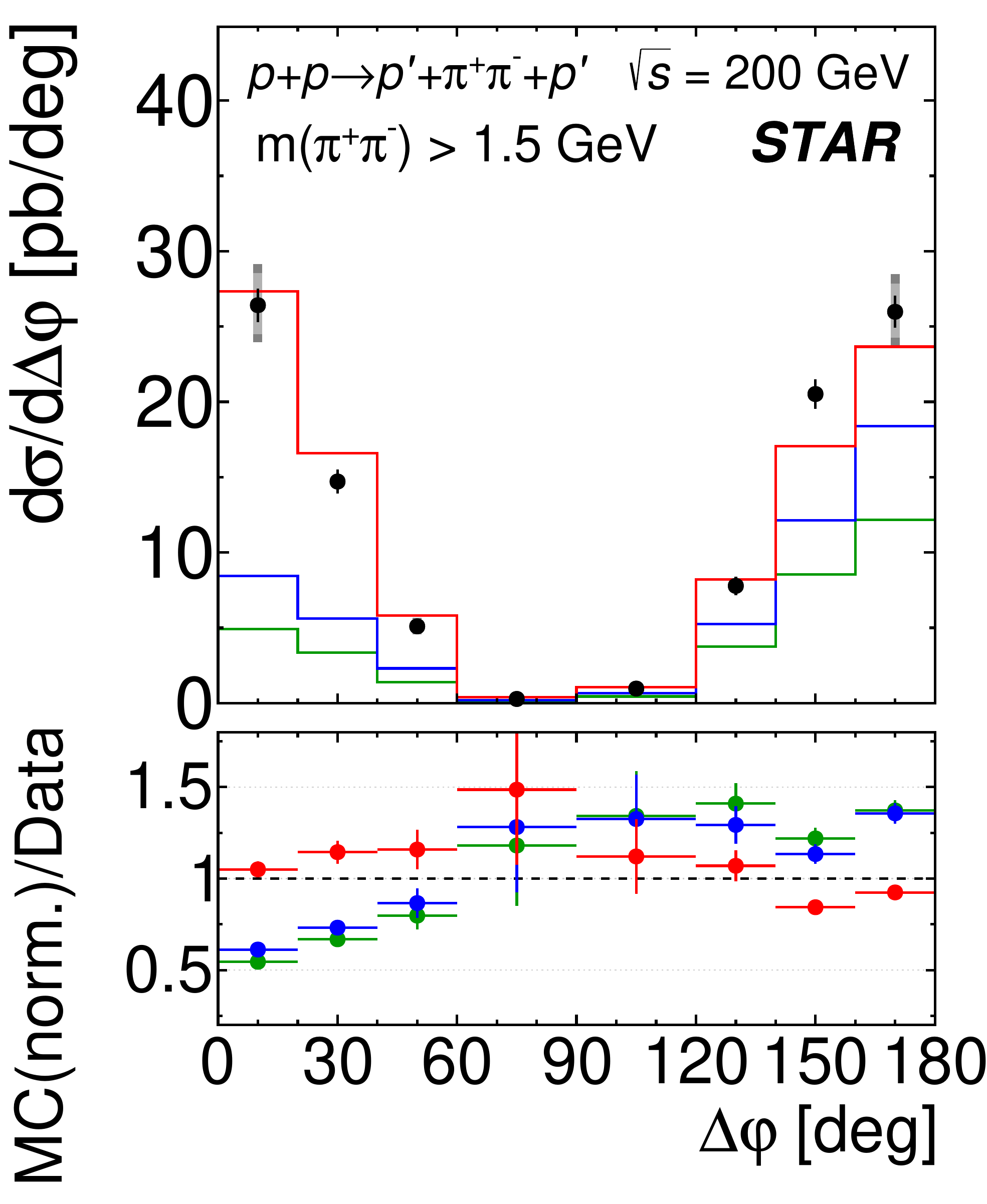}%
\hfill%
\includegraphics[width=.32\textwidth,page=1]{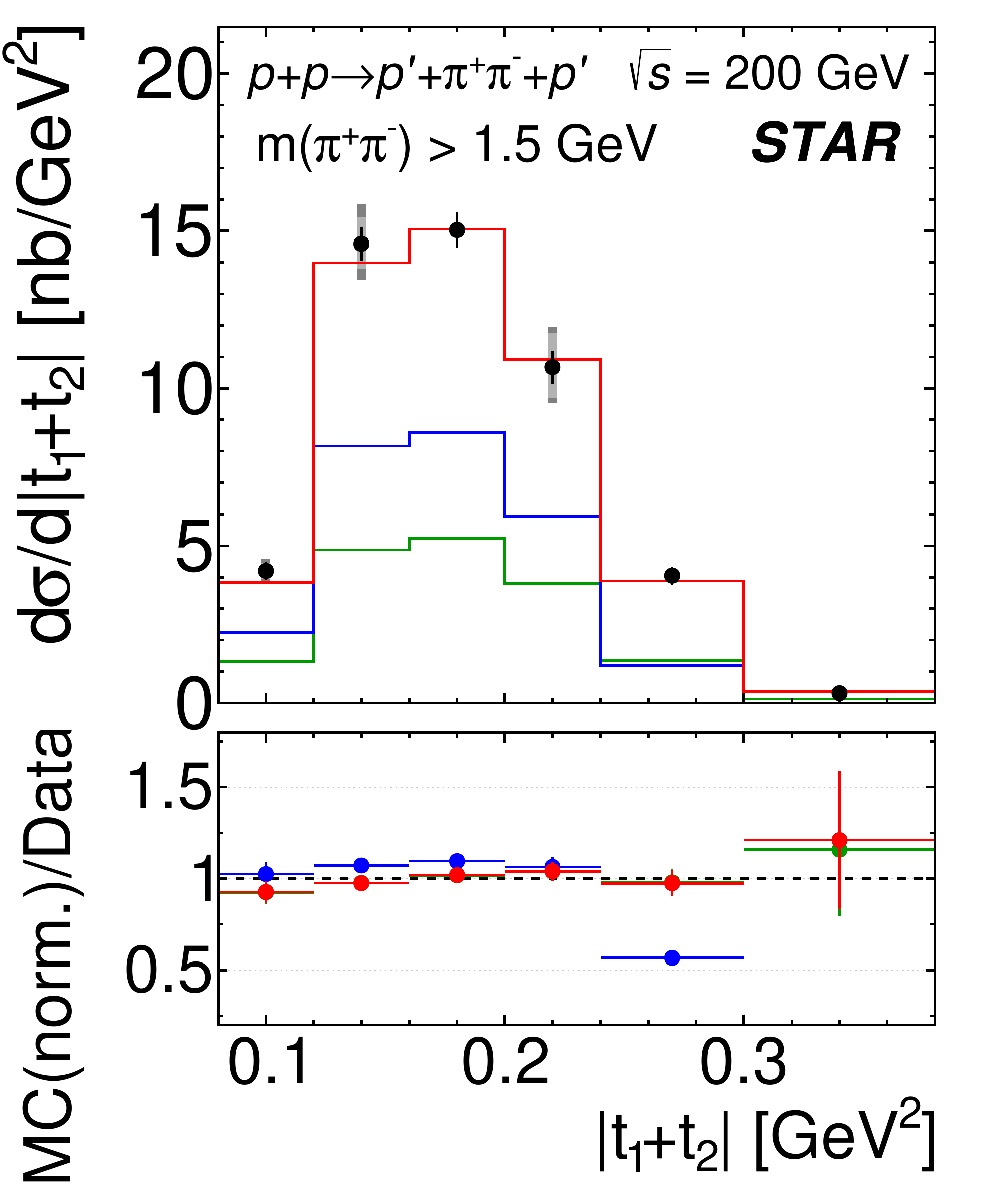}%
\caption{Differential cross sections for CEP of $\pi^+\pi^-$ pairs as a function of the rapidity of the pair (left column), the difference in azimuthal angles of the forward-scattered protons (middle column) and the sum of the squares of the four-momentum losses in the proton vertices (right column) measured in the fiducial region explained in the Sec.~\ref{sec:analysis}, separately for three ranges of the $\pi^+\pi^-$ pair invariant mass: $m<1$~GeV (top), $1\,\mbox{GeV}<m<1.5\,\mbox{GeV}$ (middle) and $m>1.5$~GeV (bottom). Data are shown as points with error bars representing the statistical uncertainties. The typical systematic uncertainties are shown as gray boxes for only a few data points as they are almost fully correlated between neighboring bins. Predictions from three MC models, GenEx, DiMe and MBR, are shown as histograms. In the lower panels, the ratios between the MC predictions (scaled to data) and \mbox{the data are shown.}}
\label{results_4}
\end{figure}

\begin{figure}[ht]%
\centering%
\includegraphics[width=.317\textwidth,page=1]{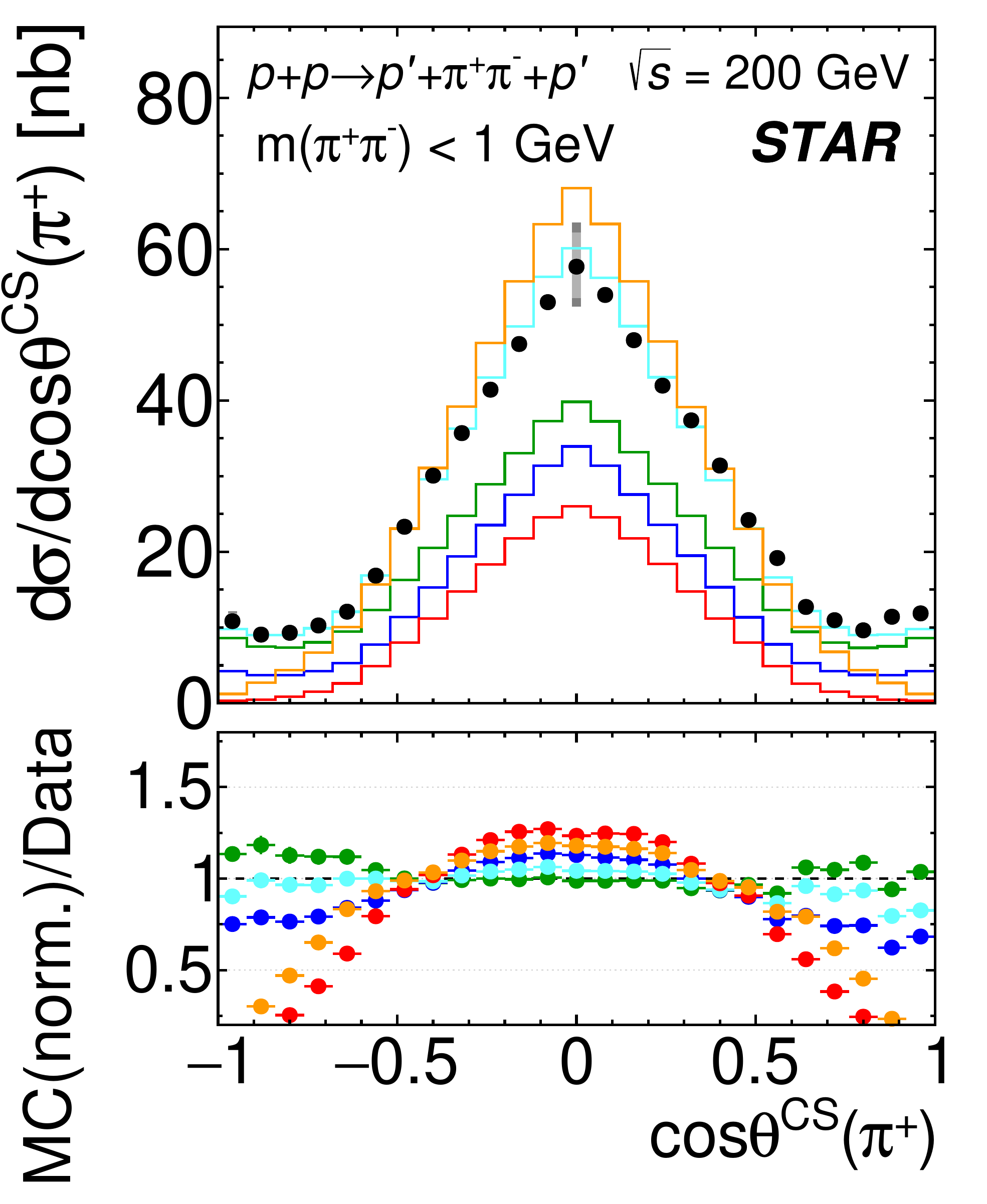}%
\hfill%
\includegraphics[width=.317\textwidth,page=1]{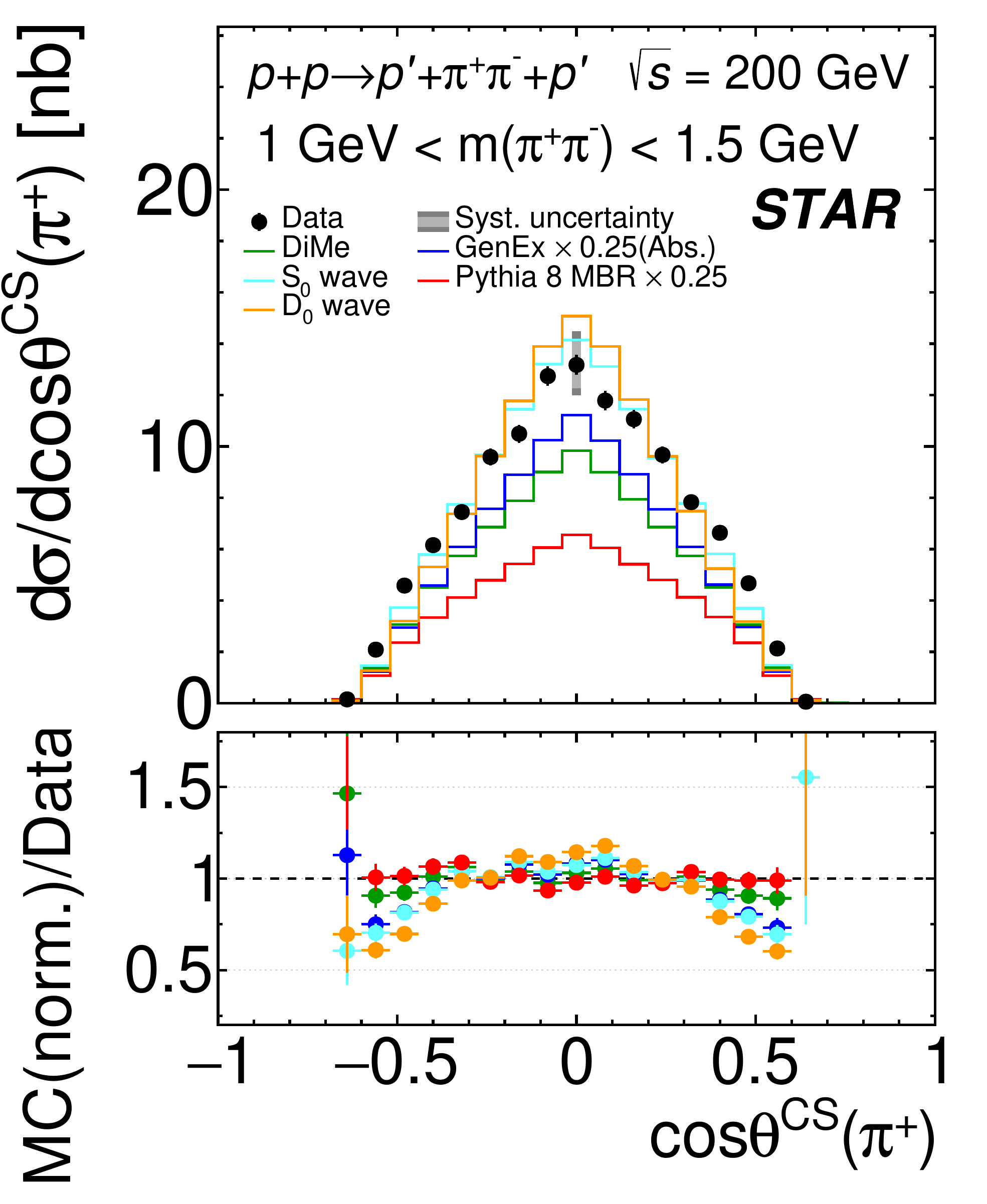}%
\hfill%
\includegraphics[width=.317\textwidth,page=1]{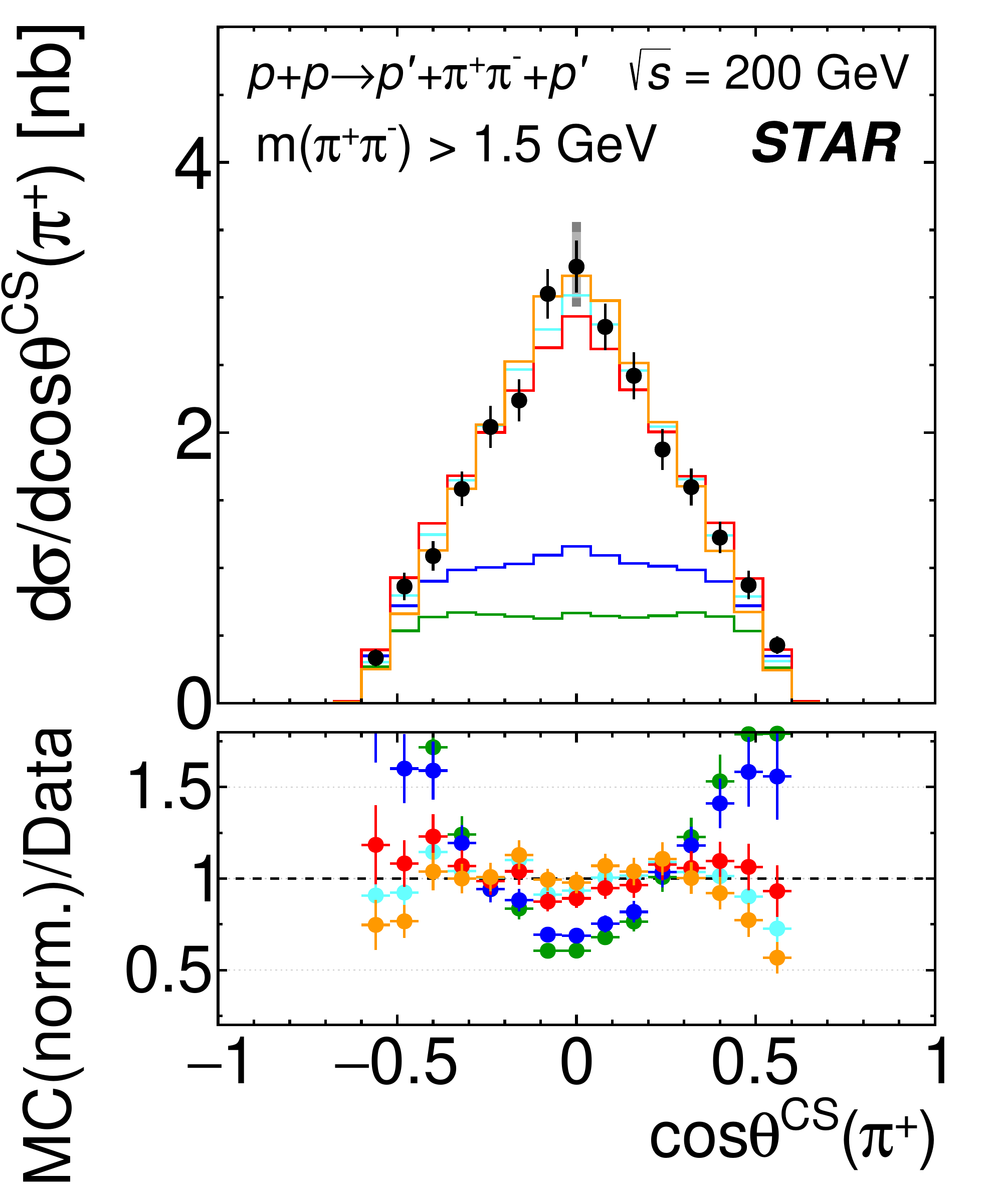}\\%
\includegraphics[width=.32\textwidth,page=1]{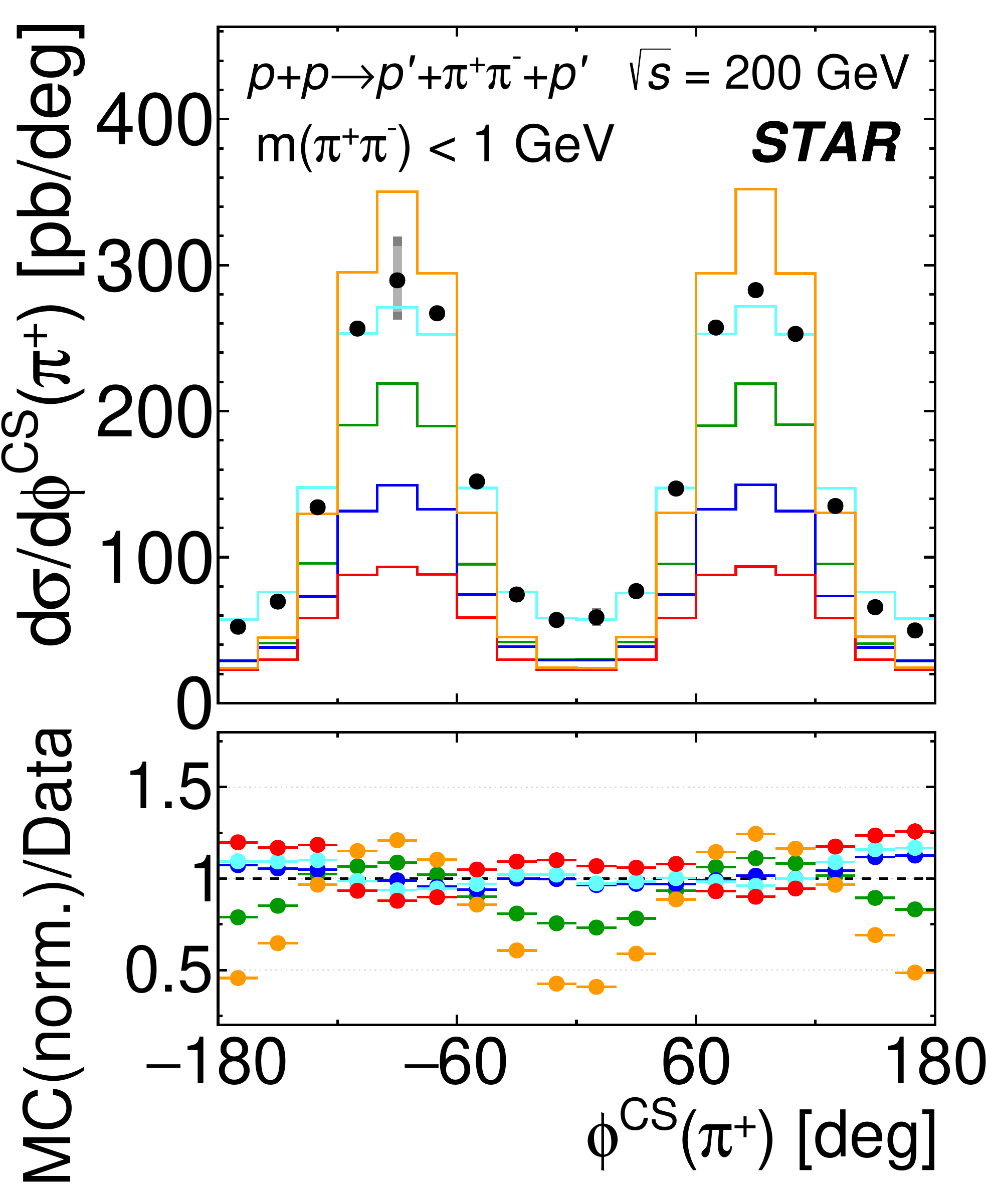}%
\hfill%
\includegraphics[width=.32\textwidth,page=1]{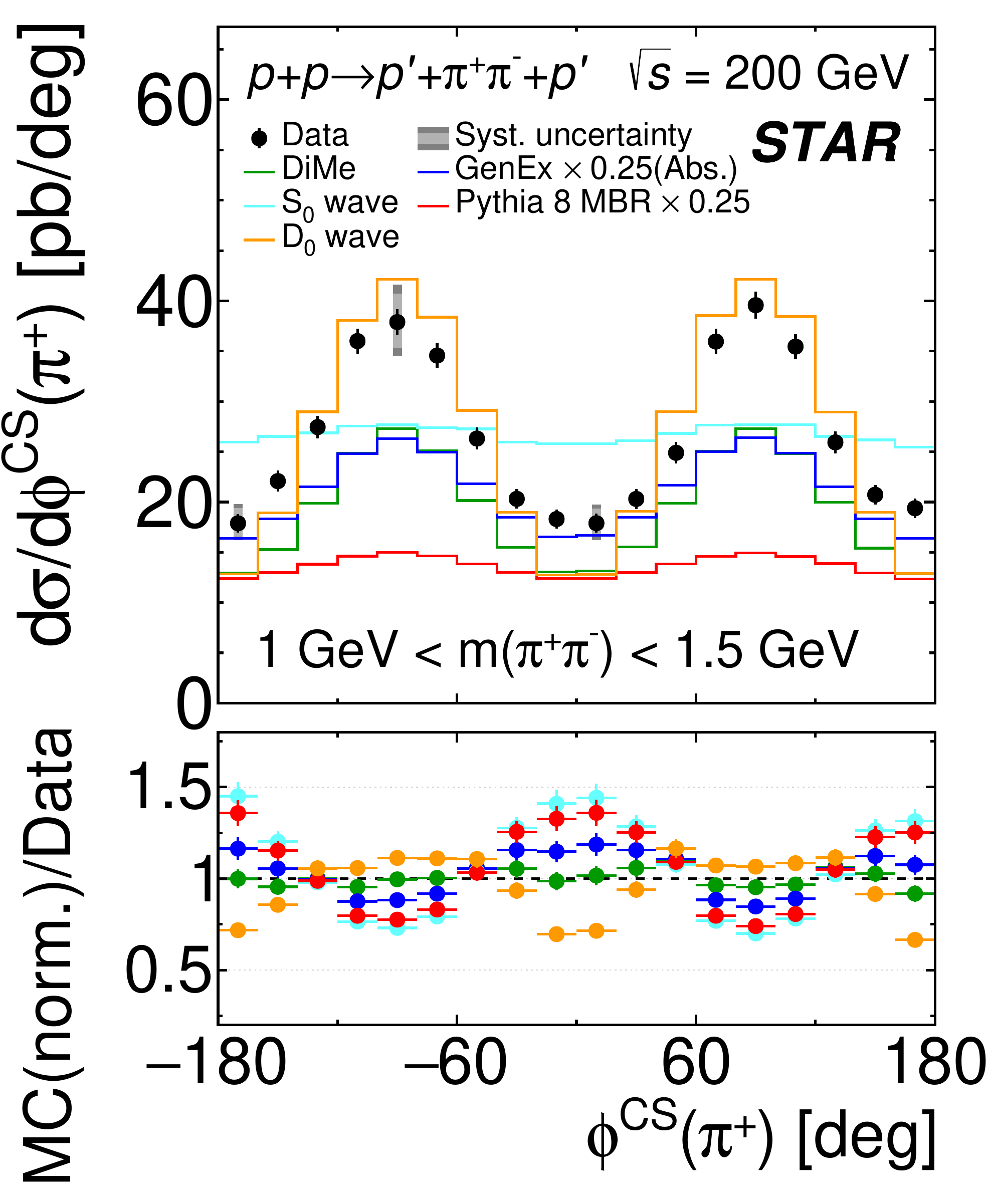}%
\hfill%
\includegraphics[width=.32\textwidth,page=1]{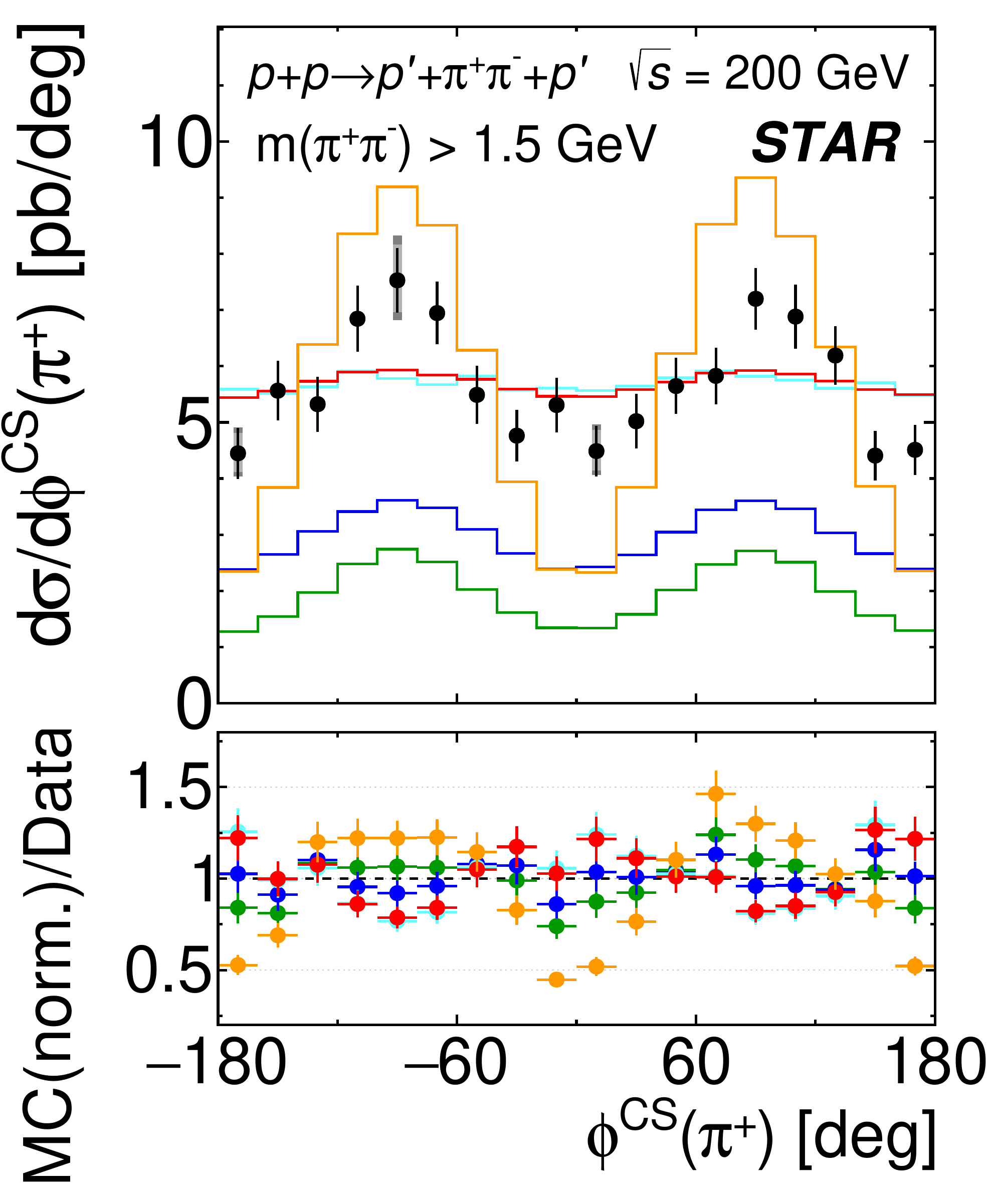}%
\caption{Differential cross sections for CEP of $\pi^+\pi^-$ pairs as a function of $\cos{\uptheta^\mathrm{CS}}$ (top) and of $\upphi^\mathrm{CS}$ (bottom) measured in the fiducial region is shown in three ranges of the $\pi^+\pi^-$ pair invariant mass: $m<1$ GeV (left column), $1\,\mbox{GeV}<m<1.5\,\mbox{GeV}$ (middle column) and $m>1.5$ GeV (right column). Data are shown as solid points with error bars representing their statistical uncertainties. The typical systematic uncertainties are shown as gray boxes for only a few data points as they are almost fully correlated between neighboring bins. Predictions from three MC models, GenEx, DiMe and MBR, as well as from pure $S_{0}$ and $D_{0}$ waves are shown as histograms. In the lower panels the ratios between the MC predictions (scaled to data) and the data are shown.}
\label{results_7}\vspace*{-10pt}
\end{figure}

In Fig.~\ref{results_02}, the differential fiducial cross sections for CEP of $K^+K^-$ and $p\bar{p}$ pairs are shown. The measured differential cross section, $d\sigma/dm(K^+K^-)$, shows significant enhancement in the $f_2^\prime(1525)$ mass region and a possible smaller resonant signal in the mass region of $f_2(1270)$. Both structures are expected to be produced in the Pomeron-Pomeron fusion process. The ratio of the cross sections for $\pi^+\pi^-$ to $K^+K^-$ production in the $f_2(1270)$ mass region is roughly 18, consistent with the PDG ratio of the $f_2(1270)$ branching fractions for its decays into $\pi^+\pi^-$ and $K^+K^-$~\cite{pdg}, assuming similar contributions from non-resonant production under the $f_2(1270)$ peaks and similar STAR acceptance. The DiMe and GenEx predictions roughly describe the non-resonant contribution to the data in the resonance region. The data are also consistent with the ratio of the non-resonant exclusive production of $\pi^+\pi^-$ to $K^+K^-$ pairs expected by GenEx and DiMe.
In the case of the differential cross section $d\sigma/dm(p\bar{p})$, only predictions from the MBR model are available and they overestimate the data by a factor of 8.\\
\indent
Figure~\ref{results_1} shows the differential cross sections for CEP of different particle species pairs as a function of the pair rapidity. 
The shapes of the measured distributions are generally well described by all the model predictions. \\
\begin{figure}
\centering
\includegraphics[width=.32\textwidth,page=1]{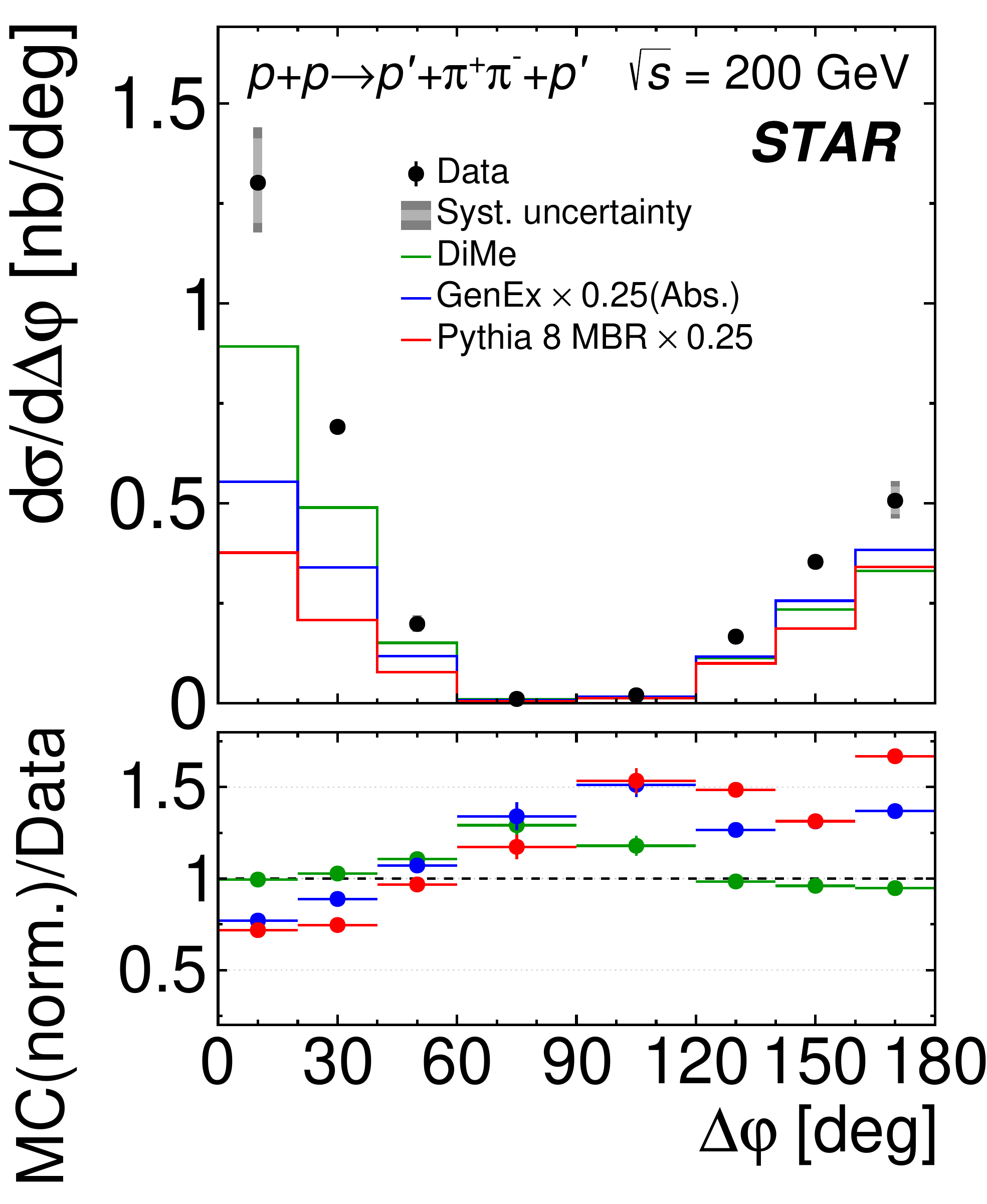}
\hfill
\includegraphics[width=.32\textwidth,page=1]{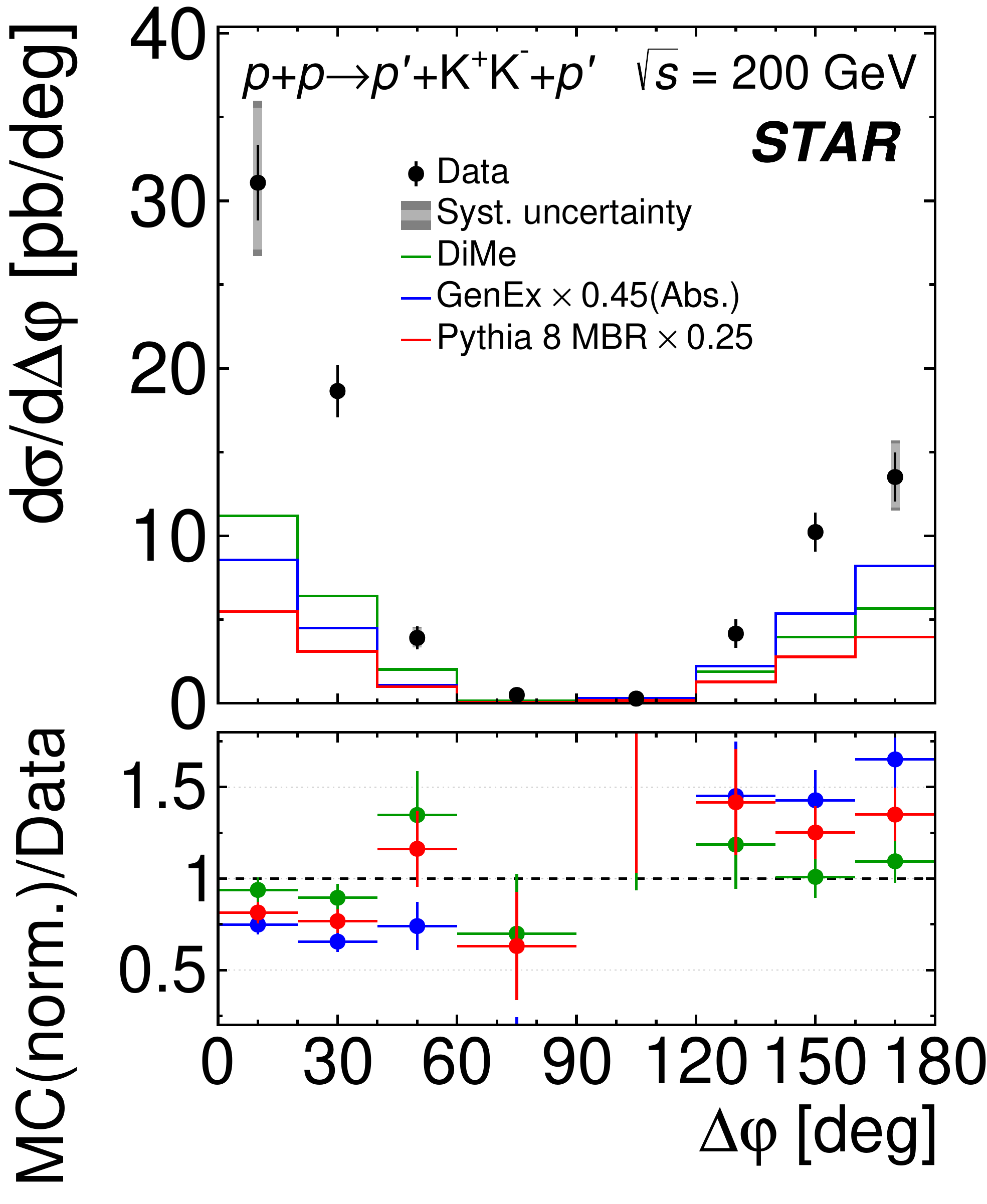}
\hfill
\includegraphics[width=.32\textwidth,page=1]{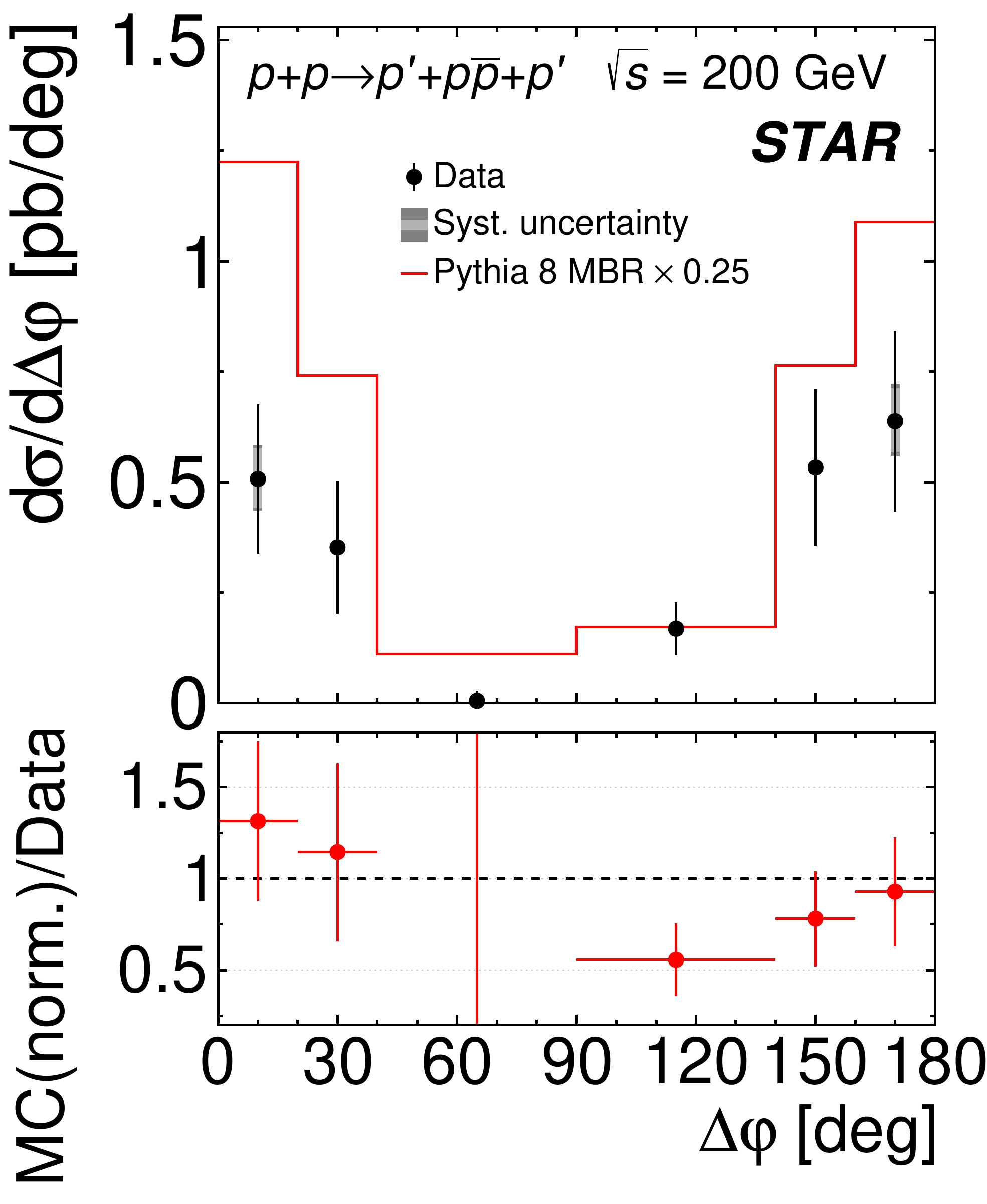}
\newline
\includegraphics[width=.32\textwidth,page=1]{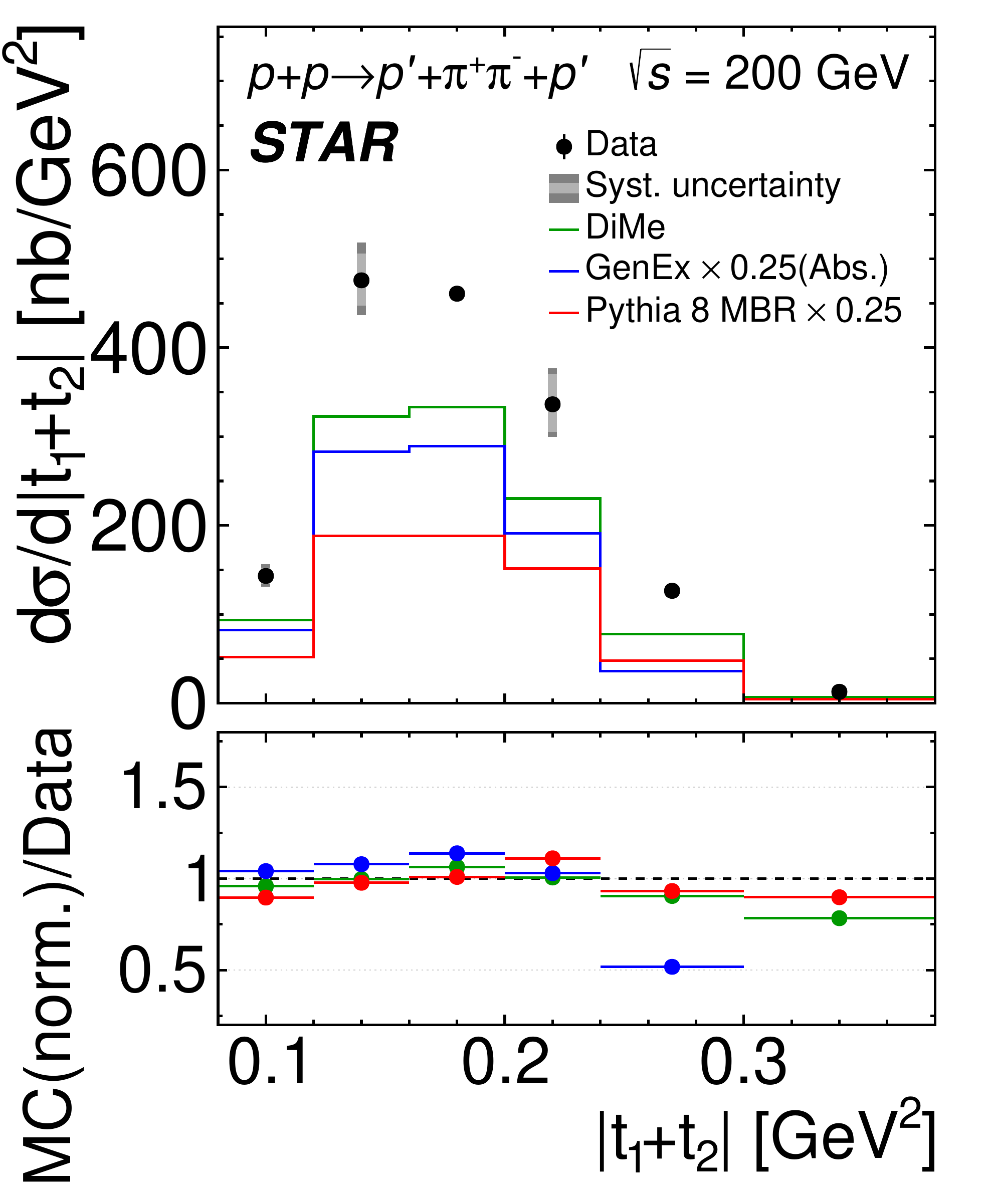}
\hfill
\includegraphics[width=.32\textwidth,page=1]{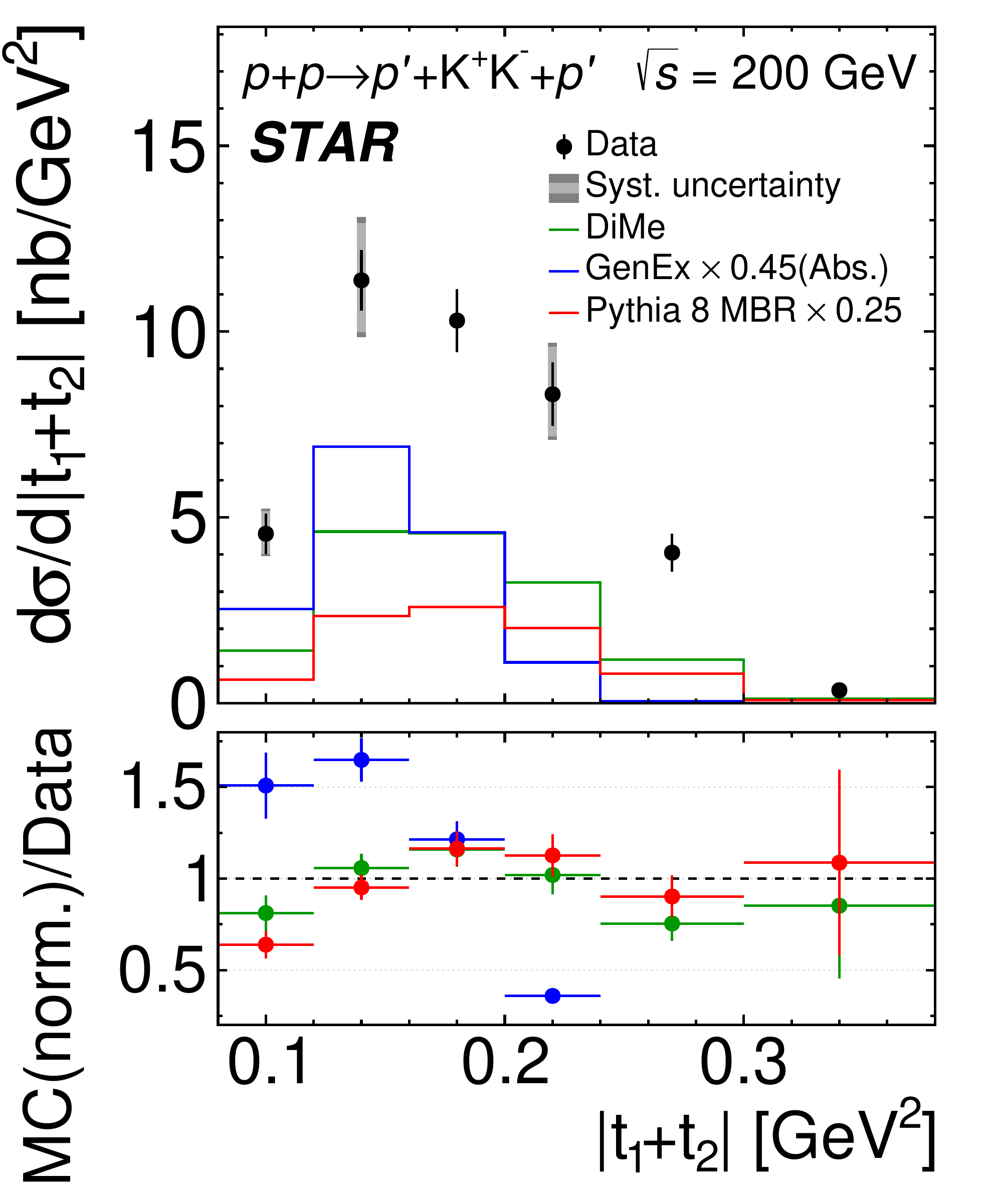}
\hfill
\includegraphics[width=.32\textwidth,page=1]{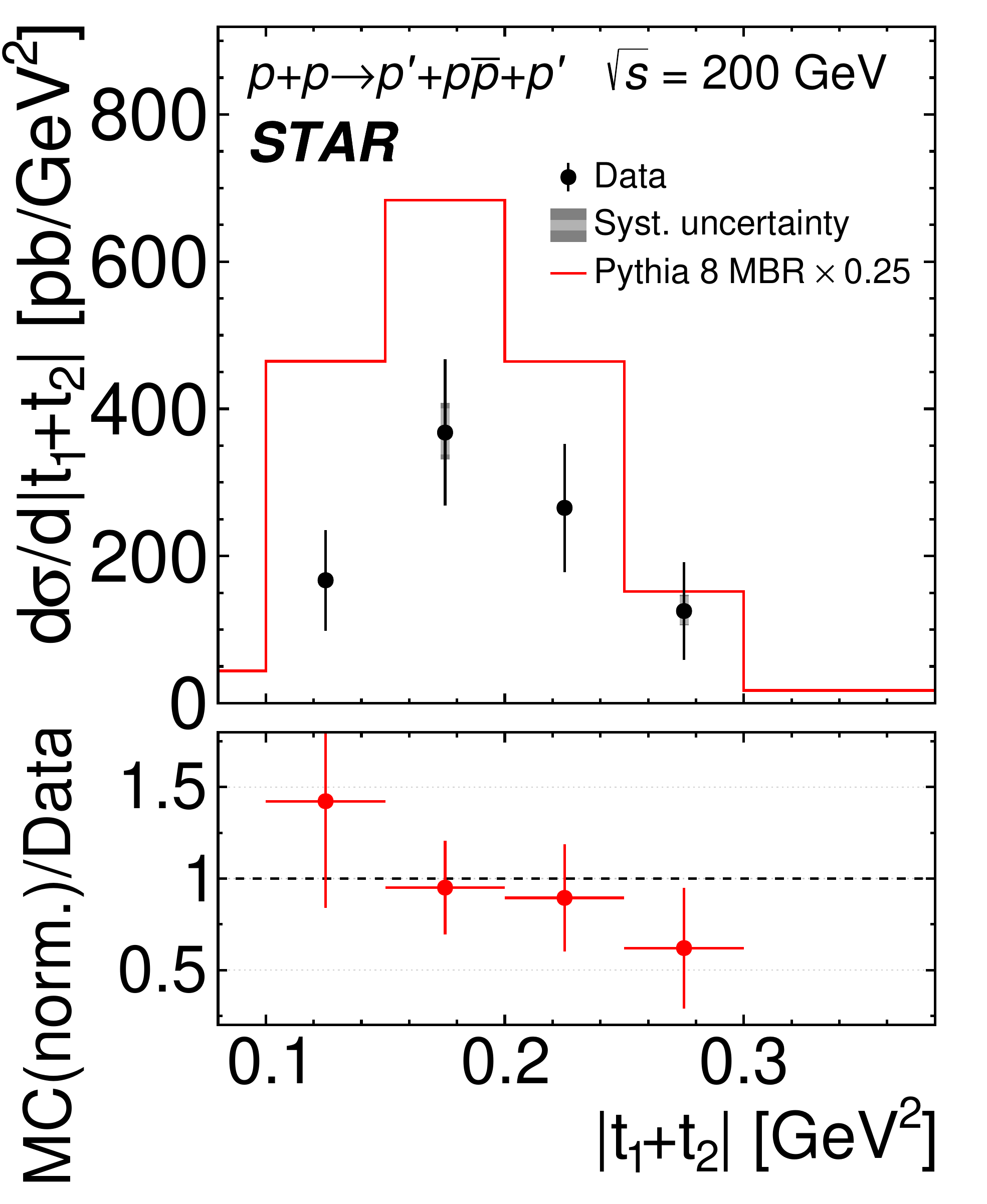}%
\caption{Differential cross sections for CEP of charged particle pairs $\pi^+\pi^-$ (left column), $K^+K^-$ (middle column) and $p\bar{p}$ (right column) as a function of the difference of azimuthal angles of the forward-scattered protons (top) and of the sum of the squares of the four-momenta losses in the proton vertices (bottom) measured in the fiducial region explained in the Sec.~\ref{sec:analysis}. Data are shown as points with error bars representing the statistical uncertainties. The typical systematic uncertainties are shown as gray boxes for only a few data points as they are almost fully correlated between neighboring bins. Predictions from three MC models, GenEx, DiMe and MBR, are shown as histograms. In the lower panels the ratios between the MC predictions (scaled to data) and the data are shown.}
\vspace*{-10pt}
\label{results_2}
\end{figure}
\indent
In Fig.~\ref{results_2}, the differential fiducial cross sections related to the forward-scattered protons are presented.
Figure~\ref{results_2} (top) shows the differential cross sections for CEP of different particle species pairs as a function of $\Delta\upvarphi$.
Strong suppression of the differential cross sections close to $90^{\circ}$ is due to the fiducial cuts applied to the forward \mbox{scattered protons.}
The shape of DiMe model prediction agrees with data for $\pi^+\pi^-$ and $K^+K^-$. The model implemented in GenEx does not describe the data. The MBR model implemented in PYTHIA8 describes the data fairly well in shape for $K^+K^-$ and $p\bar{p}$.
Figure~\ref{results_2} (bottom) shows the differential cross sections for CEP of different particle species pairs as a function of $|t_1+t_2|$. The shapes of the measured cross sections are strongly affected by the fiducial cuts applied to the forward-scattered protons.
The shapes of the differential cross sections for both $\pi^+\pi^-$ and $K^+K^-$ pair production are better described by the DiMe and MBR models than by the GenEx model.
For $p\bar{p}$ pair production, the MBR model predicts a steeper slope.\\
\indent
The STAR detector acceptance naturally splits the fiducial region into two ranges of $\Delta\upvarphi$, which are differently sensitive to absorption effects. Figure~\ref{results_3} shows the differen\-tial %
\noindent cross sections for CEP of different particle species pairs as a function of the pair invariant mass in two $\Delta\upvarphi$ regions: $\Delta\upvarphi<90^\circ$ (left column) and $\Delta\upvarphi>90^\circ$  (right column). 
Sharp drops in the measured cross sections at $m(\pi^+\pi^-) < 0.6$~GeV and at $m(K^+K^-) < 1.3$~GeV for the $\Delta\upvarphi>90^\circ$ range are due to the fiducial cuts applied to the forward-scattered protons. 
In the case of the cross section for CEP of $\pi^+\pi^-$ pairs in the $\Delta\upvarphi<90^\circ$ range, the peak around the $f_2(1270)$ resonance in data is significantly suppressed while the peak at $f_0(980)$, as well as possible resonances in the mass ranges $1.3-1.5$ GeV and $2.2-2.3$ GeV, is enhanced compared to the $\Delta\upvarphi>90^\circ$ range. 
Such correlations, between resonances observed in the mass spectrum and in azimuthal angle between outgoing protons, indicate factorisation breaking between the two proton vertices. In the range $\Delta\upvarphi<90^\circ$, the DiMe model describes well both the normalisation and the shape of the mass spectrum at $m(\pi^+\pi^-)<$ 0.5~GeV.
In the cross section for CEP of $K^+K^-$ pairs, the data do not show any significant $\Delta\upvarphi$ asymmetry except for a possible widening of the peak at $f_2^\prime(1520)$ in the region $\Delta\upvarphi<90^\circ$. This widening may indicate an enhancement of additional resonances around 1.7~GeV in this configuration.
In the cross section for CEP of $p\bar{p}$ pairs, the data do not show a significant $\Delta\upvarphi$ asymmetry except for a possible enhancement in the $2.2-2.4$~GeV mass range for the $\Delta\upvarphi>90^\circ$ region.\\
\begin{figure}[t]
\centering
\includegraphics[width=.48\textwidth,page=1]{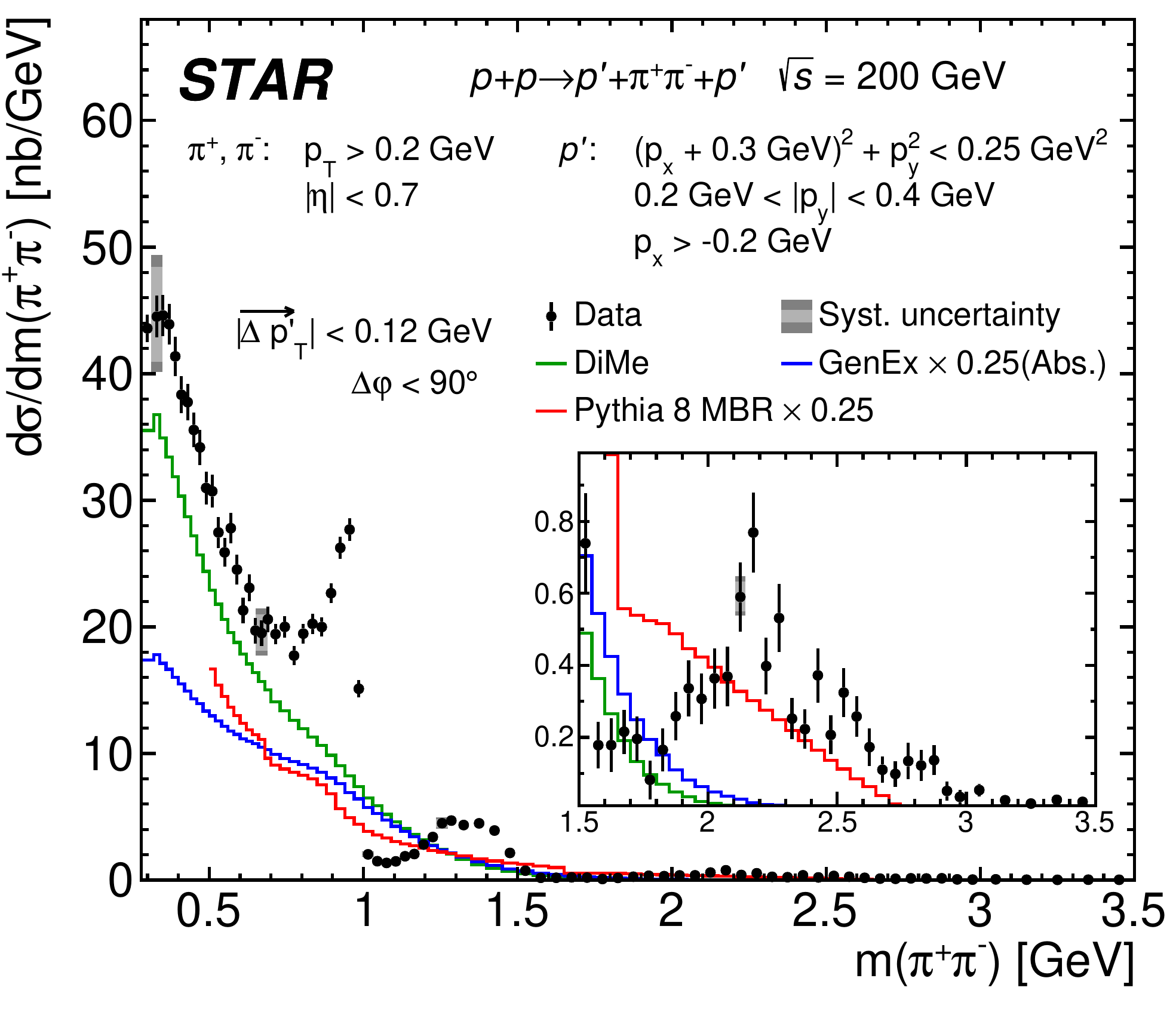}
\hfill
\includegraphics[width=.48\textwidth,page=1]{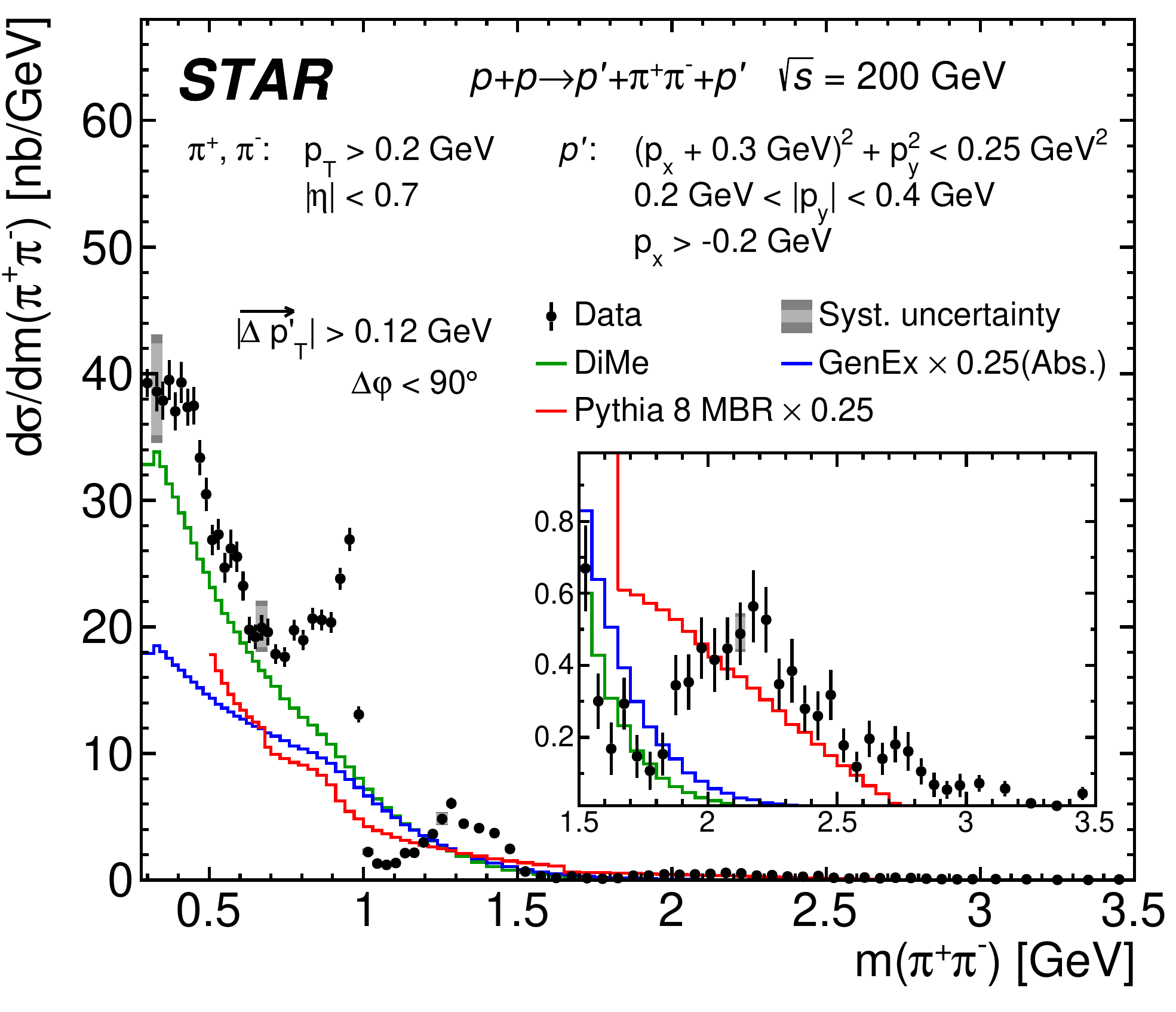}
%
\vspace{-10pt}\caption{Differential cross sections $d\sigma/dm(\pi^+\pi^-)$ for CEP of $\pi^+\pi^-$ pairs in two $|\vec{p}_{1,T}^{\,\prime}-\vec{p}_{2,T}^{\,\prime}|$ regions, $|\vec{p}_{1,T}^{\,\prime}-\vec{p}_{2,T}^{\,\prime}|<0.12$ GeV (left) and $|\vec{p}_{1,T}^{\,\prime}-\vec{p}_{2,T}^{\,\prime}|>0.12$ GeV (right), in the fiducial region and with $\Delta\upvarphi<90^{\circ}$. There is no significant difference between the two $|\vec{p}_{1,T}^{\,\prime}-\vec{p}_{2,T}^{\,\prime}|$ regions. Data are shown as solid points with error bars representing the statistical uncertainties. The typical systematic uncertainties are shown as gray boxes for only a few data points as they are almost fully correlated between neighboring bins. Predictions from three MC models, GenEx, DiMe and MBR, are shown as histograms.}
\vspace{-19pt}
\label{results_5}
\end{figure}
\indent
Experimental observation of vertex factorisation breaking in the $pp$ 
collisions \mbox{motivated} Close and Kirk in Ref.~\cite{close} to propose a method for filtering glueballs from their $q\bar{q}$ counter\-parts.
The $gg$ configurations were proposed to be enhanced in the limit \mbox{$|\vec{p}_{1,T}^{\,\prime}-\vec{p}_{2,T}^{\,\prime}| \rightarrow 0$}. Such a configuration is already enhanced in the $\Delta\upvarphi<90^\circ$ region. To further enhance a possible $gg$ configuration, the data are studied as a function of $|\vec{p}_{1,T}^{\,\prime}-\vec{p}_{2,T}^{\,\prime}|$ in the $\Delta\upvarphi<90^\circ$ region. Figure~\ref{results_5} shows the differential cross section for CEP of $\pi^+\pi^-$ as a function of the pair invariant mass separately in two $|\vec{p}_{1,T}^{\,\prime}-\vec{p}_{2,T}^{\,\prime}|$ regions: $|\vec{p}_{1,T}^{\,\prime}-\vec{p}_{2,T}^{\,\prime}|< 0.12$~GeV(left) and $|\vec{p}_{1,T}^{\,\prime}-\vec{p}_{2,T}^{\,\prime}|> 0.12$~GeV(right).
The data do not show any changes in the shape of the $\pi^+\pi^-$ mass spectrum for the two ranges of $|\vec{p}_{1,T}^{\,\prime}-\vec{p}_{2,T}^{\,\prime}|$ after filtering events with $\Delta\upvarphi<90^\circ$.\\
\begin{figure}[t]%
\centering%
\includegraphics[width=.305\textwidth,page=1]{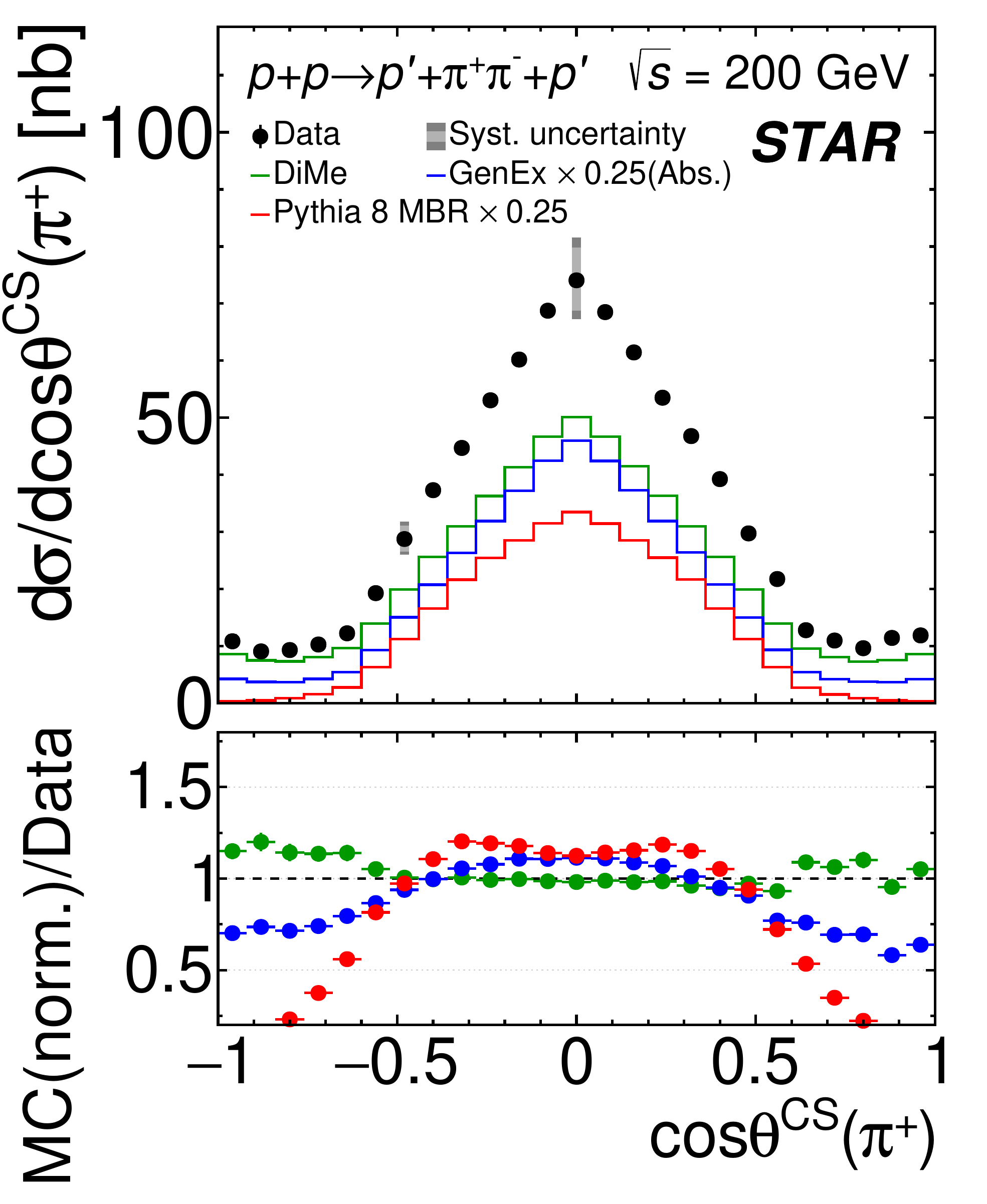}%
\hfill%
\includegraphics[width=.30\textwidth,page=1]{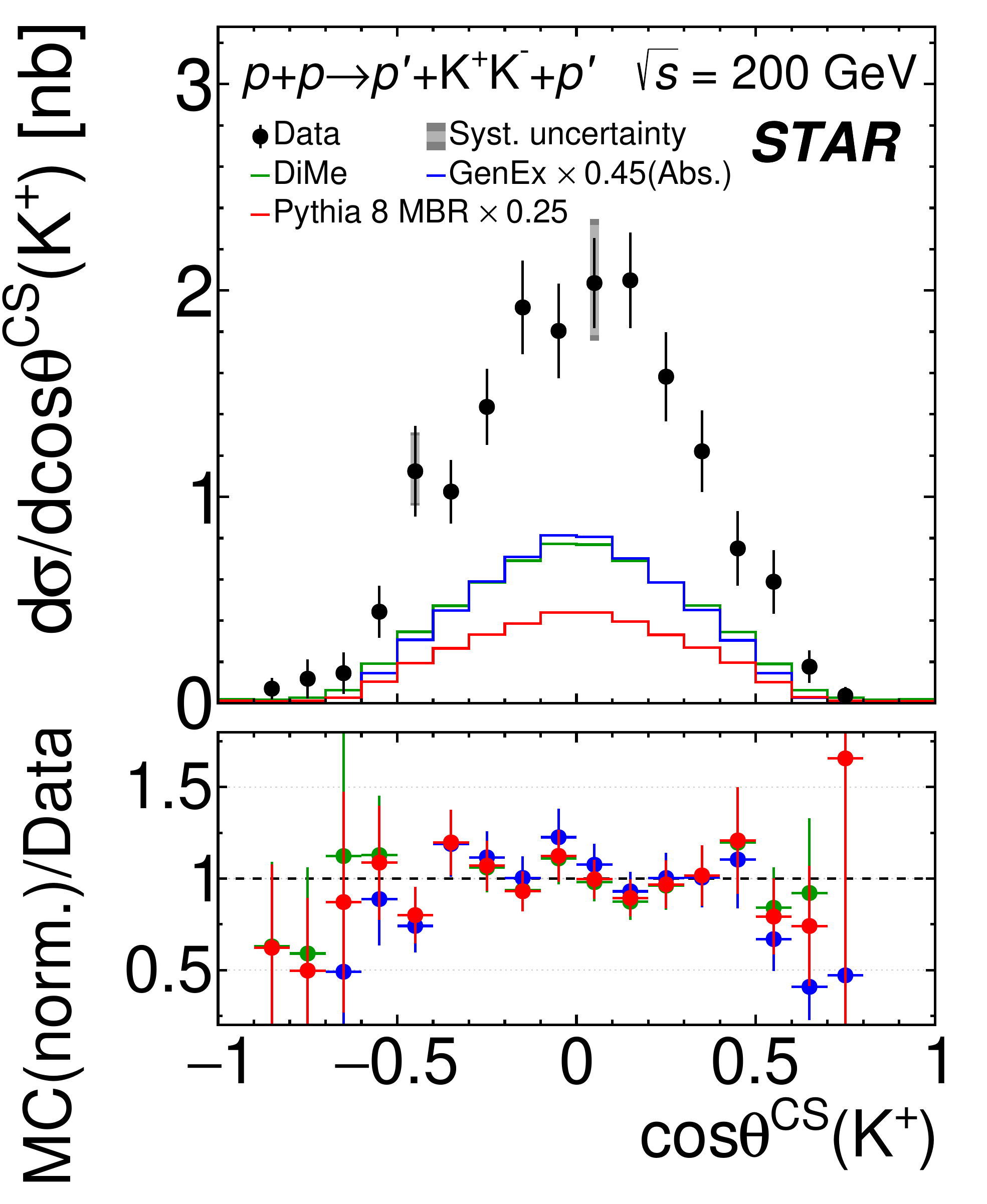}%
\hfill%
\includegraphics[width=.30\textwidth,page=1]{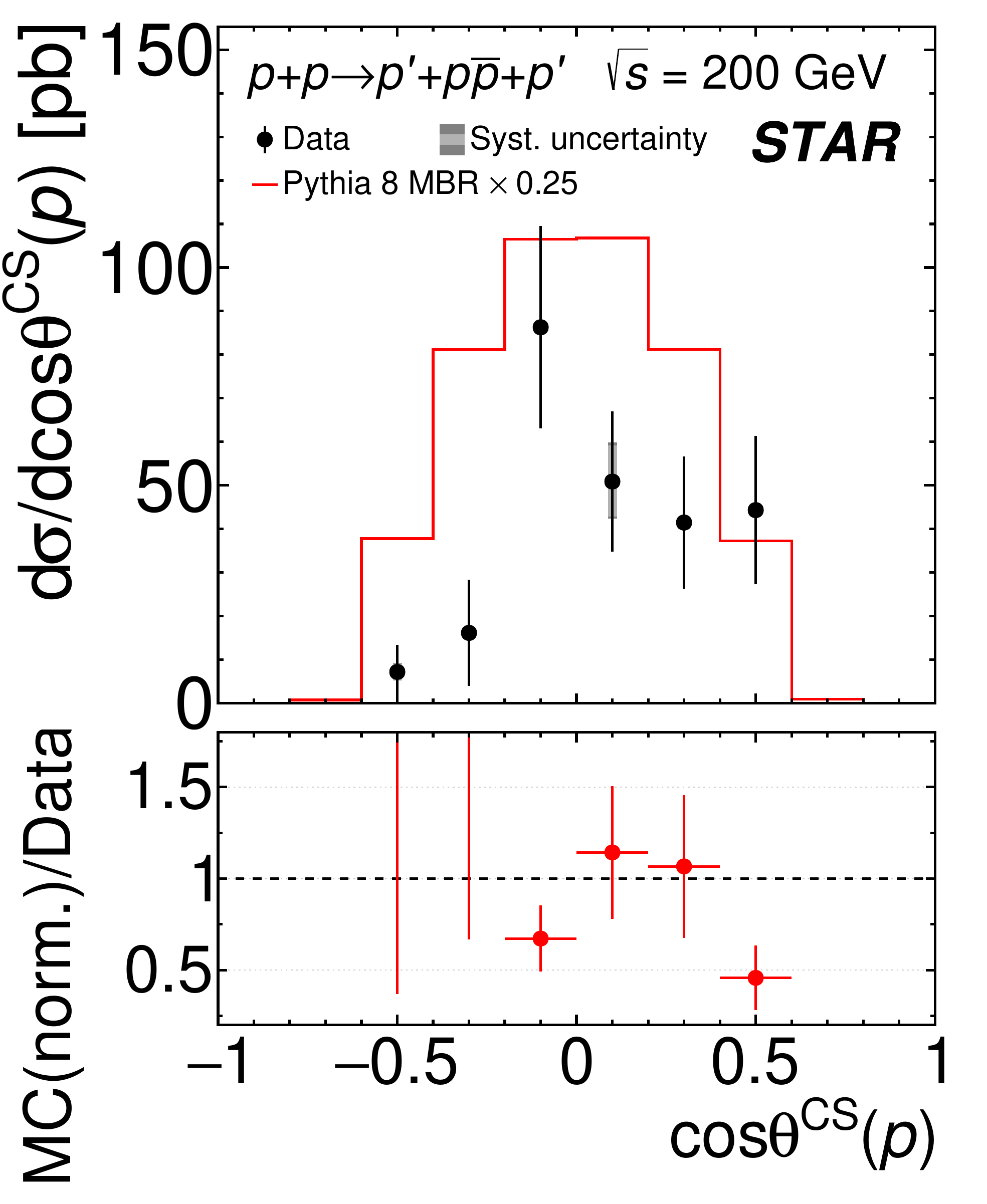}\\%
\includegraphics[width=.30\textwidth,page=1]{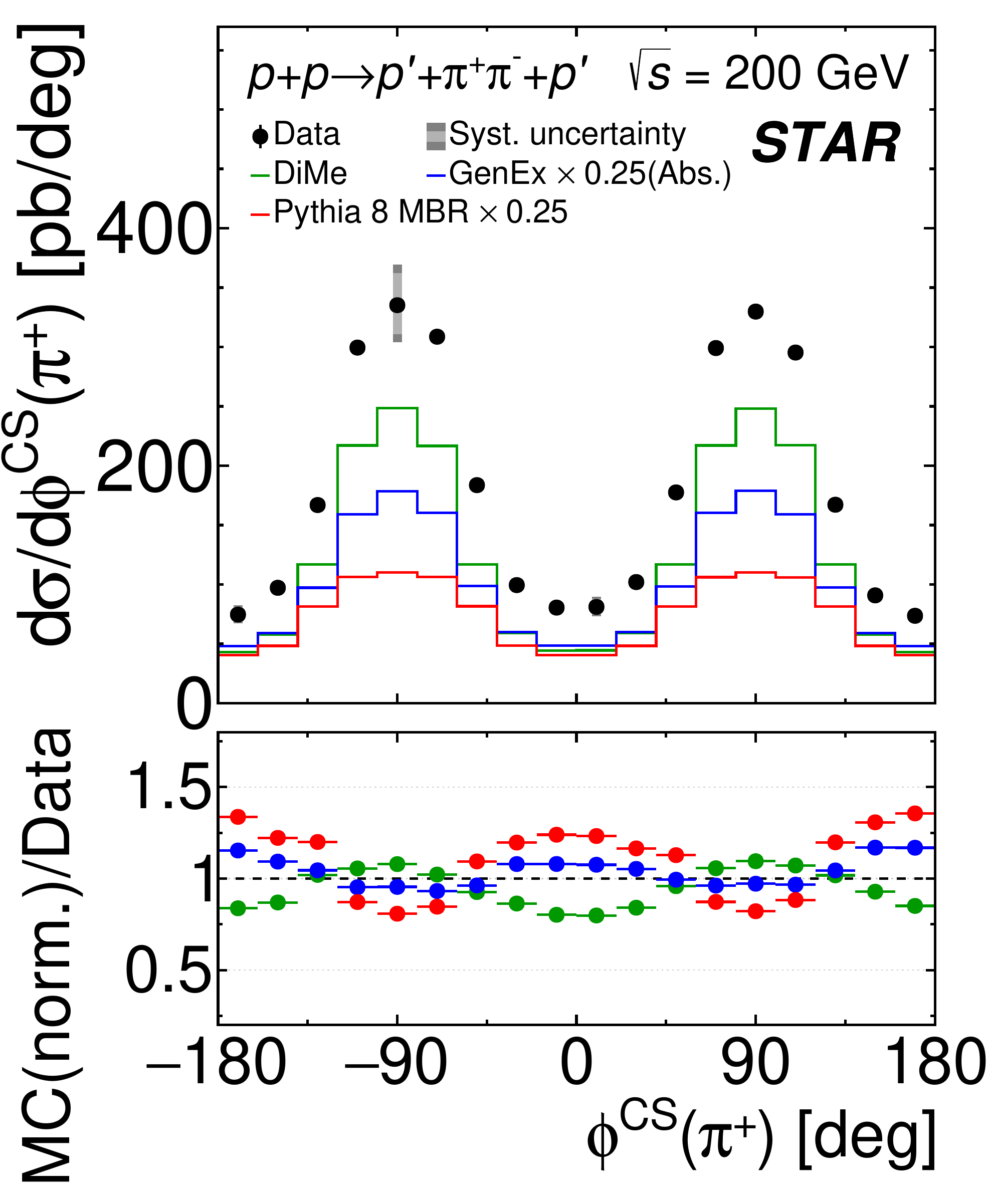}%
\hfill%
\includegraphics[width=.30\textwidth,page=1]{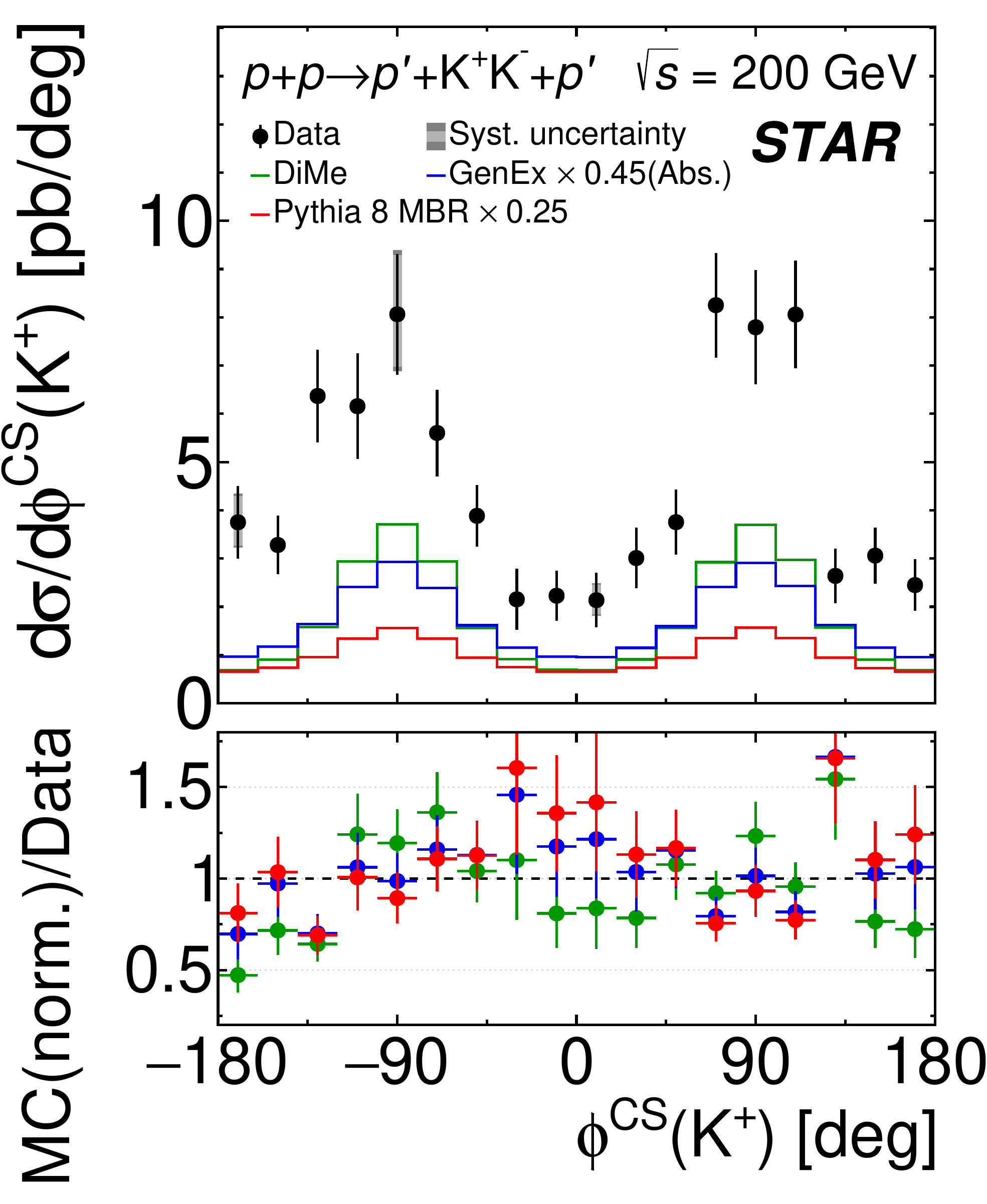}%
\hfill%
\includegraphics[width=.30\textwidth,page=1]{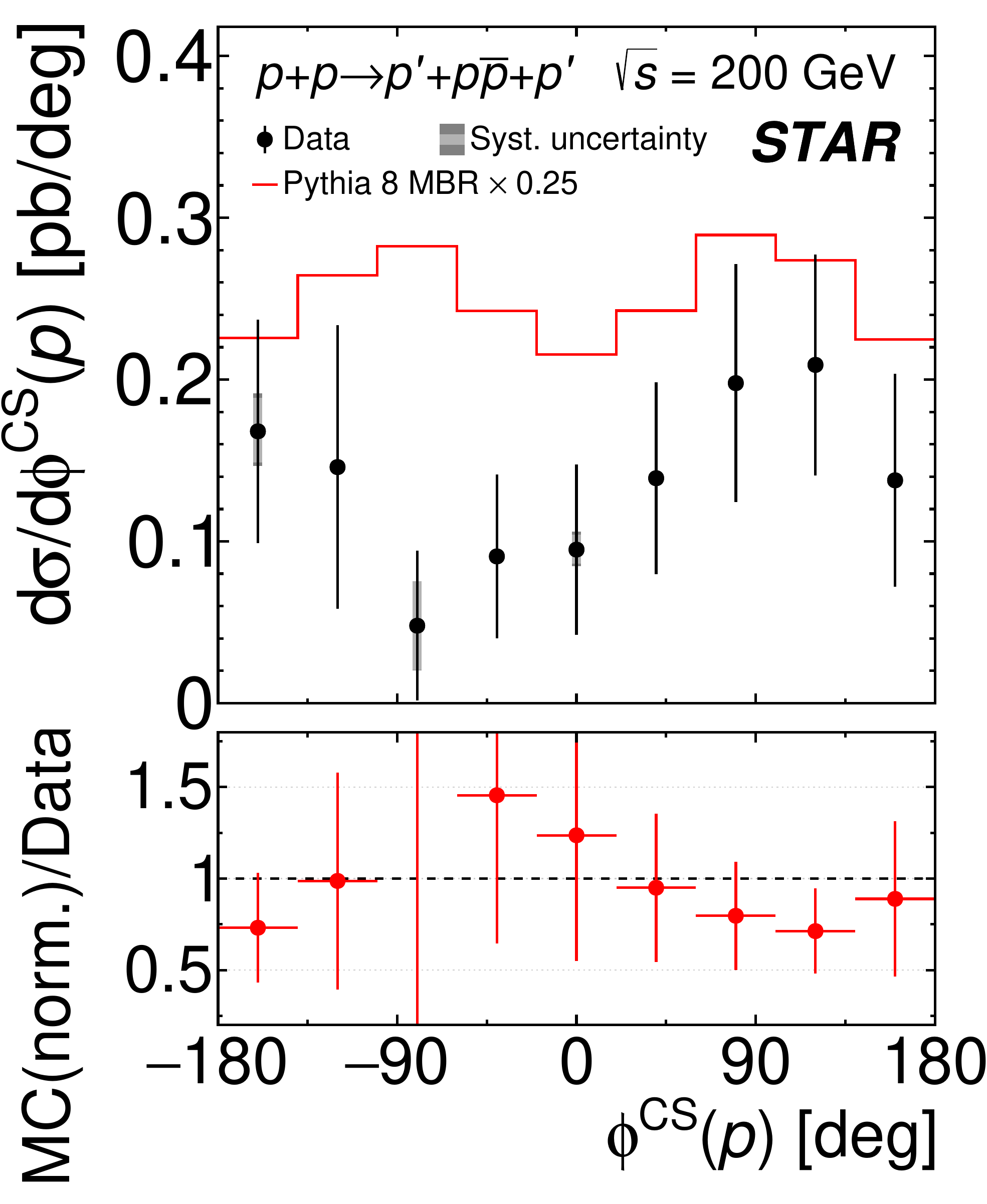}%
\vspace{-10pt}\caption{Differential cross sections for CEP of charged particle pairs $\pi^+\pi^-$ (left column), $K^+K^-$ (middle column) and $p\bar{p}$ (right column) as a function of $\cos{\uptheta^\mathrm{CS}}$ (top) and of $\upphi^\mathrm{CS}$ (bottom), measured in the fiducial region explained in the Sec.~\ref{sec:analysis}. Data are shown as solid points with error bars representing the statistical uncertainties. The typical systematic uncertainties are shown as gray boxes for only a few data points as they are almost fully correlated between neighboring bins. Predictions from three MC models, GenEx, DiMe and MBR, are shown as histograms. In the lower panels the ratios of the MC predictions scaled to data and the data are shown.}
\vspace*{-15pt}
\label{results_6}
\end{figure}
\indent
We have also studied the angular distributions of the charged particles produced in the final state, which may help to constrain the underlying reaction mechanism. This can be done in various reference frames. However, for an easy comparison with theoretical predictions, we use here the Collins-Soper~\cite{cs_frame} reference frame which is also used, e.g., in Ref.~\cite{lebiedowicz_3}. Figure~\ref{results_6} (top) shows the differential cross sections for CEP of different particle species pairs as a function of $\cos{\uptheta^\mathrm{CS}}$. In general, the model predictions are narrower than  the data for all particle species pairs. The only exception is the DiMe prediction for $\pi^+\pi^-$ production, which fits the data much better than other models. 
Figure~\ref{results_6} (bottom) shows the differential cross sections for CEP of different particle species pairs as a~function of $\upphi^\mathrm{CS}$. None of the models is able to describe the data. The double peak structure observed in the data is due to the STAR TPC acceptance.\\
\indent
High statistics of the two-pion sample allow to study the CEP of $\pi^+\pi^-$ pairs in greater detail.
Figure~\ref{results_4} shows the differential cross sections for CEP of $\pi^+\pi^-$ pairs as a function%
\noindent of the pair rapidity (left column), $\Delta\upvarphi$ (middle column) and $|t_1+t_2|$ (right column) in three characteristic ranges of the invariant mass of the pair: $m(\pi^+\pi^-)<1\,\mbox{GeV}$ (mainly non-resonant production), $1\,\mbox{GeV}< m(\pi^+\pi^-) <1.5\,\mbox{GeV}$ ($f_2(1270)$ mass range) and $m(\pi^+\pi^-)>1.5$~GeV (higher invariant masses).
In the case of the cross section, $d\sigma/dy$, all the models agree with the shape of the data in all three mass ranges except for the GenEx and DiMe predictions in the highest mass range, where the predictions are narrower than the data.
Strong suppression of the fiducial cross section close to $\Delta\upvarphi=90^\circ$ is due to the STAR RP acceptance, while the asymmetry of $\Delta\upvarphi = 0^\circ$ vs. $\Delta\upvarphi = 180^\circ$ in the lowest mass region is due to the STAR TPC acceptance. The DiMe model agrees with data only in the lowest mass range. The model implemented in GenEx does not describe the data in any of the three mass regions. Both DiMe and GenEx show similar shapes in the  $\Delta\upvarphi$ distribution except in the lowest mass region. The MBR model predicts symmetric $\Delta\upvarphi$ distributions in all mass ranges, which is not supported by the data except in the highest mass region. The slope of the cross section as a function of $|t_1+t_2|$ is less steep in the $f_2(1270)$ mass range compared to other mass ranges. A comparison between the model predictions of the $|t_1+t_2|$ distributions and the data does not show a significant mass dependence. The best description is given by the MBR model, and the worst by GenEx.\\
\indent
Figure~\ref{results_7} shows the differential cross sections for CEP of $\pi^+\pi^-$ pairs as a function of $\cos{\uptheta^\mathrm{CS}}$ (top row) and $\upphi^\mathrm{CS}$ (bottom row) in three characteristic ranges of the invariant mass of the pair: $m(\pi^+\pi^-)<1\,\mbox{GeV}$, $1\,\mbox{GeV}< m(\pi^+\pi^-) <1.5\,\mbox{GeV}$ and $m(\pi^+\pi^-)>1.5$~GeV.
To help in interpreting the data and in understanding the STAR acceptance effects, the data are compared with expectations (like angular distributions of pions in the $\pi^+\pi^-$ rest frame) from models with pure $S_0$ and $D_0$ waves. The $S_0$ wave predicts a uniform distribution of polar angle $\upphi$, in contrast to the $D_0$ wave. The angular distributions are generated in the most natural Gottfried-Jackson frame~\cite{GJ} with the Pomeron-Pomeron direction taken as the $z$-axis. The transformation to the Collins-Soper frame changes the angular distributions for the $D_0$ wave but not for the $S_0$ wave. Therefore, the shape of the $\upphi^\mathrm{CS}$ distribution for $S_0$ wave, after applying fiducial cuts, represents also the $\upphi^\mathrm{CS}$ shape of the STAR acceptance.
The $S_0$ and $D_0$ predictions are normalised to data. The double-peak structure observed in the $\upphi^\mathrm{CS}$ distribution in the lowest mass region, where data are reasonably well described by $S_0$ prediction, is due to the STAR acceptance.
In contrast, at higher masses, where prediction from the $S_0$ wave model is flat, the double-peak structure does not come from the STAR acceptance. Both $\cos{\uptheta^\mathrm{CS}}$ and $\upphi^\mathrm{CS}$ in the lowest mass region agree very well with the $S_0$ wave suggesting that this mass region is dominated by spin-0 contribution. At higher masses, pure $S_0$ or $D_0$ waves are not able to describe the data.\\
\indent
In the case of the differential cross section $d\sigma/d\cos{\uptheta^\mathrm{CS}}$, the DiMe predictions fit the data only in the lowest mass region. In contrast, the MBR predictions fail to describe the shape of the $\cos{\uptheta^\mathrm{CS}}$ distribution in this mass range only. The GenEx prediction does not describe the data in any mass range.
In the case of the differential cross section $d\sigma/d\upphi^\mathrm{CS}$, in the lowest mass region only GenEx predicts the shape of the $\upphi^\mathrm{CS}$ distribution. The DiMe prediction fits the data well in the middle mass range. Both GenEx and DiMe predictions describe the shape of the $\upphi^\mathrm{CS}$ distribution fairly well in the highest mass region.\\
\indent
The cross sections, integrated over the full fiducial range of the analysis, are shown in %
\begin{table}[H]\centering
\begin{tabular}{c|c|rll|rll}
\multicolumn{1}{c}{~} & \multicolumn{1}{c}{~} & \multicolumn{6}{c}{$\bm{ \sigma_{\text{\bf{fid}}} \pm \delta_{\text{\bf{stat}}} \pm \delta_{\text{\bf{syst}}}}$} \\
 \bf{Particle species} & \bf{unit} & \multicolumn{3}{c}{$\bm{\Delta\upvarphi<90^{\circ}}$} & \multicolumn{3}{c}{$\bm{\Delta\upvarphi>90^{\circ}}$} \\ \hline\hline
 $\bm{\pi^{+}\pi^{-}}$ & \bf{nb} & $44.1$ & $\pm0.2$ & $^{+4.6}_{-4.2}$ & $21.1$ & $\pm0.2$ & $^{+2.1}_{-1.9}$ \\ 
 $\bm{K^{+}K^{-}}$ & \bf{pb} & $1090$ & $\pm60$ & $^{+170}_{-150}$ & $570$ & $\pm40$ & $^{+100}_{-90}$ \\ 
 $\bm{p\bar{p}}$ & \bf{pb} & $17.4$ & $\pm4.7$ & $^{+2.9}_{-2.7}$ & $31.8$ & $\pm6.2$ & $^{+4.6}_{-4.3}$\\ 
\end{tabular}
\caption{Integrated fiducial cross sections with statistical and systematic uncertainties for CEP of $\pi^{+}\pi^{-}$, $K^{+}K^{-}$ and $p\bar{p}$ pairs in two ranges of azimuthal angle difference, $\Delta\upvarphi$, between the two forward-scattered protons.}
\label{tab:xSec}\vspace*{-5pt}
\end{table}
\noindent Table~\ref{tab:xSec} in two ranges of $\Delta\upvarphi$. The largest contribution to the uncertainty of $\pi^+\pi^-$ production arises from the luminosity measurement. For $K^+K^-$ production, the largest contribution to the uncertainty arises from the TOF efficiency. In case of $p\bar{p}$ production, the uncertainty is dominated by statistical fluctuations.
%
%
%
%
\subsection{Extrapolated \texorpdfstring{$\pi^{+}\pi^{-}$}{pi+pi-} differential cross sections \texorpdfstring{$d\sigma/dm$ and $d^2\sigma/dt_1dt_2$}{dsigma/dm and d2sigma/dt1dt2}}%
Invariant mass distributions in the fiducial region of the measurement cannot be directly used to extract yields of possible resonances without extrapolation to the full kinematic region of the central pion pair, given by $p_\mathrm{T}\rightarrow 0$ and $|\eta|\rightarrow\infty$ (full solid angle in the central system rest frame). Extrapolation to an unmeasured region is always model dependent. In this section, we present the cross section corrected to the full phase space using a flat angular approximation which distributes scalar decays uniformly over the solid angle in the rest frame of the central system. This choice is supported by the generally good description of the pion angular distribution by the $S_0$ wave distribution shown in Fig.~\ref{results_7}, and by the expected dominant production of 0-spin states. However, other scenarios are also considered.
To limit the corrections, the measurement is restricted to $|y(\pi^+\pi^-)|<0.4$. This keeps scalar decays uniform, by Lorentz invariance. In the correction calculation, the factorisation of the phase space of the central system and forward protons is assumed. For the forward protons' phase space, a uniform distribution of azimuthal angles is assumed, while polar angles are generated according to an exponential $t$ distribution with $t$-slope of 6~GeV$^{-2}$. 
The measurement is extrapolated from the part of the fiducial region given by Eq.~\eqref{fp_fiducial}, covering $0.05\,\mbox{GeV}^2 \leq -t_1 , -t_2 \leq 0.16\,\mbox{GeV}^2$, to the Lorentz-invariant phase space region defined by the same $t$ interval and full azimuthal angle of forward-protons.
The measurement is further restricted to two ranges of $\Delta\upvarphi$,  $\Delta\upvarphi<45^\circ$ and $\Delta\upvarphi>135^\circ$, which reduces the extrapolation correction and the systematic uncertainties related to their \mbox{modelling.}\\
\indent
A minimal model of the $\pi^+\pi^-$ invariant mass spectrum was fitted to the extrapolated differential cross section.
In this model, we assume contributions from direct pair production and only three resonances in the mass range of $0.6-1.7$ GeV: $f_0(980)$, $f_2(1270)$ and $f_0(1500)$. The total amplitude for the exclusive $\pi^{+}\pi^{-}$ production is then given by
\begin{equation}
\label{eq:amplitude}
\begin{aligned}
A(m) = & \;A_{\textrm{cont}}\times f_{\textrm{cont}}(m)+ \\
        & \;\sqrt{\sigma_{f_0(980)}} \times \exp{\left(i\phi_{f_0(980)}\right)} \times \mathcal{R}_{\textrm{F}}\left(m;M_{f_0(980)},\Gamma_{0,f_0(980)}\right)+ \\
        & \;\sqrt{\sigma_{f_2(1270)}} \times \exp{\left(i\phi_{f_2(1270)}\right)} \times \mathcal{R}_{\textrm{BW}}\left(m;M_{f_2(1270)},\Gamma_{0,f_2(1270)}\right) +\\
        & \;\sqrt{\sigma_{f_0(1500)}} \times \exp{\left(i\phi_{f_0(1500)}\right)} \times \mathcal{R}_{\textrm{BW}}\left(m;M_{f_0(1500)},\Gamma_{0,f_0(1500)}\right).\\
\end{aligned}
\end{equation}
Thus all states are added coherently and can interfere with each other. The amplitude for continuum production is chosen to be real, while the amplitudes for the production cross sections for resonances in the $\pi^{+}\pi^{-}$ channel are allowed to have non-zero phase shifts, $\phi$.
The shape of the continuum amplitude in the fitted mass range is assumed to have the form
\begin{equation}\label{eq:contAmp}
f_{\textrm{cont}}(m) = \sqrt{\frac{q}{m}}\times \exp{\left[-\frac{B}{2}\cdot q\right]},
\end{equation}
with the break-up momentum $q(m) = \frac{1}{2}\sqrt{m^{2}-4m_{\pi}^{2}}$. This continuum effectively includes the production of other wide resonant states, e.g. $f_0(500)$, with a phase of amplitude slowly varying within the fit range, as described below in the discussion of the fit result.
For the $f_2(1270)$ and $f_0(1500)$ resonances, we use the relativistic Breit-Wigner form of the production amplitude with mass-dependent width:
\begin{equation}
\label{eq:BW}
\!\!\!\!\mathcal{R}_{\textrm{BW}}(m;M,\Gamma_{0}) = \frac{1}{\sqrt{\mathcal{I}}} \times \frac{M\sqrt{\Gamma_{0}}\sqrt{\Gamma(m)}}{M^{2}-m^{2}-i M\Gamma(m)},\;\;\;\;\;\Gamma(m) = \Gamma_{0}\frac{q}{m}\frac{M}{q_{0}}\left(\frac{B_{J}(q^{2}R^{2})}{B_{J}(q_{0}^{2}R^{2})}\right)^{\!2}\!\!.
\end{equation}
A factor $\mathcal{I}$ is introduced to provide proper normalisation,  $\int_{2m_{\pi}}^{+\infty}dm|\mathcal{R}_{\textrm{BW}}|^{2}=1$. As a~result, the total cross sections, $\sigma_{f_0(980)}$, $\sigma_{f_2(1270)}$ and $\sigma_{f_0(1500)}$, for resonance production are directly obtained from the fit.
The centrifugal effects are accounted for in Eq.~\eqref{eq:BW} through the Blatt-Weisskopf barrier factors, $B_{J}$~\cite{BarrierFactors1,BarrierFactors}, with the empirical interaction radius, $R$, set to 1~fm and $q_{0} = q(M)$. $J=2$ and $J=0$ are used for $f_2(1270)$ and $f_0(1500)$, respectively.\\
\indent
The $f_0(980)$ meson requires a different treatment due to the large branching ratio to the $K\bar{K}$ channel, which opens in the vicinity of the mass peak. This changes the resonance shape and is accounted for in the parameterisation of the amplitude via the Flatt\'e formula~\cite{Flatte}:
\begin{equation}
\label{eq:Flatte}
\mathcal{R}_{\textrm{F}}(m;M,\Gamma_{0}) = \frac{1}{\sqrt{\mathcal{I}}} \times \frac{M\sqrt{\Gamma_{0}}\sqrt{\Gamma_{\pi}(m)}}{M^{2}-m^{2}-i M\left(\Gamma_{\pi}(m)+\Gamma_{K}(m)\right)}.
\end{equation}
The partial widths, $\Gamma_{j}$ ($j=\pi, K$), are described by the product of the coupling parameter $g_{j}$ and the break-up momentum $q_j$ 
for particle $j$:
\begin{equation}
    \Gamma_{j}(m) = g_{j}q_{j}(m) = \frac{g_{j}}{2}\sqrt{m^{2}-4m_{j}^{2}}.
\end{equation} 
$\Gamma_{0} \equiv \Gamma_{\pi}(M)$ is the partial width in the $\pi^{+}\pi^{-}$ channel at the resonance mass.
In the fit, the ratio $g_{K}/g_{\pi}$ is fixed to 4.21, the value well-constrained experimentally through the measurement of $J/\psi$ decays into $\phi$ mesons and $\pi^{+}\pi^{-}$/$K^{+}K^{-}$ pairs~\cite{BES_JPsi}.\\
\indent
To fit model parameters to the data, the binning in the invariant mass distribution must be adjusted to the expected structures in the cross section. This requires narrower binning than for the measurement  of the fiducial cross section and taking account of the impact of detector resolution. 
The squared amplitude from Eq.~\eqref{eq:amplitude}, $|A|^{2}$, is convoluted for the purpose of the fit with the normal distribution, $\mathcal{N}(0, \sigma_{\text{res}})$, representing the finite measurement resolution of the invariant mass of the pion pair. The resolution parameter, $\sigma_{\text{res}}(m)$, is provided to the fitting algorithm; it is set to grow linearly with increasing invariant mass according to MC simulation of the STAR TPC detector. The $m(\pi^{+}\pi^{-})$ resolution at the lower and upper limit of the fit range is equal to 4~MeV and 13~MeV, respectively. The final form of the function fitted to the extrapolated $d\sigma/dm(\pi^{+}\pi^{-})$ distribution is given by the convolution of the total amplitude squared with the normal distribution:
\begin{equation}
    \mathcal{F}(m) = \left( |A|^{2} \otimes \mathcal{N}\left(0,\sigma_{\text{res}}\right) \right) (m) 
    =
    \int\limits_{2m_\pi}^{+\infty}\! dm'\,\mathcal{N}\left(m'-m;0, \sigma_{\text{res}}(m')\right)|A(m')|^{2}.
\end{equation}
The fitting is performed using the Minuit2 toolkit~\cite{Minuit2} within the ROOT analysis software~\cite{ROOT}. The standardly-defined $\chi^{2}$ is minimised simultaneously in two $\Delta\upvarphi$ ranges. For each of the two $f_0$ resonances the fitted values of mass and width in the two $\Delta\upvarphi$ subsets are forced to be equal, while the phases and absolute values of amplitudes of all resonances are left independent. The mass and width of the $f_2(1270)$ resonance is fixed to the well-known Particle Data Group values~\cite{pdg}.\\
\indent
The experimental systematic uncertainties of the model parameters are estimated through the independent fits to the extrapolated $d\sigma/dm(\pi^{+}\pi^{-})$ distributions with each of the systematic variations described in Sec.~\ref{sec:systematics} applied. In addition to this, we take into consideration the sensitivity of the fit result to the modelling of the extrapolation to the full kinematic region. We check the effect of the extrapolation to the full solid angle in the $\pi^{+}\pi^{-}$ rest frame, assuming a smooth transition from the angular distributions for pure $S_{0}$-wave (up to 1~GeV) to the angular distributions for pure $D_{0}$-wave (starting at 1.2~GeV). We also check the effect of using the extrapolation calculated with predictions from the DiMe and GenEx generators, both for the central state and for the forward-scattered protons. We also check the result of the fit with the ratio $g_{K}/g_{\pi}$ varied within its uncertainties.
The systematic uncertainty on a parameter is separated into two parts. The first one is related to the experimental uncertainties and is calculated as the quadratic sum of the differences between the nominal fit result and the result of the fit to $d\sigma/dm(\pi^{+}\pi^{-})$ with each systematic effect. The second part is related to the extrapolation and is quoted as the largest deviation from the nominal result. \\
\begin{figure}
\centering
\includegraphics[width=\textwidth,page=1]{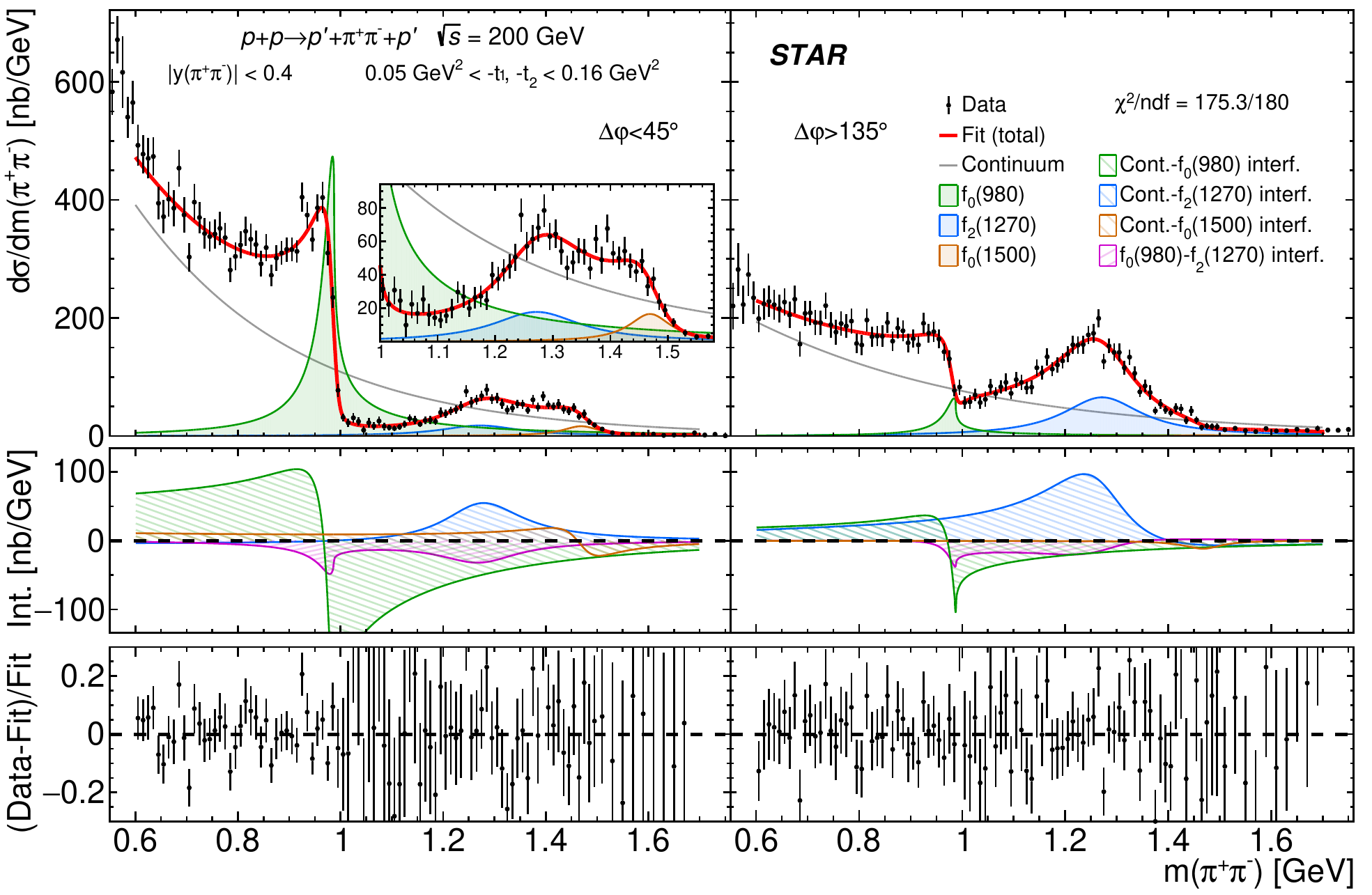}
\vspace*{-20pt}
\caption{Differential cross section $d\sigma/dm(\pi^{+}\pi^{-})$ extrapolated from the fiducial region to the Lorentz-invariant phase space given by the central-state rapidity, $|y(\pi^{+}\pi^{-})|<0.4$, and squared four-momentum transferred in forward proton vertices, $0.05\,\text{GeV}^{2} < -t_1, -t_2 < 0.16\,\text{GeV}^{2}$. The left and right columns show the cross sections for $\Delta\upvarphi<45^\circ$ and $\Delta\upvarphi>135^\circ$, respectively. The data are shown as black points with error bars representing statistical uncertainties. The result of the fit, $\mathcal{F}(m)$, is drawn with a solid red line. The squared amplitudes for the continuum and resonance production are drawn with lines of different colours, as explained in the legend. The most significant interference terms are plotted in the middle panels, while the relative differences between each data point and the fitted model is shown in the bottom panels.}
\label{invMassFit}
\end{figure}
\indent
The extrapolated cross sections are shown in Fig.~\ref{invMassFit}, together with the result of the fit described above. The model parameters providing the minimum $\chi^{2}$ are listed in Tab.~\ref{tab:fitRes}. The fit, with a total of 20 free parameters, gives $\chi^2$/ndf = 175/180 which shows that the data and the model are in excellent agreement in the fit region. Alternative extrapolation models show a similar fit quality, although some parameters change significantly as can be noted from the model-related uncertainties in Tab.~\ref{tab:fitRes}.
The fitted model shows a small deviation from the extrapolated data around $1.37$~GeV. This might result from the presence of the $f_0(1370)$, however the inclusion of the $f_{0}(1370)$ is not necessary to describe the data. 
The cross section for $f_0(1500)$ production differs from zero by 5 and 2 standard deviations in the $\Delta\upvarphi<45^\circ$ and $\Delta\upvarphi>135^\circ$ regions, respectively. Removing the $f_0(1500)$ from the fit makes the $\chi^2$/ndf change to 352/186, a 7.0 standard-deviation effect. From this, we infer that the shape of $d\sigma/dm(\pi^{+}\pi^{-})$ around $1.4-1.6$~GeV, the high-mass part of the $f_{2}(1270)$ region, is primarily determined by the $f_0(1500)$ interfering with $\pi^{+}\pi^{-}$ continuum.\\
\begin{table}\centering
\begin{tabular}{ccc rP{0.6cm}P{0.6cm}P{0.6cm} rP{0.6cm}P{0.6cm}P{0.6cm}}
~ & ~ & \bf{unit} & \multicolumn{4}{c}{$\bm{\Delta\upvarphi<45^{\circ}}$} & \multicolumn{4}{c}{$\bm{\Delta\upvarphi>135^{\circ}}$} \\ \hline\hline 
\multirow{2}{*}{\bf{Continuum}} & $\bm{A}$ & $\bm{\left(\text{\bf{nb/GeV}}\right)^{\frac{1}{2}}}$ & $69$ & $ \pm 5 $ & $\pm 4 $ & $^{+1}_{-12}$ & $39 $ & $\pm 3$ & $ \pm 3 $ & $^{+2}_{-10}$ \\ 
& $\bm{B}$ & $\bm{\text{\bf{GeV}}^{-1}}$ & $6.4$ & $ \pm 0.4 $ & $\pm 0.1 $ & $^{+0.1}_{-0.9}$ & $4.7 $ & $\pm 0.3 $ & $\pm 0.2 $ & $^{+0.2}_{-1.1}$ \\ \hline
\multirow{4}{*}{$\bm{f_{0}(980)}$} & $\bm{\sigma}$ & \bf{nb} & $43.3 $ & $\pm 4.7 $ & $^{+4.6}_{-4.1} $ & $^{+2.7}_{-4.4}$ & $5.8$ & $ \pm 1.0$ & $ ^{+0.6}_{-0.5} $ & $^{+0.3}_{-1.7}$ \\ & $\bm{\phi}$ & \bf{rad} & $0.66 $ & $\pm 0.08$ & $ ^{+0.01}_{-0.02} $ & $^{+0.02}_{-0.06}$ & $0.56 $ & $\pm 0.10 $ & $\pm 0.01 $ & $^{+0.01}_{-0.09}$ \\ 
& $\bm{M}$ & \bf{MeV} & \multicolumn{8}{c}{$956 \pm \phantom{1}7 \pm 1 ~^{+4\phantom{1}}_{-6\phantom{1}}$} \\ 
& $\bm{\Gamma_{0}}$ & \bf{MeV} & \multicolumn{8}{c}{$163 \pm 26 \pm 3 ~^{+17}_{-20}$} \\ \hline
\multirow{2}{*}{$\bm{f_{2}(1270)}$} & $\bm{\sigma}$ & \bf{nb} & $4.9 $ & $\pm 1.1 $ & $^{+0.6}_{-0.5} $ & $^{+0.3}_{-2.0}$ & $17.9 $ & $\pm 1.6 $ & $^{+1.9}_{-1.7} $ & $^{+0.2}_{-5.2}$ \\ 
& $\bm{\phi}$ & \bf{rad} & $-1.83 $ & $\pm 0.12 $ & $\pm 0.01$ & $^{+0.03}_{-0.12}$ & $-0.92 $ & $\pm 0.05 $ & $\pm 0.03$ & $^{+0.06}_{-0.23}$ \\ \hline
\multirow{4}{*}{$\bm{f_{0}(1500)}$} & $\bm{\sigma}$ & \bf{nb} & $2.3 $ & $\pm 0.5$ & $ \pm 0.2 $ & $^{+1.1}_{-0.7}$ & $0.2 $ & $\pm 0.1 $ & $\pm 0.0 $ & $^{+0.1}_{-0.0}$ \\ 
& $\bm{\phi}$ & \bf{rad} & $0.16 $ & $\pm 0.17 $ & $^{+0.04}_{-0.03} $ & $^{+0.04}_{-0.15}$ & $1.59$ & $\pm 0.31 $ & $\pm 0.04 $ & $^{+0.04}_{-0.07}$ \\ 
& $\bm{M}$ & \bf{MeV} & \multicolumn{8}{c}{$1469 \pm\phantom{1}9\pm 1 \pm 2$} \\ 
& $\bm{\Gamma_{0}}$ & \bf{MeV} & \multicolumn{8}{c}{$\phantom{1}\phantom{1}89 \pm 14 \pm 2 ~^{+4\phantom{1}}_{-3\phantom{1}}$} \\ \hline
\end{tabular}
\caption{Results of the fit described in the text in two ranges of azimuthal angle difference $\Delta\upvarphi$ between forward-scattered protons. Statistical, systematic and model uncertainties are provided for each parameter, in that order.}\label{tab:fitRes}\vspace{-5pt} 
\end{table}
\indent
Since the masses and widths of the $f_0(980)$ and the $f_0(1500)$ are free parameters, we can compare their fitted values with the PDG data~\cite{pdg}. In the case of the $f_0(980)$, the mass and width are found to be $M_{f_0(980)}=956 \pm 7 (\text{stat.}) \pm 1 (\text{syst.})\,^{+4}_{-6} (\text{mod.})$~MeV and $\Gamma_{0,f_0(980)} = 163 \pm 26 (\text{stat.}) \pm 3 (\text{syst.})\,^{+17}_{-20} (\text{mod.})$~MeV, respectively. These values differ from the PDG estimates of mass ($990 \pm 20$~MeV) and width (from 10~MeV to 100~MeV). However, the PDG emphasises a strong dependence of the resonance parameters on the model of the amplitude. Some measurements listed in Ref.~\cite{pdg} are in reasonable agreement with our measured numbers. In addition to this, the mass and width of the $f_0(980)$ resulting from the fit with the Breit-Wigner form (instead of the Flatt\'e form) of the amplitude gives a result $M_{f_0(980)}=974 \pm 1 (\text{stat.}) \pm 1 (\text{syst.})$~MeV and $\Gamma_{0,f_0(980)} = 65 \pm 3 (\text{stat.}) \pm 1 (\text{syst.})$~MeV (albeit with a notably worse $\chi^{2}$/ndf of 226/180 providing evidence for a significant branching fraction for the decay into $K\bar{K}$, which needs to be accounted for in the resonance parameterisation). These values are in excellent agreement with PDG estimates and $f_0(980)$ parameters from other measurements that assume a Breit-Wigner resonance shape~\cite{pdg}. \\
\indent
For the $f_0(1500)$, we obtain from the fit $M_{f_0(1500)} = 1469\,\pm 9 (\text{stat.})\,\pm 1 (\text{syst.})\,\pm 2 (\text{mod.})$~MeV and $\Gamma_{0,f_0(1500)} = 89 \pm 14 (\text{stat.})\,\pm 2 (\text{syst.})\,^{+4}_{-3} (\text{mod.})$~MeV. These values also deviate from the PDG averages $1505\pm 6$~MeV for the mass and $109\pm 7$~MeV for the width. However, numerous measurements on $f_{0}(1500)$ referenced in the PDG (and not used for the average calculation) report masses below 1500~MeV and widths below 100~MeV that are consistent with our result. \\
\indent
We have tested the possibility of the existence of an additional resonance produced in the mass range $1.2-1.5$ GeV. With an $f_0$-like component added to the model in Eq.~\eqref{eq:amplitude}, the best fit is achieved for $M_{f_0}=1372 \pm 13 (\text{stat.})$~MeV and $\Gamma_{0,f_0} = 44 \pm 24 (\text{stat.})$~MeV. In that case, the $\chi^{2}$/ndf is equal to 158/174 ($p$-value: 0.8), compared to the nominal value 175/180 ($p$-value: 0.6). The cross section $d\sigma/dm(\pi^{+}\pi^{-})$ around 1.37~GeV for $\Delta\upvarphi<45^{\circ}$ is better described than with the nominal fit. Other parameters in the fit change slightly but remain compatible with their original values. The fitted content of the additional $f_0$ resonance is several times lower than the extracted yield of $f_0(1500)$ for $\Delta\upvarphi<45^{\circ}$, while for $\Delta\upvarphi>135^{\circ}$ it is consistent with 0. The mass agrees with that of the $f_{0}(1370)$ resonance, however the measured width is much narrower than PDG estimates of about $200-500$~MeV. \\
%
\begin{table}[b]\centering
\begin{tabular}{c|P{0.4cm}P{0.75cm}P{0.75cm}P{0.75cm}|P{0.4cm}P{0.75cm}P{0.75cm}P{0.75cm}|P{0.4cm}P{0.75cm}P{0.75cm}P{0.75cm}}%
\multicolumn{1}{c}{~} & \multicolumn{8}{c}{$\bm{\displaystyle\sigma/\sigma_{f_{2}(1270)}}$} & \multicolumn{4}{c}{\multirow{2}{*}{$\bm{\displaystyle\frac{\sigma(\Delta\upvarphi<45^{\circ})}{\sigma(\Delta\upvarphi>135^{\circ})}}$}} \\
 & \multicolumn{4}{c|}{$\bm{\Delta\upvarphi<45^{\circ}}$} & \multicolumn{4}{c|}{$\bm{\Delta\upvarphi>135^{\circ}}$} & \\
\hline\hline $\bm{f_{0}(980)}$ & $8.9$ &  $\pm 2.3$ &  $^{+0.4}_{-0.5}$ & $^{+7.2}_{-0.3}$ & $0.33$ & $\pm 0.06$ & $\pm 0.01$ & $^{+0.13}_{-0.08}$ & $7.4$ & $\pm 1.6$ & $\pm 0.2$ & $^{+2.4}_{-0.2}$\\ 
$\bm{f_{2}(1270)}$ & \multicolumn{4}{c|}{1} & \multicolumn{4}{c|}{1} & $0.27$ & $\pm 0.07$ & $\pm 0.01$ & $^{+0.02}_{-0.05}$\\
$\bm{f_{0}(1500)}$ & $0.47$ & $\pm 0.15$ & $\pm0.03$ & $^{+0.24}_{-0.05}$ & $0.01$ & $\pm 0.01$ & $\pm 0.00$ & $\hspace*{-3pt}\pm0.00$ & $12.3$ & $\pm 8.6$ & $^{+0.7}_{-0.8}$ & $^{+2.3}_{-3.6}$\\ 
\end{tabular}
\caption{Ratios of integrated cross sections of resonance production. For each ratio statistical, systematic and model uncertainties are provided, in that order.}\label{tab:resonanceRatio}
\end{table}
%
%
\indent
We calculated the ratios of the total cross sections $\sigma_{f_0(980)}/\sigma_{f_2(1270)}$ and $\sigma_{f_0(1500)}/\sigma_{f_2(1270)}$ in two $\Delta\upvarphi$ regions, as well as the ratio $\sigma(\Delta\upvarphi<45^{\circ})/\sigma(\Delta\upvarphi>135^{\circ})$ for all resonances, as listed in Tab.~\ref{tab:resonanceRatio}. In the ratios, many of the systematic uncertainties cancelled out.
We observe a~significant dependence of the resonance production cross sections on the azimuthal separation of the forward-scattered protons. The two scalar mesons, $f_{0}(980)$ and $f_{0}(1500)$, are produced predominantly at $\Delta\upvarphi<45^{\circ}$, whereas the tensor meson $f_{2}(1270)$ is produced predominantly at $\Delta\upvarphi>135^{\circ}$. This $\Delta\upvarphi$ dependence is consistent with the observation made by the WA102 Collaboration~\cite{WA102}.\\
\begin{figure}[t]
\centering
\includegraphics[width=\textwidth,page=1]{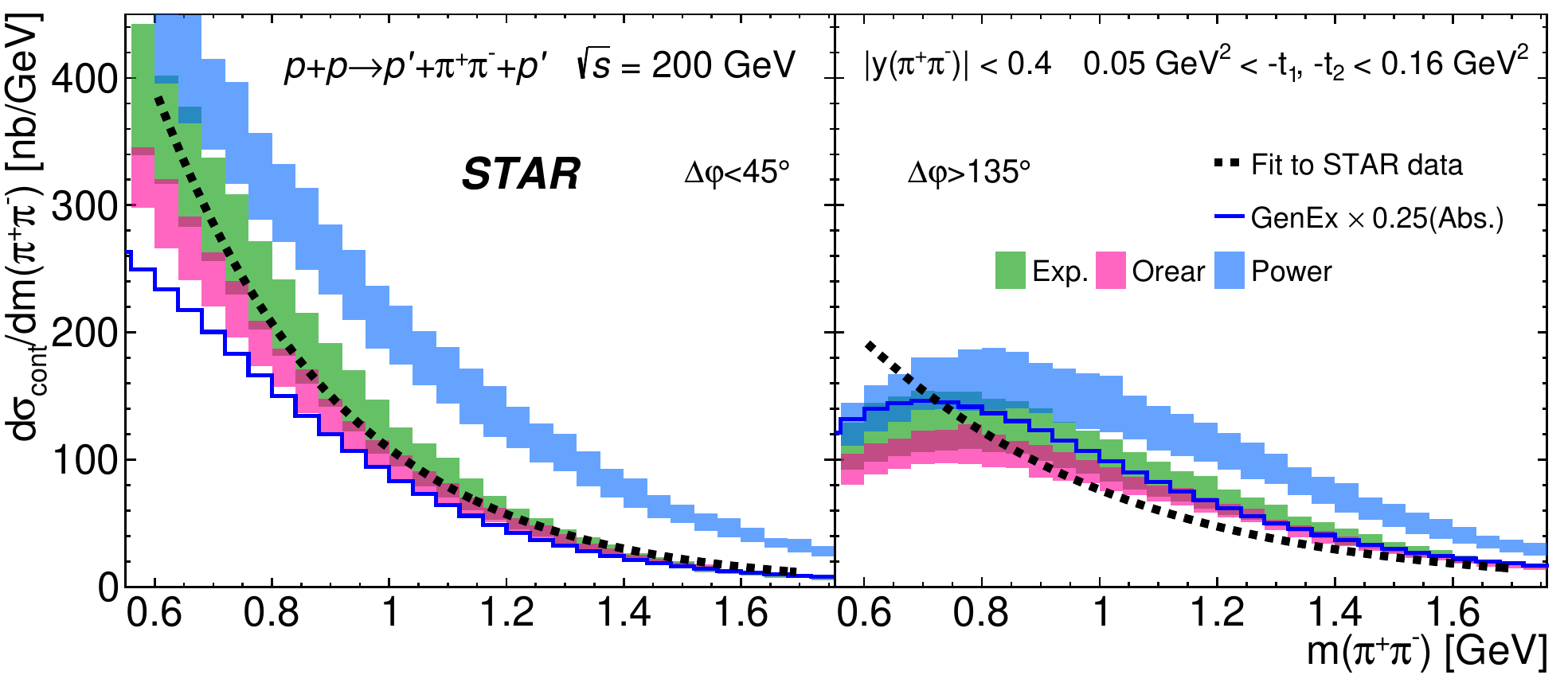}
\vspace*{-20pt}
\caption{Comparison of the continuum production cross section obtained from the fit to the extrapolated $d\sigma/dm(\pi^{+}\pi^{-})$ (dotted black line) with predictions from the continuum models. Scaled GenEx predictions are shown with the blue histogram. Three predictions from DiMe representing different meson form factors are shown as coloured bands, each spanning between minimum and maximum predicted cross sections from four available absorption models. Left and right panels show cross sections for $\Delta\upvarphi<45^\circ$ and $\Delta\upvarphi>135^\circ$, respectively.}
\label{continuum}
\end{figure}
\indent
The differential cross section for the $\pi^{+}\pi^{-}$ continuum production, $d\sigma_{\text{cont}}/dm(\pi^{+}\pi^{-}) = A_{\textrm{cont}}^{2}f_{\textrm{cont}}^{2}(m)$, is extracted from the fit to the extrapolated $d\sigma/dm(\pi^{+}\pi^{-})$ distribution. It is then compared with expectations from GenEx and DiMe models using all three forms of meson form factors ($\Lambda^2_{of\!f}=1.0$ GeV$^2$, $b=2$ GeV$^{-1}$ and $a^2=0.5$ GeV$^2$, and $a_0=1.0$ GeV$^2$) and four models of absorption~\cite{dime}. All models predict a shape for $d\sigma_{\text{cont}}/dm(\pi^{+}\pi^{-})$ that is different from the assumed form, $f_{\text{cont}}^{2}$,
as shown in Fig.~\ref{continuum}. The shape of the continuum predicted by the models can be achieved by changing the factor $\sqrt{q/m}$ in Eq.~\eqref{eq:contAmp} to $(\sqrt{q/m})^{P}$, with the parameter $P$ taking values between 2 and 10. %
Using such a modified continuum amplitude in the fit one obtains
a $P$ parameter consistent with 1 in both $\Delta\upvarphi$ ranges, and the remaining parameters are consistent with the results of the nominal fit. The deviation of the fitted $d\sigma_{\text{cont}}/dm(\pi^{+}\pi^{-})$ from all the model predictions is most evident at the lower edge of the fit range. A possible explanation of this observation is the presence of the $f_{0}(500)$ state, expected in DPE, and the photo-produced $\rho^0$ vector meson, generally suppressed within the kinematic region of this measurement.
One should therefore treat the continuum obtained from the fit as an effective description of the coherent sum of the continuum and other states not explicitly included in the Eq.~\eqref{eq:amplitude}.\\
\indent
Apart from the extrapolation and modelling of the $\pi^{+}\pi^{-}$ invariant mass cross section, we have also applied geometrical corrections to the $d^2\sigma/dt_{1}dt_{2}$ distribution in the same Lorentz-invariant phase space given by $|y(\pi^+\pi^-)|<0.4$ and $0.05 \leq -t_1 , -t_2 \leq 0.16$ GeV$^2$, and in two $\Delta\upvarphi$ ranges ($\Delta\upvarphi<45^{\circ}$ and $\Delta\upvarphi>135^{\circ}$). 
These cross sections were fitted in two dimensions with an exponential function $\propto \exp{\left[\beta t_{1}\right]}\cdot\exp{\left[\beta t_{2}\right]}$. The fits are done separately in three selected ranges of $m(\pi^{+}\pi^{-})$. The obtained values for $\beta$ are provided in Tab.~\ref{tab:slopes}. 
For these values, we do not separate modelling uncertainties since they are generally much smaller than experimental uncertainties. This is a consequence of the uniform $\upvarphi$ distribution in all the models and the rather weak dependence of the cross section on $\Delta\upvarphi$  within the measured ranges. Such approximations are well-founded and in good agreement with the data.
Variations of the slope, $\beta$, with $m(\pi^{+}\pi^{-})$ and $\Delta\upvarphi$ can give important constraints for model developers.  For example, it was pointed out in Ref.~\cite{lebiedowicz_1} that the $f_2(1270)$ cross section may show an enhancement when  $t_1\rightarrow 0$ and $t_2\rightarrow 0$ for some couplings, while for others a suppression is expected. This enhancement or suppression results in a larger or smaller value of $\beta$, respectively.
\begin{table}[t]\centering
\begin{tabular}{c|ll}%
~ & \multicolumn{1}{c}{$\bm{\Delta\upvarphi<45^{\circ}}$} & \multicolumn{1}{c}{$\bm{\Delta\upvarphi>135^{\circ}}$}\\ \hline\hline
$\bm{0.6~\text{\bf{GeV}} < m <1~\text{\bf{GeV}}}$ & \phantom{1}$8.9 \pm 0.3 \phantom{1}^{+0.9}_{-0.6}$ & $14.1 \pm 0.5 \phantom{1}^{+0.5}_{-0.9}$\\ 
$\bm{1~\text{\bf{GeV}} < m < 1.5~\text{\bf{GeV}}}$ & $10.1 \pm 0.7 \pm 0.7$ & \phantom{1}$4.5 \pm 0.4 \pm 0.7$\\ 
$\bm{m>1.5~\text{\bf{GeV}}}$ & \phantom{1}$8.3 \pm 1.2 \pm 0.7$ & \phantom{1}$5.0 \pm 1.0 \pm 0.7$\\ 
\end{tabular}
\caption{Slope $\beta$ (in GeV$^{-2}$) of the $-t$ distribution in three ranges of $m(\pi^{+}\pi^{-})$ and two two ranges of $\Delta\upvarphi$. For each number, statistical and systematic uncertainties are provided, in that order.}\label{tab:slopes}
\end{table}
\FloatBarrier
\section{Summary}
\label{sec:summary}
We have studied the CEP of charged particle pairs ($\pi^+\pi^-$, $K^+K^-$ and $p\bar{p}$) in events with forward protons tagged in the RP detectors, using a sample of 14.2~pb$^{-1}$ data collected with the STAR detector at RHIC in proton-proton collisions at $\sqrt{s} = 200$ GeV. The centre-of-mass energy of $pp$ collisions in the present measurement is three times larger than the previous highest-energy measurement of the DPE process with forward-proton tagging performed at CERN at the ISR in the AFS experiment~\cite{afs}. The uncertainty of the absolute normalisation of the STAR measurement is a factor of four better compared to measurements at the ISR, giving much stronger constraints for phenomenological models. The fits to the extrapolated cross sections as a function of the $\pi^+\pi^-$ invariant mass of the minimal model, including the $f_0(980), f_2(1270)$ and $f_0(1500)$ resonances and the non-resonant contribution, provide for the first time measurements of the relative phases between all the production amplitudes.
In the glueball sector, there is no evidence except for a small enhancement around 1.7 GeV for production of $f_0(1710)$ decaying to a pair of kaons, and only weak evidence for $f_0(1370)$ production decaying to a pair of pions. However, we measured the production of a scalar meson which could have a gluonic contribution mixed in ($f_0(1500)$). Measured masses and widths of the $f_0(980)$ and $f_0(1500)$, together with extensive studies of the non-resonant "background", may provide constraints for better understanding the role of the gluonic component in the family of scalar mesons.  
\subsection*{Cross sections in fiducial region}
The mass spectrum of the $\pi^{+}\pi^{-}$ pairs shows an order-of-magnitude drop at 1.0~GeV, a clear peak around 1.3~GeV and possible further structures at higher masses. This is consistent with expectations for the DPE process and with experiments at lower energies investigating the production of $f_0(980)$, $f_2(1270)$ and also higher-mass resonances. 
The mass spectrum of $K^{+}K^{-}$ shows a clear peak above 1.5~GeV and possible enhancement around 1.3~GeV. This is also consistent with expectations for the DPE process and with experiments at lower energies investigating the production of the  $f_2(1270)$ and $f_2^\prime(1520)$, and possibly also $f_0(1710)$ resonances. The ratio of the measured cross sections for $\pi^+\pi^-$ to $K^+K^-$ production in the $f_2(1270)$ mass region is roughly 18, assuming a similar contribution from non-resonant production under the $f_2(1270)$ peak and similar STAR acceptance. This is consistent with the PDG ratio of $f_2(1270)$ branching fractions for decays into \mbox{$\pi^+\pi^-$ and $K^+K^-$ pairs.}\\
\indent
The mass spectrum of $\pi^{+}\pi^{-}$ shows that, for $\Delta\upvarphi<90^\circ$, the peak around the $f_2(1270)$ resonance is significantly suppressed while the peak at $f_0(980)$, as well as possible resonances in the mass range $1.3-1.5$ GeV, are enhanced compared to the region of $\Delta\upvarphi>90^\circ$. Such a correlation between resonances seen in the mass spectrum and the azimuthal angle between outgoing forward protons indicates factorisation breaking between the two proton vertices, reported for the first time by the WA91 experiment~\cite{wa91}. The present data do not show significant changes in the shape of the $\pi^+\pi^-$ mass spectrum as a function of $|\vec{p}_{1,T}^{\,\prime}-\vec{p}_{2,T}^{\,\prime}|$ after filtering events with $\Delta\upvarphi<90^\circ$. The mass spectra of the $K^{+}K^{-}$ and $p\bar{p}$ pairs also show some indications of factorisation breaking.\\
\indent
The $|t_1+t_2|$ distribution in the case of the $\pi^{+}\pi^{-}$ pair production in the $f_2(1270)$ mass region is less steep compared to other mass ranges, suggesting that the $t$-slope of $f_2(1270)$ production is smaller than that of other states. \\
\indent
The angular distributions of pions in the $\pi^+\pi^-$ rest frame indicate that, for invariant masses below 1~GeV, the data are dominated by spin-0 contributions. In the higher-mass region, the data show significant contributions from higher-spin states.\\
\indent 
The measured cross sections are compared to phenomenological predictions from DPE models implemented in the form of the Monte Carlo generators GenEx, DiMe and PYTHIA8 (MBR model). The cross sections for CEP of $\pi^{+}\pi^{-}$ and $K^{+}K^{-}$ pairs are significantly above the GenEx and DiMe predictions. This is expected, as neither of these models includes contributions from resonant production. The shapes of distributions other than the mass spectra show generally good agreement between the data and predictions, especially for DiMe. GenEx predicts shapes slightly different from the data for the $\Delta\upvarphi$, $\cos(\uptheta^\mathrm{CS})$ and $y(\pi^+\pi^-)$ distributions. This might be attributed to absorption effects, which are treated fully differentially in DiMe and only on average in GenEx.\\
\indent
The shapes of several distributions for $p\bar{p}$ production are reasonably well described by PYTHIA8 (MBR model) but the prediction overestimates the data by a factor 8. The limited statistics do not allow any significant conclusions about the expected resonances below $m(p\bar{p})\approx 2.5$ GeV.\\
\indent
The phenomenological interpretation of the data requires improvements of the DPE models to consistently include the continuum and resonance-production mechanisms, and the interference between the two, as well as absorption and rescattering effects.
\subsection*{Extrapolated differential cross sections} 
The fiducial measurement of $\pi^+\pi^-$ production was extrapolated to the  Lorentz-invariant phase space given by $|y(\pi^+\pi^-)|<0.4$ and $0.05\,\mbox{GeV}^2< -t_1,-t_2< 0.16$ GeV$^2$ in two ranges of $\Delta\upvarphi<45^\circ$ and $\Delta\upvarphi>135^\circ$. This allows us to fit the extrapolated differential cross section with a minimal model of the $\pi^+\pi^-$ invariant mass spectrum consisting of $f_0(980), f_2(1270)$ and $f_0(1500)$ resonances and direct non-resonant $\pi^+\pi^-$ production. The masses and widths of the $f_0(980)$ and $f_0(1500)$ resonances obtained from the fit are in good agreement with the PDG values.
The two scalar mesons, $f_0(980)$ and $f_0(1500)$, are predominantly \mbox{produced at} \mbox{$\Delta\upvarphi<45^\circ$}, whereas the tensor meson $f_2(1270)$ is predominantly produced at \mbox{$\Delta\upvarphi>135^\circ$}.
We observed weak evidence for an additional resonant state with a mass of $1372\pm 13$(stat.)~MeV and a width of $44\pm 24$(stat.)~MeV. This mass agrees with that of the $f_0(1370)$, but the width is much narrower than the PDG value of $200-500$ MeV.  The extrapolated cross sections of continuum production within the mass range $0.6\,\mbox{GeV} < m < 1.7\,\mbox{GeV}$ show an expected $\Delta\upvarphi$ asymmetry caused by absorption. 
Fits to an exponential function of the form $\propto \exp{\left[\beta t_{1}\right]}\exp{\left[\beta t_{2}\right]}$ were performed on the differential cross sections $d^2\sigma/dt_1dt_2$ to extract the slope of the $-t$ distributions. Variations in the slope with $m(\pi^{+}\pi^{-})$ and $\Delta\upvarphi$ can give important constraints for construction of phenomenological models of CEP of $\pi^+\pi^-$ pairs. 
\FloatBarrier
\acknowledgments
We thank the RHIC Operations Group and RCF at BNL, the NERSC Center at LBNL, and the Open Science Grid consortium for providing resources and support. This work was supported in part by the Office of Nuclear Physics within the U.S. DOE Office of Science, the U.S. National Science Foundation, the Ministry of Education and Science of the Russian Federation, National Natural Science Foundation of China, Chinese Academy of Science, the Ministry of Science and Technology of China and the Chinese Ministry of Education, the Higher Education Sprout Project by Ministry of Education at NCKU, the National Research Foundation of Korea, Czech Science Foundation and Ministry of Education, Youth and Sports of the Czech Republic, Hungarian National Research, Development and Innovation Office, New National Excellency Programme of the Hungarian Ministry of Human Capacities, Department of Atomic Energy and Department of Science and Technology of the Government of India, the National Science Centre of Poland, the Ministry of Science, Education and Sports of the Republic of Croatia, RosAtom of Russia and German Bundesministerium f\"ur Bildung, Wissenschaft, Forschung and Technologie (BMBF), Helmholtz Association, Ministry of Education, Culture, Sports, Science, and Technology (MEXT) and Japan Society for the Promotion of Science (JSPS).
%
%

\appendix

\setcounter{secnumdepth}{0}
\section{The STAR Collaboration}

\bfseries\raggedright\sffamily{
J.~Adam$^{6}$,
L.~Adamczyk$^{2}$,
J.~R.~Adams$^{39}$,
J.~K.~Adkins$^{30}$,
G.~Agakishiev$^{28}$,
M.~M.~Aggarwal$^{41}$,
Z.~Ahammed$^{61}$,
I.~Alekseev$^{3,35}$,
D.~M.~Anderson$^{55}$,
A.~Aparin$^{28}$
% ...
}

\afterAuthorSpace


\normalfont\small\itshape
\begin{list}{}{%
\setlength{\leftmargin}{0.28cm}%
\setlength{\labelsep}{0pt}%
\setlength{\itemsep}{\affiliationsSep}%
\setlength{\topsep}{-\parskip}}%
\item{$^{1}$Abilene Christian University, Abilene, Texas   79699}
\item{$^{2}$AGH University of Science and Technology, FPACS, Cracow 30-059, Poland}
\item{$^{3}$Alikhanov Institute for Theoretical and Experimental Physics NRC "Kurchatov Institute", Moscow 117218, Russia}
\item{$^{4}$Argonne National Laboratory, Argonne, Illinois 60439}
\item{$^{5}$American University of Cairo, New Cairo 11835, New Cairo, Egypt}
\item{$^{6}$Brookhaven National Laboratory, Upton, New York 11973}
\item{$^{7}$University of California, Berkeley, California 94720}
\item{$^{8}$University of California, Davis, California 95616}
\item{$^{9}$University of California, Los Angeles, California 90095}
\item{$^{10}$University of California, Riverside, California 92521}
\item{$^{11}$Central China Normal University, Wuhan, Hubei 430079 }
\item{$^{12}$University of Illinois at Chicago, Chicago, Illinois 60607}
\item{$^{13}$Creighton University, Omaha, Nebraska 68178}
\item{$^{14}$Czech Technical University in Prague, FNSPE, Prague 115 19, Czech Republic}
\item{$^{15}$Technische Universit\"at Darmstadt, Darmstadt 64289, Germany}
\item{$^{16}$ELTE E\"otv\"os Lor\'and University, Budapest, Hungary H-1117}
\item{$^{17}$Frankfurt Institute for Advanced Studies FIAS, Frankfurt 60438, Germany}
\item{$^{18}$Fudan University, Shanghai, 200433 }
\item{$^{19}$University of Heidelberg, Heidelberg 69120, Germany }
\item{$^{20}$University of Houston, Houston, Texas 77204}
\item{$^{21}$Huzhou University, Huzhou, Zhejiang  313000}
\item{$^{22}$Indian Institute of Science Education and Research (IISER), Berhampur 760010 , India}
\item{$^{23}$Indian Institute of Science Education and Research (IISER) Tirupati, Tirupati 517507, India}
\item{$^{24}$Indian Institute Technology, Patna, Bihar 801106, India}
\item{$^{25}$Indiana University, Bloomington, Indiana 47408}
\item{$^{26}$Institute of Modern Physics, Chinese Academy of Sciences, Lanzhou, Gansu 730000 }
\item{$^{27}$University of Jammu, Jammu 180001, India}
\item{$^{28}$Joint Institute for Nuclear Research, Dubna 141 980, Russia}
\item{$^{29}$Kent State University, Kent, Ohio 44242}
\item{$^{30}$University of Kentucky, Lexington, Kentucky 40506-0055}
\item{$^{31}$Lawrence Berkeley National Laboratory, Berkeley, California 94720}
\item{$^{32}$Lehigh University, Bethlehem, Pennsylvania 18015}
\item{$^{33}$Max-Planck-Institut f\"ur Physik, Munich 80805, Germany}
\item{$^{34}$Michigan State University, East Lansing, Michigan 48824}
\item{$^{35}$National Research Nuclear University MEPhI, Moscow 115409, Russia}
\item{$^{36}$National Institute of Science Education and Research, HBNI, Jatni 752050, India}
\item{$^{37}$National Cheng Kung University, Tainan 70101 }
\item{$^{38}$Nuclear Physics Institute of the CAS, Rez 250 68, Czech Republic}
\item{$^{39}$Ohio State University, Columbus, Ohio 43210}
\item{$^{40}$Institute of Nuclear Physics PAN, Cracow 31-342, Poland}
\item{$^{41}$Panjab University, Chandigarh 160014, India}
\item{$^{42}$Pennsylvania State University, University Park, Pennsylvania 16802}
\item{$^{43}$NRC "Kurchatov Institute", Institute of High Energy Physics, Protvino 142281, Russia}
\item{$^{44}$Purdue University, West Lafayette, Indiana 47907}
\item{$^{45}$Rice University, Houston, Texas 77251}
\item{$^{46}$Rutgers University, Piscataway, New Jersey 08854}
\item{$^{47}$Universidade de S\~ao Paulo, S\~ao Paulo, Brazil 05314-970}
\item{$^{48}$University of Science and Technology of China, Hefei, Anhui 230026}
\item{$^{49}$Shandong University, Qingdao, Shandong 266237}
\item{$^{50}$Shanghai Institute of Applied Physics, Chinese Academy of Sciences, Shanghai 201800}
\item{$^{51}$Southern Connecticut State University, New Haven, Connecticut 06515}
\item{$^{52}$State University of New York, Stony Brook, New York 11794}
\item{$^{53}$Instituto de Alta Investigaci\'on, Universidad de Tarapac\'a, Chile}
\item{$^{54}$Temple University, Philadelphia, Pennsylvania 19122}
\item{$^{55}$Texas A\&M University, College Station, Texas 77843}
\item{$^{56}$University of Texas, Austin, Texas 78712}
\item{$^{57}$Tsinghua University, Beijing 100084}
\item{$^{58}$University of Tsukuba, Tsukuba, Ibaraki 305-8571, Japan}
\item{$^{59}$United States Naval Academy, Annapolis, Maryland 21402}
\item{$^{60}$Valparaiso University, Valparaiso, Indiana 46383}
\item{$^{61}$Variable Energy Cyclotron Centre, Kolkata 700064, India}
\item{$^{62}$Warsaw University of Technology, Warsaw 00-661, Poland}
\item{$^{63}$Wayne State University, Detroit, Michigan 48201}
\item{$^{64}$Yale University, New Haven, Connecticut 06520}
\end{list}


\begin{thebibliography}{99}
\bibitem{brodsky} S.J. Brodsky, G.P. Lepage, {\it Large-angle two-photon exclusive channels in quantum chromodynamics}, \href{http://dx.doi.org/10.1103/PhysRevD.24.1808}{Phys. Rev. D 24 (1981) 1808}.
%
\bibitem{terazawa} H. Terazawa, {\it Pion-pair production by two photons}, \href{https://doi.org/10.1103/PhysRevD.51.R954}{Phys. Rev. D 51 (1995) 954-957}.
%
\bibitem{boyer} J. Boyer et al., {\it Two-photon production of pion pairs}, \href{http://dx.doi.org/10.1103/PhysRevD.42.1350}{Phys. Rev. D 42 (1990) 1350-1367}.
%
\bibitem{tpc} TPC/Two-Gamma Collaboration, H. Aihara et al., {\it Pion and kaon pair production in photon-photon collisions}, \href{http://dx.doi.org/10.1103/PhysRevLett.57.404}{Phys. Rev. Lett. 57 (1986) 404}.
%
\bibitem{topaz} TOPAZ Collaboration, I. Adachi et al., {\it A study of pion pair production in two-photon process}, \href{http://dx.doi.org/10.1016/0370-2693(90)92026-F}{Phys. Lett. B 234 (1990) 185}.
%
\bibitem{cleo} CLEO Collaboration, J. Dominick et al., {\it Two-photon production of charged pion and kaon pairs}, \href{http://dx.doi.org/10.1103/PhysRevD.50.3027}{Phys. Rev. D 50 (1994) 3027-3037}.
%
\bibitem{aleph} ALEPH Collaboration, A. Heister et al., {\it Exclusive production of pion and kaon meson pairs in two photon collisions at LEP}, \href{http://dx.doi.org/10.1016/j.physletb.2003.07.031}{Phys. Lett. B 569 (2003) 140-150}.
%
\bibitem{rho_STAR} STAR Collaboration, L. Adamczyk at al., {\it Coherent diffractive photoproduction of $\rho^0$ mesons on gold nuclei at 200 GeV/nucleon-pair at the Relativistic Heavy Ion Collider}, \href{https://doi.org/10.1103/PhysRevC.96.054904}{Phys. Rev. C 96 (2017) 054904}, \href{http://arxiv.org/pdf/1702.07705.pdf}{arXiv:1702.07705} [nucl-ex].
%
\bibitem{rho_CMS} CMS Collaboration, A.~M.~Sirunyan at al., {\it Measurement of exclusive $\rho(770)^0$ photoproduction in ultraperipheral pPb collisions at $\sqrt{s_\mathrm{NN}} =$ 5.02 TeV}, \href{https://doi.org/10.1140/epjc/s10052-019-7202-9}{Eur. Phys. J. C 79 (2019) 702 }, \href{http://arxiv.org/pdf/1902.01339.pdf}{arXiv:1902.01339} [hep-ex].
%
\bibitem{pho_1} H1 Collaboration, V. Andreev et al., {\it Exclusive $\rho^0$ meson photoproduction with a leading neutron at HERA}, \href{https://doi.org/10.1140/epjc/s10052-015-3863-1}{Eur. Phys. J. C 76 (2016) 41}, \href{http://arxiv.org/pdf/1508.03176.pdf}{arXiv:1508.03176} [hep-ex].
%
\bibitem{pho_2} ZEUS Collaboration, S. Chekanov et al., {\it Exclusive photoproduction of upsilon mesons at HERA}, \href{https://doi.org/10.1016/j.physletb.2009.07.066}{Phys. Lett. B 680 (2009) 4-12}, \href{http://arxiv.org/pdf/0903.4205.pdf}{arXiv:0903.4205} [hep-ex].
%
\bibitem{dis_1} ZEUS Collaboration, H. Abramowicz et al., {\it Production of exclusive dijets in diffractive deep inelastic scattering at HERA}, \href{https://doi.org/10.1140/epjc/s10052-012-1869-5}{Eur. Phys. J. C 76 (2016) 16}, \href{http://arxiv.org/abs/arXiv:1505.05783}{arXiv:1505.05783} [hep-ex].
%
\bibitem{dis_2} ZEUS Collaboration, H. Abramowicz et al., {\it Exclusive electroproduction of two pions at HERA}, \href{https://doi.org/10.1140/epjc/s10052-012-1869-5}{Eur. Phys. J. C 72 (2012) 1869}, \href{http://arxiv.org/abs/arXiv:1111.4905}{arXiv:1111.4905} [hep-ex].
%
\bibitem{book_1} S. Donnachie, H.G. Dosch, O. Nachtmann, P. Landshoff, {\it Pomeron physics and QCD}, Cambridge Univ. Press, 2002.
%
\bibitem{book_2} V. Barone and E. Predazzi, {\it High-Energy Particle Diffraction}, Springer, 2002.
%
\bibitem{book_3} P.D.B. Collins, {\it An Introduction to Regge Theory and High Energy Physics}, Cambridge Univ. Press, 2009.
%
\bibitem{albrow_1} M.G. Albrow, T.D. Coughlin, J.R. Forshaw, {\it Central exclusive particle production at high energy hadron colliders}, \href{https://doi.org/10.1016/j.ppnp.2010.06.001}{Prog. Part. Nucl. Phys. 65 (2010) 149}, \href{http://arxiv.org/abs/arXiv:1006.1289}{arXiv:1006.1289} [hep-ph].
%
\bibitem{harland_lang_1} L.A. Harland-Lang, V.A. Khoze, M.G. Ryskin, W.J. Stirling, {\it The phenomenology of central exclusive production at hadron colliders}, \href{https://doi.org/10.1140/epjc/s10052-012-2110-2}{Eur. Phys. J. C 72 (2012) 2110},  \href{http://arxiv.org/abs/arXiv:1204.4803}{arXiv:1204.4803} [hep-ph].
%
\bibitem{dime} L.A. Harland-Lang, V.A. Khoze, M.G. Ryskin, {\it Modelling exclusive meson pair production at hadron colliders}, \href{https://doi.org/10.1140/epjc/s10052-014-2848-9}{Eur. Phys. J. C 74 (2014) 2848},  \href{http://arxiv.org/abs/arXiv:1312.4553}{arXiv:1312.4553} [hep-ph].
%
\bibitem{CEPatLHC} J.~R.~Forshaw, {\it Central Exclusive Production at the LHC}, \href{https://doi.org/10.1016/j.nuclphysbps.2009.03.132}{Nucl.\ Phys.\ B\ Proc.\ Suppl.\ 191 (2009) 247}, \href{https://arxiv.org/abs/0901.3040}{arXiv:0901.3040}
[hep-ph].
%
\bibitem{cms_pipi} CMS Collaboration, V. Khachatryan et al., {\it Exclusive and semi-exclusive $\pi^+\pi^-$ production in proton-proton collisions at $\sqrt{s}=7$ TeV}, \href{http://arxiv.org/abs/arXiv:1706.08310}{arXiv:1706.08310} [hep-ex].
%
\bibitem{cms_pipi_2} CMS Collaboration, A.M. Sirunyan et al., {\it Study of central exclusive $\pi^+\pi^-$ production in proton-proton collisions at $\sqrt{s}=5.02$ and $13$~TeV}, \href{https://arxiv.org/abs/2003.02811}{arXiv:2003.02811} [hep-ex].
%
\bibitem{lhcb} LHCb Collaboration, R. Aaij et al., {\it Central exclusive production of $J/\psi$ and $\psi(2S)$ mesons in pp collisions at $\sqrt{s} =13$ TeV}, \href{https://doi.org/10.1007/JHEP10(2018)167}{JHEP 10 (2018) 167}, \href{http://arxiv.org/abs/arXiv:1806.04079}{arXiv:1806.04079} [hep-ex].
%
\bibitem{lhcb_2} LHCb Collaboration, R. Aaij et al., {\it Measurement of the exclusive $\Upsilon$ production cross-section in pp collisions at $ \sqrt{s}=7 $ TeV and 8 TeV}, \href{https://doi.org/10.1007/JHEP09(2015)084}{JHEP 09 (2015) 84}, \href{http://arxiv.org/abs/arXiv:1505.08139}{arXiv:1505.08139} [hep-ex].
%
\bibitem{cdf} CDF Collaboration, T.A. Aaltonen et al., {\it Measurement of central exclusive $\pi^+\pi^-$ production in $p\bar{p}$ collisions at $\sqrt{s}=0.9$ and $1.96$ TeV at CDF}, \href{https://doi.org/10.1103/PhysRevD.91.091101}{Phys. Rev. D 91 (2015) 091101},  \href{http://arxiv.org/abs/arXiv:1502.01391}{arXiv:1502.01391} [hep-ex].
%
\bibitem{afs} AFS Collaboration, T. Akesson et al., {\it A Search for Glueballs and a Study of Double Pomeron Exchange at the CERN Interacting Storage Rings}, \href{http://dx.doi.org/10.1016/0550-3213(86)90477-3}{Nucl. Phys. B 264 (1986) 154}.
%
\bibitem{wa91} WA91 Collaboration, F. Antinori et al., {\it A further study of the centrally produced $\pi^+\pi^-$ and $\pi^+ \pi^- \pi^+ \pi^-$ channels in $pp$ interactions at $300$ GeV/c and $450$ GeV/c}, \href{http://dx.doi.org/10.1016/0370-2693(95)00629-Y}{Phys. Lett. B 353 (1995) 589}.
%
%
\bibitem{lebiedowicz_2} P. Lebiedowicz, A. Szczurek, {\it Exclusive $pp\to pp\pi^+\pi^-$ reaction: From the threshold to LHC}, \href{http://dx.doi.org/10.1103/PhysRevD.81.036003}{Phys. Rev. D 81 (2010) 036003}, \href{http://arxiv.org/abs/arXiv:0912.0190}{arXiv:0912.0190} [hep-ph].
%
\bibitem{LSModelKK} P. Lebiedowicz, A. Szczurek, {\it $pp \to pp K^{+}K^{-}$ reaction at high energies}, \href{https://doi.org/10.1103/PhysRevD.85.014026}{Phys. Rev. D 85 (2012) 014026}, \href{http://arxiv.org/pdf/1110.4787.pdf}{arXiv:1110.4787} [hep-ph].
%
\bibitem{LSAbsorption} P. Lebiedowicz, A. Szczurek, {\it Revised model of absorption corrections for the $pp \to pp \pi^{+} \pi^{-}$ process}, \href{https://doi.org/10.1103/10.1103/PhysRevD.92.054001}{Phys. Rev. D 92 (2015) 054001}, \href{http://arxiv.org/pdf/1504.07560.pdf}{arXiv:1504.07560} [hep-ph].
%
\bibitem{lebiedowicz_4} P. Lebiedowicz, O. Nachtmann, A. Szczurek, {\it Towards a complete study of central exclusive production of $K^+K^-$ pairs in proton-proton collisions within the tensor Pomeron approach},  \href{https://doi.org/10.1103/PhysRevD.98.014001}{Phys. Rev. D 98 (2018) 014001}, \href{http://arxiv.org/abs/arXiv:1804.04706}{arXiv:1804.04706} [hep-ph].
%
\bibitem{lebiedowicz_5} P. Lebiedowicz, O. Nachtmann, A. Szczurek, {\it Central exclusive diffractive production of $p\bar{p}$ pairs in proton-proton collisions at high energies}, \href{https://doi.org/10.1103/PhysRevD.97.094027}{Phys. Rev. D 97 (2018), 094027}, \href{http://arxiv.org/abs/arXiv:1801.03902}{arXiv:1801.03902} [hep-ph].
%
\bibitem{lebiedowicz_1} P. Lebiedowicz, O. Nachtmann, A. Szczurek, {\it Central exclusive diffractive production of $\pi^+\pi^-$ continuum, scalar and tensor resonances in $pp$ and $p\bar{p}$ scattering within tensor pomeron approach}, \href{http://dx.doi.org/10.1103/PhysRevD.93.054015}{Phys. Rev. D 93 (2016) 054015},  \href{http://arxiv.org/abs/arXiv:1601.04537}{arXiv:1601.04537} [hep-ph].
%
\bibitem{schiecker} R. Fiore, L. Jenkovszky, R. Schicker, {\it Exclusive diffractive resonance production in proton-proton collisions at high energies}, \href{https://doi.org/10.1140/epjc/s10052-018-5907-9}{Eur. Phys.J. C 78 (2018) 468}, \href{http://arxiv.org/pdf/1711.08353.pdf}{arXiv:1711.08353} [hep-ph].
%
\bibitem{gpd} M. Diehl, {\it Generalized parton distributions}, \href{https://doi.org/10.1016/j.physrep.2003.08.002}{Phys. Rept. 388 (2003) 41-277}, \href{http://arxiv.org/abs/hep-ph/0307382}{arXiv:hep-ph/0307382}.
%
\bibitem{updf} A.D. Martin and M.G. Ryskin, {\it Unintegrated generalized parton distributions}, \href{https://doi.org/10.1103/PhysRevD.64.094017}{Phys. Rev. D 64 (2001) 094017}, \href{http://arxiv.org/abs/hep-ph/0107149}{arXiv:hep-ph/0107149}.
%
\bibitem{glueball_2} W. Ochs, {\it The status of glueballs}, \href{https://doi.org/10.1088/0954-3899/40/4/043001}{J. Phys. G 40 (2013) 043001},  \href{http://arxiv.org/abs/arXiv:1301.5183}{arXiv:1301.5183} [hep-ph].
%
\bibitem{glueball_3} Crystal Barrel Collaboration, C. Amsler et al., {\it Proton-antiproton annihilation into eta eta pi: Observation of a scalar resonance decaying into eta eta}, \href{http://dx.doi.org/10.1016/0370-2693(92)91057-G}{Phys. Lett. B 291 (1992) 347}.
%
\bibitem{glueball_4} U. Wiedner, {\it Future prospects for hadron physics at PANDA}, \href{https://doi.org/10.1016/j.ppnp.2011.04.001}{Prog. Part. Nucl. Phys. 66 (2011) 477-518}, \href{http://arxiv.org/abs/arXiv:1104.3961}{arXiv:1104.3961} [hep-ex].
%
\bibitem{glueball_5} MARK-III Collaboration, R.M. Baltrusaitis et al., {\it Study of the radiative decay $J/\Psi\to\gamma\rho\rho$}, \href{http://dx.doi.org/10.1103/PhysRevD.33.1222}{Phys. Rev. D 33 (1986) 1222}.
%
\bibitem{harland_lang_3} L.A. Harland-Lang, V.A. Khoze, M.G. Ryskin, W.J. Stirling, {\it Central exclusive production within the Durham model: a review}, \href{http://dx.doi.org/10.1142/S0217751X14300312}{Int. J. Mod. Phys. A 29 (2014) 1430031},  \href{http://arxiv.org/abs/arXiv:1405.0018}{arXiv:1405.0018} [hep-ph].
%
\bibitem{rhic} H. Hahn et al., {\it The RHIC design overview}, \href{https://doi.org/10.1016/S0168-9002(02)01938-1}{Nucl. Instrum. Meth. A 499 (2003) 245}.
%
\bibitem{star} STAR Collaboration, K.H. Ackermann et al., {\it STAR detector overview}, \href{https://doi.org/10.1016/S0168-9002(02)01960-5}{Nucl. Instrum. Meth. A 499 (2003) 624}.
%
\bibitem{pp2pp} S. B\"ultmann et al., {\it The PP2PP experiment at RHIC: silicon detectors installed in Roman Pots for forward proton detection close to the beam}, \href{https://doi.org/10.1016/j.nima.2004.07.162}{Nucl. Instrum. Meth. A 535 (2004) 415-420}.
%
\bibitem{star_tpc} M. Anderson et al., {\it The STAR time projection chamber: a unique tool for studying high multiplicity events at RHIC}, \href{https://doi.org/10.1016/S0168-9002(02)01964-2}{Nucl. Instrum. Meth. A 499 (2003) 659}.
%
\bibitem{star_tof} J. Wu, M. Xu, {\it A barrel TOF for STAR at RHIC}, \href{https://doi.org/10.1088/0954-3899/34/8/S83}{J. Phys. G 34 (2007) S729-S732}.
%
\bibitem{hft} G.~Contin, {\it The STAR Heavy Flavor Tracker and Upgrade Plan}, \href{https://doi.org/10.1016/j.nuclphysa.2016.02.064}{Nucl. Phys. A 956 (2016) 858-861}.
%
\bibitem{star_bbc} J. Kiryluk, {\it Relative Luminosity Measurement in STAR and Implications for Spin Asymmetry Determinations}, 
\href{https://doi.org/10.1063/1.1607171}{AIP Conf. Proc. 675 (2003) 424}.
%
\bibitem{star_zdc} C. Adler et. al., {\it The RHIC zero degree calorimeters}, 
\href{https://doi.org/10.1016/S0168-9002(01)00627-1}{Nucl. Instrum. Meth. A 470 (2001) 488}, \href{https://arxiv.org/pdf/nucl-ex/0008005}{arXiv:nucl-ex/0008005}.
%
\bibitem{elastic_paper} STAR Collaboration, J. Adam et al., {\it Results on Total and Elastic Cross Sections in Proton-Proton Collisions at $\sqrt{s} = 200$ GeV}, \href{https://arxiv.org/pdf/2003.12136}{arXiv:2003.12136} [hep-ex].
%
\bibitem{Bichsel} H.~Bichsel, {\it A method to improve tracking and particle identification in TPCs and silicon detectors}, \href{https://doi.org/10.1016/j.nima.2006.03.009}{Nucl. Instrum. Meth. A 562 (2006) 154}.
%
\bibitem{genex} R.A. Kycia, J. Chwastowski, R. Staszewski, J. Turnau, {\it GenEx: A simple generator structure for exclusive processes in high energy collisions}, \href{https://doi.org/10.4208/cicp.OA-2017-0105}{Commun. Comput. Phys. 24 (2018) 860}, \href{https://arxiv.org/pdf/1411.6035.pdf}{arXiv:1411.6035}~[hep-ph].
%
\bibitem{mbr_pythia8} R. Ciesielski, K. Goulianos, {\it MBR Monte Carlo Simulation in PYTHIA8}, \href{https://arxiv.org/abs/1205.1446}{arXiv:1205.1446} [hep-ph].
%
\bibitem{pythia8}  T. Sj\"ostrand et al., {\it An Introduction to PYTHIA 8.2}, \href{https://doi.org/10.1016/j.cpc.2015.01.024}{Comp. Phys. Commun. 191 (2015) 159-177}, \href{https://arxiv.org/abs/1410.3012}{arXiv:1410.3012} [hep-ph].
%
\bibitem{geant3} R. Brun et al., {\it GEANT Detector Description and Simulation Tool}, \href{https://doi.org/10.17181/CERN.MUHF.DMJ1}{CERN-W5013}.
%
\bibitem{geant4} GEANT4 Collaboration, S. Agostinelli et al., {\it GEANT4: A Simulation toolkit}, \href{https://doi.org/10.1016/S0168-9002(03)01368-8}{Nucl. Instrum. Meth. A 506 (2003) 250}.
%
\bibitem{cs_frame} J.C. Collins, D.E. Soper, {\it Angular distribution of dileptons in high-energy hadron collisions}, \href{https://doi.org/10.1103/PhysRevD.16.2219}{Phys. Rev. D 16 (1977) 2219}.
%
\bibitem{van_der_meer} S. van der Meer, {\it Calibration of the effective beam height in the ISR}, \href{https://cdsweb.cern.ch/record/296752/files/196800064.pdf}{CERN-ISR-PO-68-31}.
%
%
%
%
%
%
\bibitem{pdg} Particle Data Group, M. Tanabashi et al., {\it Review of Particle Physics}, \href{https://doi.org/10.1103/PhysRevD.98.030001}{Phys. Rev. D 98 (2018) 030001}.
%
\bibitem{close} F.E. Close, A. Kirk, {\it Glueball-$q\bar{q}$ filter in central hadron production}, \href{https://doi.org/10.1016/S0370-2693(97)00222-0}{Phys. Lett. B 397 (1997) 333-338}, \href{http://arxiv.org/abs/hep-ph/9701222}{arXiv:hep-ph/9701222}.
%
\bibitem{lebiedowicz_3} P. Lebiedowicz, O. Nachtmann, A. Szczurek, {\it Extracting the pomeron-pomeron-$f_2(1270)$ coupling in the $pp\to pp\pi^+\pi^-$ reaction through the angular distributions of the pions}, \href{https://doi.org/10.1103/PhysRevD.101.034008}{Phys. Rev. D 101 (2020) 034008}, \href{http://arxiv.org/pdf/1901.07788.pdf}{arXiv:1901.07788} [hep-ph].
%
\bibitem{GJ} K. Gottfried, J.D. Jackson, {\it On the connection between production mechanism and decay of resonances at high energies}, \href{https://doi.org/10.1007/BF02750195}{Nuovo Cim. 33 (1964) 309-330}.
%
%
\bibitem{BarrierFactors1} J.M.~Blatt, V.F.~Weisskopf, {\it Theoretical nuclear physics}, John Wiley \& Sons, 1952.
%
\bibitem{BarrierFactors} F.~von Hippel, C.~Quigg, {\it Centrifugal-Barrier Effects in Resonance Partial Decay Widths, Shapes, and Production Amplitudes}, \href{https://doi.org/10.1103/PhysRevD.5.624}{Phys. Rev. D 5 (1972) 624-638}.
%
\bibitem{Flatte} S.M.~Flatt\'e, {\it Coupled-channel analysis of the $\pi\eta$ and $K\bar{K}$ systems near $K\bar{K}$ threshold}, \href{https://doi.org/10.1016/0370-2693(76)90654-7}{Phys. Lett. B 63 (1976) 224-227}.
%
\bibitem{BES_JPsi} BES Collaboration, M.~Ablikim et al., {\it Resonances in $J/\psi\rightarrow\phi\pi^{+}\pi^{-}$ and $\phi K^{+}K^{-}$}, \href{https://doi.org/10.1016/j.physletb.2004.12.041}{Phys. Lett. B 607 (2005) 243-253}, \href{http://arxiv.org/abs/hep-ex/0411001}{arXiv:hep-ex/0411001}.
%
\bibitem{Minuit2} M.~Hatlo, F.~James, P.~Mato, L.~Moneta, M.~Winkler, A.~Zsenei, {\it Developments of mathematical software libraries for the LHC experiments}, \href{https://doi.org/10.1109/TNS.2005.860152}{IEEE Trans. Nucl. Sci. 52 (2005) 2818-2822}.
%
\bibitem{ROOT} R.~Brun, F.~Rademakers, {\it ROOT - An object oriented data analysis framework}, \href{https://doi.org/10.1016/S0168-9002(97)00048-X}{Nucl. Instrum. Meth. A 389 (1997) 81-86}.
%
\bibitem{WA102} WA102 Collaboration, D.~Barberis et al., {\it Experimental evidence for a vector-like behaviour of Pomeron exchange}, \href{https://doi.org/10.1016/S0370-2693(99)01186-7}{Phys. Lett. B 467 (1999) 165-170}, \href{http://arxiv.org/abs/hep-ex/9909013}{arXiv:hep-ex/9909013}.
%
\end{thebibliography}
\end{document}